\documentclass[12pt,a4paper,twoside,openright]{report}
\usepackage{graphicx}
\usepackage{subfigure}
\usepackage{amssymb}
\usepackage{aastex_hack}
\usepackage{deluxetable}
\usepackage{rotating}
\usepackage{fancyhdr} 
\usepackage{amsmath}
\usepackage[usenames, dvipsnames]{xcolor}
\usepackage{gensymb} 
\usepackage{cases}

\usepackage[left=3cm,right=3cm,top=3cm,bottom=3cm]{geometry}

\bibstyle{aa}

\begin{document}

\title{
{Initial Condition of the Inflationary Universe and Its Imprint on the Cosmic Microwave Background} \\
{\large Department of Physics} \\[-7mm]
{\large College of Science} \\[-7mm]
{\large National Taiwan University} \\[-7mm]
{\large Doctoral Dissertation}
}
\author{
{Yu-Hsiang Lin} \\
{Advisor: Pisin Chen, Ph.D.}
}

\date{June 2017}

\maketitle

\pagenumbering{roman}

\addcontentsline{toc}{figure}{\bf Acknowledgments}
\chapter*{\Huge Acknowledgments\\}

	I want to thank Prof.~Pisin Chen for bringing me into the field of physics research, giving me the environment and opportunity to fully dedicate myself into physics. His support started from the time before I entered the physics department, all the way to the time I'm leaving. I also have my deepest gratitude to my mother and my father. You are very successful parents who help your son grow into a man who can face the challenges in life and pass them. I love you. To my partner and my guardian, the most beautiful Ting-An: Thank you for sharing the life with me.
	
	I thank Prof.~Jiunn-Wei Chen, Prof.~Yu-tin Huang, Dr.~Dong-han Yeom, and Dr.~Frederico Arroja for being my thesis examination committee and their effort. I would also like to thank: Dr.~Mariam Bouhmadi-L\'{o}pez and Yu-Chien Huang for giving me the opportunity to collaborate; Prof.~Keisuke Izumi and Dr.~Lance Labun for their support and many discussions, and being models of good researchers for me; Dr.~Frederico Arroja for his careful reading and thoughtful comments, from which I benefited a lot; Dr.~Dong-han Yeom for working with me and his many interesting ideas; Prof.~Teruaki Suyama, Prof.~Pei-Ming Ho, Dr.~Je-An Gu, Dr.~Christopher Gauthier, Dr.~Sean Downes, and Dr.~Jinn-Ouk Gong, for their discussion and help; Prof.~Jiwoo Nam for his teaching in physics and Korean; Prof.~Tom Abel and Prof.~Greg Madejski for having me at the KIPAC family in SLAC; Prof.~Leonardo Senatore, Prof.~Stephen Shenker, Prof.~Eva Silverstein, and Prof.~Andrei Linde for their discussion at SITP, Stanford University; Prof.~Hitoshi Murayama, Prof.~Yasunori Nomura, and Dr.~Keisuke Harigaya for their help and discussion at Berkeley. Special thanks go to all friends, faculties, and secretaries in LeCosPA, NTU, for helping me and supporting me all the time. The thesis and related works are in addition supported by Taiwan Ministry of Science and Technology, as well as Taiwan National Center for Theoretical Sciences.
	
	It has been a precious and wonderful journey through the ocean of life for me. Prof.~Cha-Ray Chu once said, ``Artistic creation is a constant progression. A creator must cherish her own work; even if it is not perfect, it is what she devoted herself to.'' I am grateful for being lucky to have the opportunity of creating something, and being able to experience what she said. As American writer Elizabeth Gilbert put it in her book, Big Magic, ``you must stubbornly walk into that room, regardless, and you must hold your head high. You make it; you get to put it out there. Never apologize for it, never explain it away, never be ashamed of it. You did your best with what you knew, and you worked with what you had, in the time that you were given. You were invited, and you showed up, and you simply cannot do more than that.'' The spirit of the Sand Hill Road days will always be a part of me.

\clearpage

\newpage
\addcontentsline{toc}{figure}{\bf Abstract}
\chapter*{Abstract}

	There is an apparent power deficit relative to the $\Lambda$CDM prediction of the cosmic microwave background spectrum at large scales, which, though not yet statistically significant, persists from WMAP to Planck data. It is well-motivated to consider such power suppression as the imprint of the preinflationary era. The observations show that about the last 60 e-folds of inflation operates at the energy scale of $10^{16}$ GeV. If at the higher energy scales the matter content is in a different phase, or that the gravity is modified, the evolution of the geometry in the preinflationary universe may not be the de Sitter expansion in a spatially flat universe, as it is deep in the inflationary era. In the scenario of ``just-enough'' inflation, the perturbations at the largest scales we observe today exit the horizon around the transition time from preinflation to inflation era, hence carrying the characteristics of the preinflationary universe. The large-scale spectrum may therefore serve as the window to peek into the preinflationary universe.
	
	We first present a simple toy model corresponding to a network of frustrated topological defects of domain walls or cosmic strings that exist previous to the standard slow-roll inflationary era of the universe. If such a network corresponds to a network of frustrated domain walls, it produces an earlier inflationary era that expands more slowly than the standard one does. On the other hand, if the network corresponds to a network of frustrated cosmic strings, the preinflationary universe would expand at a constant speed. Those features are phenomenologically modeled by a Chaplygin gas that can interpolate between a network of frustrated topological defects and a de Sitter--like or a power-law inflationary era. We show that these scenarios can alleviate the quadrupole anomaly of the cosmic microwave background spectrum, based on the approximate initial conditions for the long-wavelength perturbations. A more thorough and systematic analysis on the initial vacuum carried out later will show that the preinflationary domain wall dominated era has a different vacuum state from the approximate one and does not suppress the long-wavelength spectrum.

	We then go further to show that the large-scale spectrum at the end of inflation reflects the super-horizon spectrum of the initial state of the inflaton field. By studying the curvature perturbations of a scalar field in the Friedmann-Lema\^{i}tre-Robertson-Walker universe parameterized by the equation of state parameter $w$, we find that the large-scale spectrum at the end of inflation reflects the superhorizon spectrum of the initial state. The large-scale spectrum is suppressed if the universe begins with the adiabatic vacuum in a superinflation ($w < -1$) or positive-pressure ($w > 0$) era. In the latter case, there is however no causal mechanism to establish the initial adiabatic vacuum. On the other hand, as long as the universe begins with the adiabatic vacuum in an era with $-1 < w < 0$, even if there exists an intermediate positive-pressure era, the large-scale spectrum would be enhanced rather than suppressed. We further calculate the spectrum of a two-stage inflation model with a two-field potential and show that the result agrees with that obtained from the ad hoc single-field analysis.

	Neither of the two possibilities discovered earlier---the preinflationary superinflation and positive-pressure eras---that attempt to account for the power suppression is completely satisfactory as a realistic initial condition of the inflationary universe. This difficulty may be a hint that the origin of the power suppression does not lie in the semi-classical physics, but in the quantum theory of gravity. We consider the Hartle-Hawking no-boundary wave function, which is a solution to the Wheeler-DeWitt equation, as the initial condition of the universe. We find that the power suppression can be the consequence of a massive inflaton, whose initial vacuum is the Euclidean instanton in a compact manifold. We calculate the primordial power spectrum of the perturbations and show that, as long as the scalar field is moderately massive, the power spectrum is suppressed at the long-wavelength scales.

\vskip 2\baselineskip

\textit{Keywords} -- Cosmic microwave background, large-scale power suppression, topological defect, domain wall, cosmic string, Chaplygin gas, cosmological perturbation theory, adiabatic vacuum, inflation, initial condition, spectrum evolution, superinflation, two-field inflation, Hartle-Hawking no-boundary wave function, Wheeler-DeWitt equation, minisuperspace model.

\clearpage

\tableofcontents
\addcontentsline{toc}{figure}{\bf List of Figures}
\listoffigures
\addcontentsline{toc}{table}{\bf List of Tables}
\listoftables


\clearpage
\pagenumbering{arabic}


\chapter{Introduction}{\label{ch:Introduction}}

	The $\Lambda$CDM model of cosmology with an early inflationary era is very successful in explaining the cosmic microwave background (CMB) power spectrum. However, it has been observed in the \emph{COBE} data that the quadrupole power is lower than the model prediction \cite{Smoot1992, Tegmark1996}. This observation is further confirmed by \emph{WMAP}, reporting the quadrupole power lower than the theoretical expectation by more than $1\sigma$ but less than $2\sigma$ \cite{Bennett2013}. Although this stand-alone low quadrupole mode may be explained by the cosmic variance, the \emph{Planck} observation analyzed the low-$\ell$ ($\ell < 30$) and high-$\ell$ ($\ell \geq 30$) spectra separately, and showed that the best-fit amplitude for the low-$\ell$ spectrum is 10\% lower than that for the high-$\ell$ one at 2.5--3$\sigma$ significance \cite{Planck2013XV, Planck2016XI}.%
		\footnote{This low-$\ell$/high-$\ell$ tension is present even when the particularly low quadrupole mode is excluded from the analysis \cite{Planck2013XV}. In \cite{Planck2016XI} it is further pointed out that the low-$\ell$ power deficit is mainly caused by the low multipoles between $\ell = 20$ and 30.}

	The CMB quadruple originates from the lowest modes of the primordial power spectrum with comoving wave number $k$ of the order of $10^{-3} \, \textrm{Mpc}^{-1}$. These lowest modes are the first modes that exit the horizon during inflation and the last ones that reenter in the radiation, matter, or dark energy dominated eras. Consequently, we expect that these modes are heavily affected by the physics of the very early universe, possibly the physics prior to the slow-roll inflationary era. Based on this reasoning, there have been attempts to explain the low-$\ell$ power suppression of the CMB by introducing some preinflation era that breaks the slow-roll condition at about 60 e-folds before the end of inflation \cite{Contaldi2003, Wang2008, Scardigli2011, Bouhmadi-Lopez2013, Kouwn2015, Jain2009, Dudas2012, Namjoo2012, White2014, Biswas2014}. The basic argument about how an era that deviates from the slow-roll dynamics could suppress the power is that the amplitude of the curvature perturbation, $\mathcal{R} \sim H \delta \phi / \dot{\phi}$, would decrease as $|\dot{\phi}|$ increases. This scenario is first realized in the single-field chaotic inflation with potential $V = m^2 \phi^2 / 2$, where $m$ is the mass of the inflaton. If the inflaton $\phi$ starts with large speed $\dot{\phi}^2 \gg m^2 \phi^2$, the kinetic energy dominates the preinflation universe, and the power at the horizon scale is suppressed \cite{Contaldi2003}. Other scenarios of violating the slow-roll evolution include the preinflation era filled with some radiation \cite{Wang2008}, the primordial black hole remnants \cite{Scardigli2011}, or the frustrated network of topological defects \cite{Bouhmadi-Lopez2013}. There are also preinflation models in which the universe is dominated by the spatial curvature as the emergent property from a number of moduli fields in the models of solid inflation \cite{Endlich2013, Kouwn2015}. All of the models above report power suppression at the large scales.
	
	On the other hand, the existence of the preinflation decelerating era in models that predict multi-stage inflation does not always result in power suppression \cite{Polarski1992, Ashoorioon2006, Ashoorioon2009, Jain2009, Yamauchi2011, Dudas2012, Namjoo2012, White2014}. In the presence of two fields with mass hierarchy, there are two inflationary eras connected by a decelerating era. With the second inflationary era identified as the last 60 e-folds of the inflation, it is shown that the power is enhanced, rather than suppressed, at large scales that cross the horizon during the first inflationary era or the decelerating era \cite{Polarski1992}. Similar evolution also occurs in the early times of the hybrid inflation \cite{Linde1994}, in which the heavy field is played by the ``waterfall field'' who acquires the mass through the coupling to the inflaton field. In this case, it is however inferred that when the coupling term dominates at the early times, the inflaton field rolls faster due to the coupling, and eventually leads to the power suppression \cite{Contaldi2003}. Among other models of multi-stage inflation which commonly have a decelerating era before the last inflationary era, some predict power suppression at large scales \cite{Jain2009, Dudas2012, White2014}, while some predict enhancement \cite{Polarski1992, Ashoorioon2006, Ashoorioon2009, Yamauchi2011, Namjoo2012}. One therefore naturally wonders:~What initial condition of inflation generated by the preinflation era would actually suppress the CMB power spectrum at large scales?


\section{Topological Defects}

	As a toy model to study the preinflation era, we propose a new cosmological period just before the slow-roll inflationary era. This period corresponds to an era described by a network of topological defects, which we will assume to be frustrated domain walls or cosmic strings \cite{Bucher1999, Battye1999, Spergel1997, Lopez2002, Lopez2002a}. We expect that the production of topological defects at those scales follows the prediction of high energy physics. In addition, given that the topological defects era precedes the slow-roll inflationary era, the topological networks will affect mainly the lower modes of the power spectrum of the scalar and tensorial perturbations. Afterwards, they will soon be diluted during most of the inflationary era.
	
	Topological defects such as domain walls or cosmic strings can naturally arise in the history of the universe. In a system that has spontaneous symmetry breaking, the symmetry that is broken at low temperature is restored at high temperature. The standard model of the elementary particles has such scenarios including the electro-weak and GUT (grand unified theory) scale symmetry breaking. In the early universe, when the temperature is higher than the symmetry breaking scale, the field is in a global minimum (with quantum fluctuation around the minimum). As the universe expands, the temperature will eventually drop below the critical temperature associated with the symmetry. In the low temperature, the potential of the field has multiple degenerate vacua, and the field will fall down to one of the vacua randomly. Across the space in the universe, there are therefore many spatial regions in which the field drops to different vacua. Two regions in different vacua are separated by a domain wall, usually in the case of the discrete symmetry. Cosmic strings can arise from the breaking of a local $U(1)$ symmetry in the Abelian Higgs model, which allows the solution of vortex lines \cite{Vilenkin2000}.
	
	Topological defects are interesting subjects in cosmology for several reasons. An early one is that it could possibly seed the large-scale structure we observe today. Planar domain walls, for example, repel the baryonic matter and induce inhomogeneities \cite{Kubotani1992}. However, if the domain walls are produced too early in the history of the universe, their energy density may dominate over the other radiation and baryonic matter sources and significantly alter the observed expansion history. On the other hand, if the domain walls are created at lower energy scales, while their energy density will be subdominant, the kinetic energy of them may still spoil the isotropy of the CMB we observe today.
	
	The network of frustrated domain walls and cosmic strings are considered to be viable components of the universe because they will not destroy the isotropy of CMB \cite{Nambu1991}, nor cluster at the small scale \cite{Bucher1999}. This is because the network structure makes the domain walls behave like some kind of solid, whose elastic resistance is only shear deformations. It is hard to have large kinetic movement in orders higher than the bulk velocity. Moreover, due to their topological nature, the pressure of such networks of topological defects is negative and of the same order of their energy density, so their speed of sound is close to the speed of light. As a consequence, their Jeans length is comparable to the size of the horizon, and they does not collapse at small scales. One example of the network of domain walls joined by cosmic strings is given by the complex $U(1)$ scalar field, such as axions \cite{Kubotani1992}. In this type of simple models the string and anti-directed string will soon annihilate each other within the Hubble radius, therefore unable to form the stable network structure. To maintain a network structure, one can instead consider, for example, the $O(N)$ field \cite{Kubotani1992}, which has different types of cosmic strings. In this scenario, the probability of pair-annihilations of the cosmic strings are suppressed because the same type of strings collide less frequently.


\section{Relation between initial state and power spectrum}

	One of our most important findings is that, in general, the long-wavelength spectrum at the end of inflation reflects the super-horizon spectrum of the initial state \cite{Chen2016}. To establish such relationship between the initial state and the power spectrum, we first find the spectrum of the adiabatic vacuum in the universe with a constant equation of state driven by a scalar field. The conditions of having the blue-tilted, red-tilted, or scale-invariant spectra at the super-horizon scales are found. The spectra obtained are based on the assumption that the mode solutions approach the Minkowski limit at small scales. We point out that in the decelerating universe the super-horizon modes would enter the horizon and become sub-horizon, which means that these super-horizon modes are initially causally disconnected. Such assumption in an initially decelerating universe therefore relies on the final state of the mode evolution to govern its initial state, which reverses the cause and effect. Later it was shown that the large-scale power suppression in models with preinflation decelerating era is actually a consequence of this unnatural yet widely adopted assumption.%
		\footnote{Such a choice of the initial state for inflation, often referred to as the Bunch-Davies vacuum, even if there exists a non-slow-roll preinflation phase, has been commonly assumed in the literature (see, for example, \cite{Contaldi2003, Ashoorioon2006, Wang2008, Namjoo2012, Bouhmadi-Lopez2013, Biswas2014, Kouwn2015}). There also exist numerous proposals of a non-Bunch-Davies vacuum as the basis of the initial condition for a universe that does not begin with a slow-roll phase (see, for example, \cite{Sriramkumar2005, Boyanovsky2006, Holman2008, Agullo2011}).}

	In the next step, we demonstrate that for the universe experiencing several eras with different equations of state, the large-scale spectrum is determined by the earliest era in which the universe begins. Starting with a single slow-roll era with the scale-invariant super-horizon spectrum, we find how the spectrum changes as one incrementally stacks a kinetic era, and yet another slow-roll era, into the early times.%
		\footnote{The effect due to piecewise changes of the model parameter was studied in \cite{Mukhanov1991}, where the time-dependent effective inflaton mass was considered.} %
		If the universe begins with the initial adiabatic vacuum in the kinetic era, the spectrum is suppressed at large scales, and we find that the suppression is a direct consequence of the blue-tilted super-horizon spectrum in the initial kinetic era. With another slow-roll era preceding the kinetic era, the large-scale spectrum is enhanced because the super-horizon spectrum in the initial slow-roll era is scale-invariant with the amplitude higher than that generated in the second slow-roll era. In this case, the intermediate kinetic era only serves to connect the two scale-invariant spectra of the two slow-roll eras. One sees that the power suppression stems from the initial blue-tilted super-horizon spectrum, and once the initial spectrum is different, the large-scale power may not be suppressed even if there is a preinflation kinetic phase.
	
	We also investigate the scenario that the universe starts with a superinflation era before the slow-roll inflation. The superinflation era can be induced in theories of quantum gravity \cite{Tsujikawa2004, Ashtekar2010, Biswas2014, Domenech2015}, or by a scalar field that violates the dominant energy condition \cite{Piao2004a, Baldi2005}. Models in the latter case generally suffer from quantum instabilities and should only be regarded as effective theories (for related discussions, see, for example, \cite{Carroll2003, Cline2004}). The interest here lies in the fact that, opposite to the case of a single preinflation kinetic era, it is causal to assume the initial adiabatic vacuum in the preinflation superinflation era. We found that in this case the large-scale power is also suppressed due to the blue-tilted super-horizon initial spectrum in the superinflation era. Power suppression due to an early superinflation era has also been inferred in the models of loop quantum gravity \cite{Tsujikawa2004} or bouncing cosmology \cite{Biswas2014}, while in this work a more systematic treatment to the evolution of perturbations is given.

	After understanding the character of the spectrum in the multi-stage inflation using the \emph{ad hoc} single-field analysis, we calculate the spectrum of the curvature perturbations in a two-field model with the given potential. We consider the chaotic potential with a coupling term to the second scalar field, which is similar to the effective potential in the early stage of the hybrid inflation. By numerically solving the equations of motion and using the \texttt{CAMB} code \cite{Lewis2000, Lewis2012}, we show that the large-scale spectra of curvature perturbations and CMB are indeed enhanced due to the initial inflationary era.%
		\footnote{As regards the treatment of the two-stage inflation, our approach is closest to that of \cite{Ashoorioon2006, Ashoorioon2009}, in which a more complicated string-motivated two-field model is considered. In \cite{Polarski1992} the two fields have no direct coupling, and certain approximations are used to obtain the analytical solutions in various regimes of the model parameters. In \cite{Jain2009, Dudas2012, White2014, Yamauchi2011}, the single-field models are used. In \cite{Namjoo2012} the system is also modeled by a single fluid.}


\section{Initial Condition from Quantum Cosmology}

	As we showed earlier that the power suppression can occur if one of the two following possibilities happens in the early stage of the inflation: First, the phantom equation of state (and the superinflationary expansion due to the phantomness) can induce the power suppression. Second, a positive-pressure era (with the equation-of-state parameter $w > 0$), such as the kinetic-energy-dominated era, at the early stage of inflation can cause the power suppression. Both scenarios are logically possible, but both ideas have their own problems. For the phantom inflation scenario, it is very difficult to construct a viable theory for the phantom matter. For the positive-pressure era, the power suppression highly depends on the choice of the vacuum state. In the de Sitter space, we have a canonical choice of the vacuum---the Bunch-Davies vacuum \cite{Bunch1978}, but in the positive-pressure era, there is no such a canonical vacuum. Moreover, if we consider an eternally inflating background (and the consequent Bunch-Davies vacuum), then even though the universe evolves toward a positive-pressure era, the power suppression will not be realized \cite{Chen2016}.
	
	The existing difficulties of having a consistent explanation for the power suppression may imply that its origin does not lie in the semi-classical physics, but in the quantum theory of gravity. \emph{Can we explain the power suppression by quantum gravitational effects?} Indeed, there has been several models explaining the power suppression from quantum gravity. For example, according to the loop quantum cosmology, quantum gravitational effects can induce an effective phantom matter in the deep trans-Planckian regime. The phantomness thereof can explain the CMB power suppression as well as supporting the scenario of the big bounce universe \cite{Ashtekar2015}.
	
	In order to investigate the wave function of our universe and the power suppression problem, we will rely on the Hartle-Hawking wave function, or the so-called no-boundary wave function \cite{Hartle1983}. This wave function is one of the proposals to the boundary condition of the Wheeler-DeWitt equation \cite{DeWitt1967}. It is a path integral over the Euclidean compact manifolds, and can be approximated by the method of steepest descent. Under such approximation, we can then describe the wave function as a sum of the Euclidean instantons, where each instanton should eventually be Wick-rotated into the Lorentzian signatures \cite{Hartle2008a, Hartle2008} and approach real-valued functions \cite{Hwang2012, Hwang2012a, Hwang2014, Hwang2015, Chen2016a}. By integrating the Lagrangian, one can estimate the probability for the history described by each instanton.
	
	Following the work of Halliwell and Hawking \cite{Halliwell1985}, one can introduce perturbations to the background instanton solution. These perturbations also carry their own canonical degrees of freedom. Although in general it is very difficult to track their coupled evolution, one can consistently consider various modes separately as long as the perturbations stay in the linear regime. The probability distribution of the magnitude of each perturbation mode can then be calculated, and the expectation values of these modes, or equivalently, the power spectrum, can therefore be determined.
	
	Using the method of Laflamme \cite{Laflamme1987}, we can define the wave function for the Euclidean vacuum. The Euclidean vacuum gives the scale-invariant power spectrum at short-wavelength scales, hence consistent with the choice of the Bunch-Davies vacuum \cite{Bunch1978} at small scales. On the other hand, at the long-wavelength scales, the power spectrum is enhanced due to the curvature of the manifold. All these results have been known in the literature and consistent with the independent calculations from quantum field theoretical techniques \cite{Starobinsky1996, Halliwell1990}. However, to our best knowledge, it was not emphasized that the power spectrum can be \textit{suppressed} by introducing the potential term. In chapter \ref{ch:HH}, we include analytical and numerical details for the power suppression due to the potential term of the inflaton field \cite{Chen2017a}.
	
	We adopt the Planck units ($c = \hbar = G = 1$) and the signature ($-, +, +, +$) throughout the thesis.

\clearpage


\chapter{Preinflationary Network of Frustrated Topological Defects}{\label{ch:NFTD}}

	One natural candidate that may cause the power suppression at the lowest modes of the CMB spectrum is the primordial topological defect produced during some phase transition in the preinflationary era. If the universe is populated with the topological defects before the slow-roll inflation, the expansion rate of the preinflationary universe is generally different from that of the de Sitter universe. Therefore, the curvature perturbations evolves differently in the preinflationary era from the way they do in the inflationary era. If inflation sustains for just about 60 e-folds, we can then see the imprint of the transition from the preinflationary to the inflationary era on the perturbation spectrum.
	
	In this chapter we consider two of the most common types of topological defects---domain walls and cosmic strings---arising from the phase transitions at the cosmic scale. To model the transition from the preinflationary to the inflationary era, we introduce the generalized Chaplygin gas (GCG) \cite{Kamenshchik2001, Bilic2002, Bento2002, Lopez2005, Chimento2006}. The idea of describing the early universe by the Chaplygin gas was first suggested in Refs.~\cite{Lopez2005a, Lopez2005b} (see also \cite{Bertolami2006, Lopez2009}) and later extended in Refs.~\cite{Lopez2010, Lopez2010a, Lopez2011, Lopez2012a, Bouhmadi-Lopez2013a, Lopez2013a}. The energy density of a network of frustrated topological defects (NFTD) can be described in a compact way, for example, as
\begin{equation}
\rho= \left(\frac{B_1}{a^{\beta _1\left(1+\alpha _1\right)}}+A_1\right)^{1\left/\left(1+\alpha _1\right)\right.},
\label{1.1}
\end{equation}
where $a$ is the scale factor, $B_1$ and $A_1$ are constants related to the energy scale of the NFTD and the de Sitter-like inflationary era, respectively, $\alpha_1$ and $\beta_1$ are constants such that $\beta_1=1,2$ for the network of frustrated domain walls (NFDW) and the network of frustrated cosmic strings (NFCS), respectively. We assume that $1+\alpha_1$ is positive such that the inflationary era is preceded by a topological dominance phase. Let us be reminded in this regard that the energy density of NFDW and NFCS scales as $1/a$ and $1/a^2$, respectively \cite{Vilenkin2000}. It is worthy to stress that the NFTD epoch preceding the slow-roll inflationary can in principle produce inflation as well; indeed this is the case for NFDW, but this inflation is much slower, i.e.~much \textit{lazier} than the slow-roll inflation. Moreover, for a NFCS dominated period the universe is increasing its size at a constant speed; i.e.~with no acceleration or deceleration. From now on whenever we refer to a preinflationary era, we will be referring to a pre-slow-roll inflationary era.


We consider a spatially flat Friedmann-Lema\^{i}tre-Robertson-Walker (FLRW) universe filled with the matter content described by Eq.~\eqref{1.1}. The energy conservation gives
\begin{equation}
\dot{\rho}+3H(\rho+p)=0,
\label{eq:EnergyConservation}
\end{equation}
where a dot corresponds to a derivative with respect to the cosmic time and $H$ stands for the Hubble rate. By inserting Eq.~\eqref{1.1} into Eq.~\eqref{eq:EnergyConservation}, one obtains the pressure of the matter content
\begin{equation}
p=\left(\frac{\beta_1}{3}-1\right)\rho -\frac{\beta_1}{3}\frac{A_1}{\rho ^{\alpha_1}}.
\end{equation}
Note that in terms of the equation of state,
\begin{equation}
p = w \rho,
\end{equation}
where $w$ is the equation-of-state parameter, the universe is in the state of $w = -2/3$ and $w = -1/3$ for $\beta = 1$ (NFDW dominated) and $\beta = 2$ (NFCS dominated), respectively.

In the Planck unit, the Friedmann equation reads
\begin{equation}
 H^2=\frac{\kappa^2}{3}\rho,
\label{fe}
\end{equation}
where $\kappa^2\equiv 8\pi G=8\pi$. The conformal time $\tau$ can be expressed as
\begin{equation}
\tau =\frac{\sqrt{3}}{\kappa }\frac{b}{c}A_1^{-(b+c)}\left(\frac{B_1}{A_1}\right)^by^c \; {}_2F_1(c,1-b,c+1,y),
\label{tau}
\end{equation}
where $y=\left[1+\left(B_1/A_1\right)a^{-\beta _1\left(1+\alpha _1\right)}\right]^{-1}$, $b=1\left/\left[\beta _1\left(1+\alpha _1\right)\right]\right.$, $c=(-2+\beta_1)/[2 (1+\alpha_1) \beta_1]$, and $_2F_1$ is a hypergeometric function \cite{Gradshteyn1994}.  Eq.~\eqref{tau} describes that the universe began in NFTD dominated era in the past infinity, and turned into a de Sitter-like space, in which $a \propto 1 / \tau$, at later time.


\section{Model Building and Parameters Fixing}
\label{sec:ModelBuilding}

We divide the expansion of {\color{black} the} universe into three successive periods: the preinflationary NFTD dominated era, the slow-roll inflating phase, and the standard $\Lambda$CDM epoch. The energy density of each of these periods can be modeled as
\begin{subnumcases}{\label{1m} \rho =}
	\left[
		\frac{B_1}{a^{\beta _1 \left( 1 + \alpha_1 \right)}} + A_1
	\right]^%
		{1 \left/ \left( 1 + \alpha_1 \right) \right.},
	\label{1m1} \\
	\left[
		A_2 + \frac{B_2}{a^{4 \left( 1 + \alpha_2 \right)}}
	\right]^%
		{1 \left/ \left( 1 + \alpha_2 \right ) \right.},
	\label{1m2} \\
	\rho_{\textrm{r0}} \left( \frac{a_0}{a} \right)^4
		+ \rho_{\textrm{m0}} \left( \frac{a_0}{a} \right)^3
		+ \rho_{\Lambda}.
	\label{1m3}
\end{subnumcases}
Expression (\ref{1m1}) describes the energy density of the NFTD era, which was introduced in Eq.~(\ref{1.1}), followed by the de Sitter-like inflating phase. The model described by Eq. (\ref{1m2}) was previously studied within an inflationary framework in Ref.~\cite{Lopez2010, Lopez2010a} (see also Ref.~\cite{Chimento2006}) and, under suitable constraints on $A_2$, $B_2$, and $\alpha_2$, can depict the transition from the de Sitter-like era to the radiation dominated era. The energy density (\ref{1m3}) is the standard $\Lambda$CDM model, in which $\rho_{\text{r0}}$, $\rho_{\text{m0}}$, and $\rho_\Lambda$ are the energy densities of the radiation, matter, and dark energy today, respectively. As we will show later, the parameters of the model can be constrained using observational data corresponding to the present energy density of radiation, the scalar power spectrum, and the spectral index at a given pivot scale. In addition, by requiring that the energy density is continuous at each transition, we have the conditions
\begin{align}
A_1 &= A_2^{(1+\alpha_1)/(1+\alpha_2)},\\
B_2 &= \left(\rho_{r0}a_0^4\right)^{1+\alpha_2}\label{B2}.
\end{align}

\begin{figure}[t]
\centering
\includegraphics[width=14 cm]{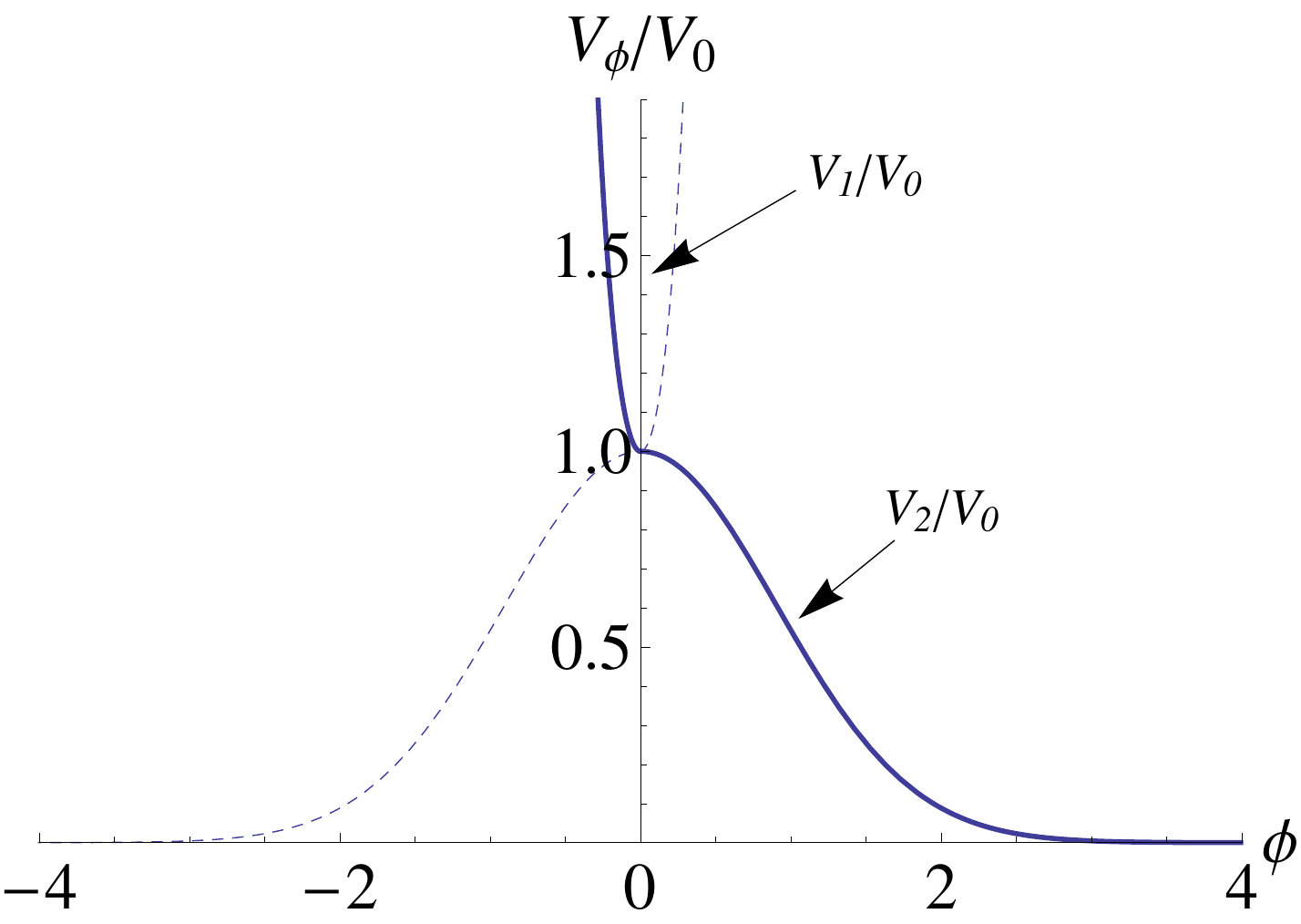}
\centering
\caption{The scalar field potential for $\beta_1=1$ (preinflationary NFDW). The upward convex curve is $V_1$ (cf.~Eq.~(\ref{v1})) and the downward cacave curve is $V_2$ (cf.~Eq.~(\ref{v2})). We can choose $\phi\leq0$ for $V_1$ and $\phi\geq0$ for $V_2$ to describe the potential of the scalar field $\phi$. In this model the scalar field rolls down too quickly in the slow-roll inflationary era and the radiation dominated phase is reached too early.}
\label{v_v0}
\end{figure}

In order to obtain the scalar power spectrum, it is useful to model the matter content in the first two periods of Eq.~(\ref{1m}); i.e.~those described by Eqs.~(\ref{1m1}) and (\ref{1m2}), through a scalar field, with the condition that the energy density and pressure of the scalar field are the same as that given by Eqs.~(\ref{1m1}) and (\ref{1m2}). The energy density and pressure of the scalar field are
\begin{equation}
\rho_\phi=\frac{\phi'^2}{2\,a^2}+V(\phi), \qquad p_\phi=\frac{\phi'^2}{2\,a^2}-V(\phi),
\label{rhop}
\end{equation}
where the primes denote the derivatives with respect to the conformal time. The scalar field and its potential in the first period is given by
%
\begin{align}
	\label{phi1} \phi (a) = &\frac{1}{\kappa \left(1+\alpha _1\right)\sqrt{\beta _1}} \notag \\
	&\times \ln \left(\frac{-1+\sqrt{1+(\frac{a}{B_1/A_1})^{\beta _1\left(1+\alpha _1\right)}}}{1+\sqrt{1+(\frac{a}{B_1/A_1})^{\beta _1\left(1+\alpha _1\right)}}}\right), \\
	\label{v1e} V_1(a) = &\frac{V_0}{6}\left\{\left(6-\beta _1\right)\left[1+\left(\frac{B_1/A_1}{a^{\beta _1}}\right)^{\left(1+\alpha _1\right)}\right]^{1\left/\left(1+\alpha _1\right)\right.}\right. \notag \\
	&\left.+\beta _1\left[1+\left(\frac{B_1/A_1}{a^{\beta _1}}\right)^{\left(1+\alpha _1\right)}\right]^{-\alpha _1/\left(1+\alpha _1\right)}\right\},
\end{align}
%
where $V_0=A_1^{1/(1+\alpha_1)}$. Similarly, the scalar field and its potential in the second period can be obtained by replacing $\alpha_1$ with $\alpha_2$ and setting $\beta_2 = 4$ in Eq.~(\ref{phi1}) and Eq.~(\ref{v1e}), giving
%
\begin{align}
	\label{phi2} \phi(a) = &\frac{1}{2 \kappa \left(1+\alpha _2\right)} \notag \\
	&\times \ln \left(\frac{-1+\sqrt{1+(\frac{a}{B_2/A_2})^{4\left(1+\alpha _2\right)}}}{1+\sqrt{1+(\frac{a}{B_2/A_2})^{4(1+\alpha _2)}}}\right),\\
	\label{v2e} V_2(a) = &\frac{V_0}{3}\left\{\left[1+\left(\frac{B_2/A_2}{a^4}\right)^{\left(1+\alpha _2\right)}\right]{}^{1\left/\left(1+\alpha _2\right)\right.}\right. \notag \\
	&\left.+2\left[1+\left(\frac{B_2/A_2}{a^4}\right)^{\left(1+\alpha _2\right)}\right]^{-\alpha _2/\left(1+\alpha _2\right)}\right\}.
\end{align}
%
We can obtain the potential of the scalar field for the two periods as functions of the scalar field by substituting the inverse function of Eq.~(\ref{phi1}) into Eq.~(\ref{v1e}) and similarly {\color{black}that of} Eq.~(\ref{phi2}) into Eq.~(\ref{v2e}), respectively, which leads to 

\begin{eqnarray}
\label{v1}
V_1(\phi )&=&\frac{V_0}{6}\left\{\left(6-\beta _1\right)\cosh \left[\kappa \left(1+\alpha _1\right)\sqrt{\beta _1}\frac{\phi }{2}\right]^\frac{2}{1+\alpha_1}\right. \nonumber \\ &\;&\left.+\;\beta _1\cosh \left[\kappa \left(1+\alpha _1\right)\sqrt{\beta _1}\frac{\phi }{2}\right]^\frac{-2\alpha _1}{1+\alpha _1}\right\},\\
V_2(\phi)&=&\frac{V_0}{3}\left\{\cosh  \left[\kappa \left(1+\alpha _2\right)\phi \right]^{\frac{2}{1+\alpha _2}} \right. \nonumber \\ &\;&\left. +2\cosh \left[\kappa \left(1+\alpha _2\right)\phi \right]^{\frac{-2\alpha _2}{1+\alpha _2}}\right\}.
\label{v2}
\end{eqnarray}
Eq.~(\ref{v2}) coincides with the potential in Ref.~\cite{Lopez2010, Lopez2010a}, as it should be. The form of Eq.(\ref{v1}) and Eq.(\ref{v2}) has been chosen such that the two periods are connected at $\phi=0$ with the potential and its first derivative with respect to $\phi$ being analytically continuous at the connecting point. The result is shown in Figure \ref{v_v0}.

\begin{figure}[t]
\centering
\includegraphics[width=14cm]{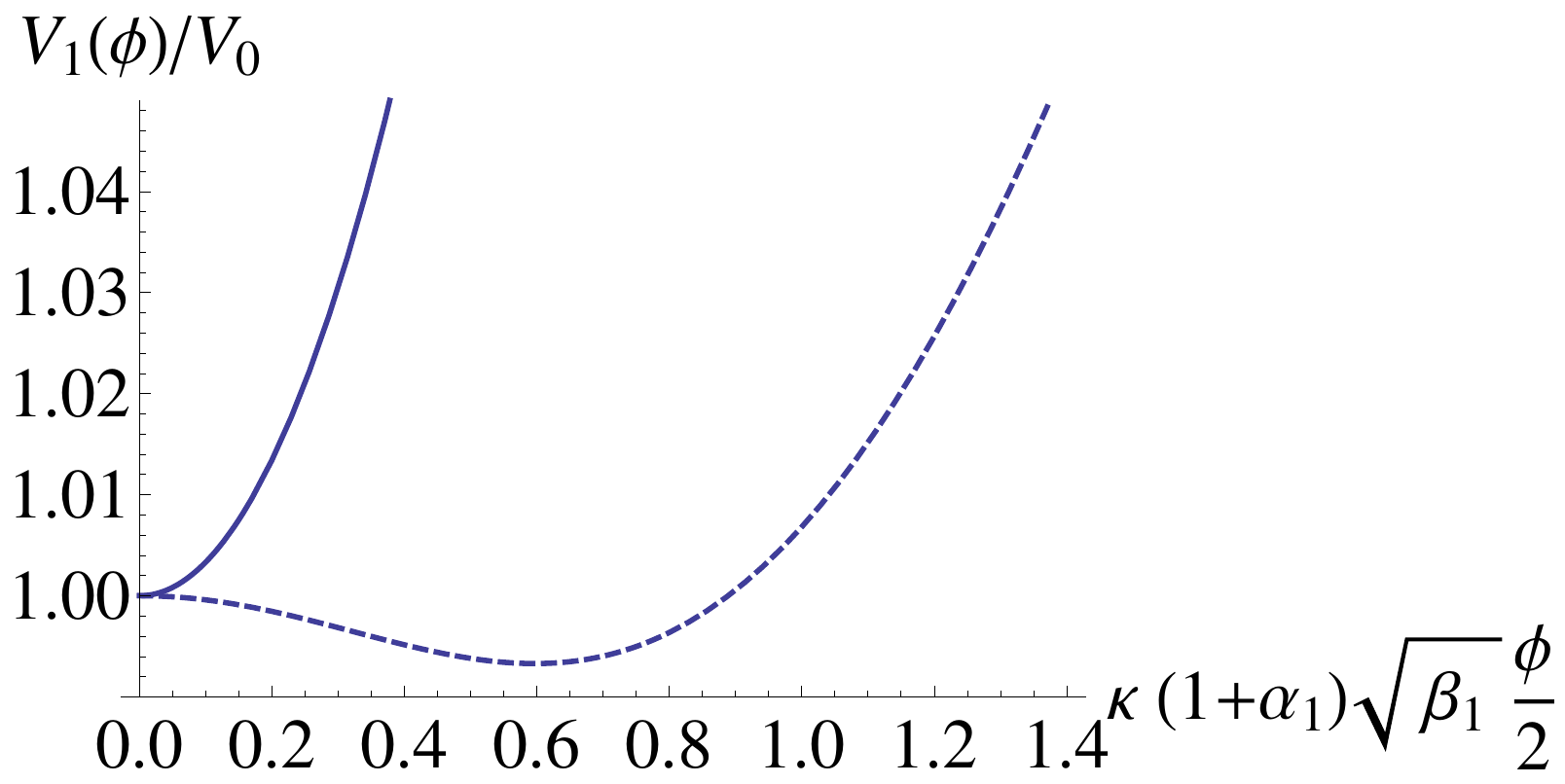}
\centering
\caption{The scalar field potential for the NFDW era ($\beta_1 = 1$) for different values of $\alpha_1$. The solid curve corresponds to $\alpha_1 = 1$ (has unique minimum at $\phi = 0$) and the dashed curve corresponds to $\alpha_1 = 7$ (has minimum at $\phi > 0$). The situation is similar for the case of NFCS ($\beta_1 = 2$), which we omit here.}
\label{alpha}
\end{figure}

Next we tackle the issue of analyzing the potentials (\ref{v1}) and (\ref{v2}).
First of all, we consider that the scalar field potential (\ref{v1}) has a unique minimum at $\phi=0$ to maximize the amount of inflation during the first period. Notice that unless this condition is imposed, the scalar field might roll down the potential till it reaches the minimum of $V_1(\phi)$ and then would have to climb up to reach the local maximum located at $\phi=0$, as shown in Figure \ref{alpha}. Imposing that the potential (\ref{v1}) has a unique minimum reached at $\phi=0$ implies a condition on the parameters $\alpha_1$ and $\beta_1$ that
\begin{equation}
\alpha_1 < \frac{6-\beta_1}{\beta_1}.
\label{conditionalpha_1}
\end{equation}
Therefore, bearing in mind that (i) $\beta_1=1,2$ for NFDW and NFCS, respectively, and (ii)  $0<1+\alpha_1$ so that the phase of FNTD precedes the inflationary phase, we conclude that $-1<\alpha_1<5$ for NFDW and $-1<\alpha_1<2$ for NFCS. We show the shape of the potential $V_1(\phi)$ for different cases when the condition (\ref{conditionalpha_1}) is fulfilled and violated in Figure \ref
{alpha}.


In addition, we can constrain our model, potentials (\ref{v1}) and (\ref{v2}), using the methodology in Ref.~\cite{Lopez2010, Lopez2010a}. More precisely,  we can use the WMAP7 observation of the power spectrum of the comoving curvature perturbation, $P_s=2.45\times10^{-9}$, and the spectral index, $n_s=0.963$, at the pivot scale $k_0=0.002$ Mpc$^{-1}$ to fix the parameters in our model \cite{Komatsu2011}. We can as well impose a bound on the number of e-folds, $N_c$, since a given mode exits the horizon until the end of inflation as done in Ref.~\cite{Lopez2010, Lopez2010a}.
This gives the best-fit values for $\alpha_2$, $V_0$, and, therefore, $A_2$. Notice that once $V_0$ is fixed, the parameter $A_1$ is fixed for a given $\alpha_1$ as well, since $V_0=A_1^{1/(1+\alpha_1)}$.

The parameter $B_1$ in Eq.~(\ref{1m1}) fixes the energy density of the NFTD, which strongly affects the lowest modes that exited the horizon around the onset of inflation, and causes a significant drop on the lowest modes of the primordial spectrum of the curvature perturbation. Although we expect that the NFTD would affect the lowest modes, we must make sure that the curvature power spectrum $P_s$ and the spectral index $n_s$ at the pivot scale $k_0$ are consistent with the observations.
Therefore, we choose the value of $B_1$ such that $P_s$ and $n_s$ match the observed values at $k_0$, and that the amplitude of $P_s$ drops at the scales whose comoving wave numbers are smaller than $k_0$. Roughly speaking, the parameter $B_1$ controls the horizontal shift of the curvature power  spectrum.

However, following this procedure, it turns out that we can not find a set of values for the parameter $B_1$ that satisfies the constraint on the spectral index. In fact, in this model $B_1$ turns out to be always smaller than $0.9$. This drawback originates from the second period described by Eq.~(\ref{1m2}), which corresponds to the transition from the slow-roll inflationary era to the radiation dominated period. More precisely, the model described by Eq.~(\ref{1m}) does not give enough e-folds during the slow-roll inflationary era. We show, as an example, in Figure \ref{v_v0} how the scalar field rolls too quickly and the radiation dominated phase is reached too early in the case corresponding to a NFDW.

We therefore suggest an alternative model described by
\begin{subnumcases}{\label{2m}\rho =}
 \frac{B_1}{a^{\beta_1}}+\left(\frac{A_2}{a^{1+\beta_2}}\right)^{1 / ( 1+\beta_3 )}\label{2m1}, \\
\left[\frac{A_2}{a^{1+\beta_2 }}+\frac{B_2}{a^{4(1+\beta_3 )}}\right]^{1 / ( 1+\beta_3 )}
\label{2m2}, \\
 \rho _{\text{r0}}\left(\frac{a_0}{a}\right){}^4+\rho _{\text{m0}}\left(\frac{a_0}{a}\right){}^3+\rho _{\Lambda }
\label{2m3},
\end{subnumcases}
where $\beta_1=1,2$ discriminates the NFDW and the NFCS as in the former model, and $\beta_2, \beta_3, B_1, A_2, B_2$ are constants we will explain below. The major difference of this model from out previous one is that here we model the slow-roll inflation by a power-law expansion. We choose this model because it generates an almost flat curvature spectrum for modes larger than the pivot scale $k_0=0.002$ Mpc$^{-1}$ and gives enough e-folds during the power-law inflationary period. In addition, the model introduces in a natural way that a NFTD precedes the power-law inflation as $(1+\beta_2)/(1+\beta_3)<\beta_1$ (please see also the conditions (\ref{c1}), (\ref{c2}) and (\ref{c3})).

The first period in Eq.~(\ref{2m1})  describes the matter content of the universe during a period that transits from a NFTD dominated phase to a power-law inflationary era. The parameters $B_1$ and $A_2$ are associated with the energy scale of the NFTD and that of the power-law inflation, respectively.  The second period with the energy density (\ref{2m2}) was previously studied within another inflationary framework in Ref.~\cite{Lopez2011, Lopez2012a}. It connects smoothly a power-law inflating phase with a radiation dominated universe, and the constraints on the parameters $\beta_2$ and $\beta_3$ are,\footnote{The notation is different from the one used in the work \cite{Lopez2011, Lopez2012a}. The parameters $\beta$ and $\alpha$ in Ref.~\cite{Lopez2011, Lopez2012a} are denoted as $\beta_2$ and $\beta_3$ here, respectively.}
\begin{eqnarray}
\label{c1} 1+\beta_2&<&0, \\ 
\label{c2} 1+\beta_3&<&0, \\ 
\label{c3} 2(1+\beta_3)&<&1+\beta_2 .
\end{eqnarray}
These constraints imply that (i) there is a power-law inflating phase, (ii) the inflationary era precedes the radiation dominated period, and (iii) the null energy condition is always fulfilled so that there is no superinflationary phase. Finally, the energy density  described by (\ref{2m3}) corresponds to the $\Lambda$CDM model as that described by Eq.~(\ref{1m3}).

Although there seems to be many free parameters, they can be fixed down to only one by the following procedure: (i) Fix $B_2$ by the current amount of radiation for a given $\beta_3$. (ii) Constrain the power-law expansion quantified by $A_2^{(1+\beta_3)/(1+\beta_2)}$ by the WMAP7 data of the curvature power spectrum $P_s$ and the spectral index $n_s$. Since there are three parameters ($A_2$, $\beta_2$ and $\beta_3$) to be constrained by only two conditions ($P_s$ and $n_s$), we are left with the only one free parameter, which we choose to be $\beta_3$. (iii) Fix $B_1$ such that $P_s$ and $n_s$ remain the correct values at the pivot scale $k_0$, and $P_s$ drops only at comoving wave numbers smaller than $k_0$.

Again, it is suitable to introduce a scalar field that mimics the matter content described in Eqs.~(\ref{2m1}) and (\ref{2m2}); i.e.
we describe the dynamics of the model through a scalar field with a potential whose energy density and pressure can be obtained from Eq.~(\ref{rhop}).  During the NFTD period (cf. Eq.~(\ref{2m1})),  the mapping between the scalar field $\phi$ and the perfect fluid of our model leads to
\begin{eqnarray}
\label{phia1}
\phi(a)&=&\sqrt{v}\coth ^{-1}\sqrt{\frac{\beta _1}{v }} \nonumber\\
&\,&-\sqrt{\beta _1}\tanh ^{-1}\sqrt{\zeta+(1-\zeta)\frac{v }{\beta _1}}, \\
\label{v21}
V_1(a)&=&\frac{1}{6} \left[\left(5+\frac{ \beta_3 -\beta_2 }{1+\beta_3 }\right)\left(\frac{A_2}{a^{1+\beta_2 }}\right)^{\frac{1}{1+\beta_3 }} \right. \nonumber\\
&\,\,& \left.+\, (6-\beta_1)\frac{B_1}{a^{\beta_1}}\right],
\end{eqnarray}
where $\zeta=[1+(A_2^{1\left/\left(1+\beta _3\right)\right.}/B_1)a^{\beta _1-v }]^{-1}$, $v=(1+\beta_2 )/(1+\beta_3 )$, and $V_1(\phi)$ stands for the scalar field potential during this period.
Similarly, we map the perfect fluid with the energy density (cf. Eq.~(\ref{2m2})) to the scalar field $\phi$ with  a new potential $V_2(\phi)$  \cite{Lopez2011, Lopez2012a}
\begin{eqnarray}
\phi(a)&=&\frac{1}{q \kappa }\left[4 \tanh^{-1}\sqrt{1+\frac{q}{4(1+\beta_3 )}\frac{1}{1+\xi }}-2  \right. \nonumber\\ &\,&\left.
 \sqrt{\zeta}\coth^{-1}\sqrt{\frac{4}{\zeta}\left(1+\frac{q}{4(1+\beta_3 )}\frac{1}{1+\xi }\right)}\right], \nonumber\\
\label{phia2}\\ \nonumber
\label{v22}
V_2(a)&=&A_2^{1/(1+\beta_3 )}\left(\frac{A_2}{B_2} \right)^{-\zeta /q}(1+\xi )^{1/(1+\beta_3 )}
\\ \nonumber &\,&\xi ^{-\zeta/q} \left(\frac{1}{3}-\frac{q}{6(1+\beta_3 )}\frac{1}{1+\xi }\right), \nonumber\\
\label{v22}
\end{eqnarray}
where $\xi =(B_2/A_2)a^q$ and $q=1+\beta_2 -4(1+\beta_3 )$. The potential (\ref{v22}) was previously obtained in Ref.~\cite{Lopez2011, Lopez2012a}. Such a potential, with an appropriate initial condition, drives a power-law inflation and mimics a radiation dominated universe afterwards.

\begin{figure}[t]
\centering
\includegraphics[width=14cm]{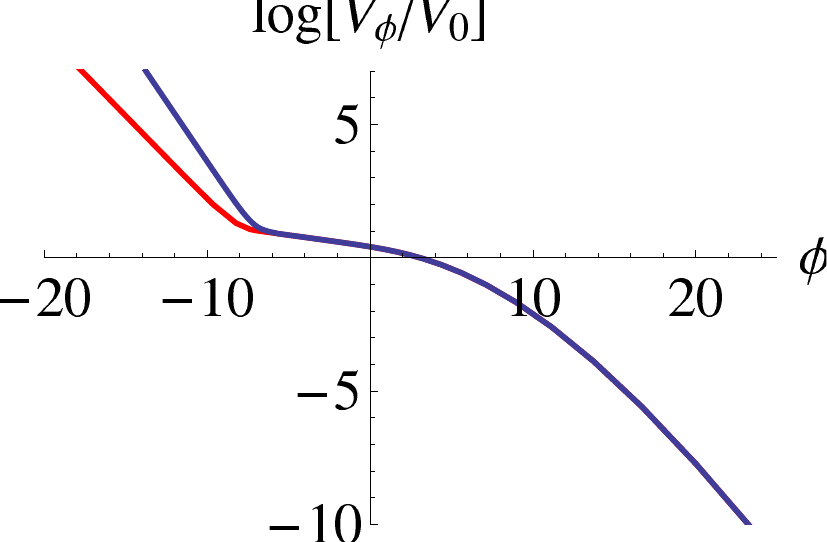}
\centering
\caption{This plot shows the rescaled potentials given in Eqs.~(\ref{v21}) and (\ref{v22}) versus the scalar field $\phi$, where $V_0=A_2^{1/(1+\beta_3 )}\left(A_2/B_2 \right)^{-(1+\beta_2 )/(q(1+\beta_3 ))}$. The blue curve corresponds to the NFCS case, and the red curve corresponds to the  NFDW case. The energy scale of inflation, $V_0$, in our model is about $10^{15}$ GeV for both NFDW and NFCS.}
\label{potential}
\end{figure}

Unlike the previous model described by Eq.~(\ref{phi1})-(\ref{v1e}) and Eq.~(\ref{phi2})-(\ref{v2e}), here it is not feasible to find analytically the inverse functions of Eq.~(\ref{phia1}) and Eq.~(\ref{phia2}), so we cannot obtain the analytical forms of the potential as functions of $\phi$. {\color{black}We thus} connect the scalar field potential numerically. $V_1(a)$ and $V_2(a)$ are connected at the intersection of the first two periods (Eq.~(\ref{2m1}) and Eq.~(\ref{2m2})), {\color{black}where} the second term of Eq.~(\ref{2m1}) dominates over {\color{black}its first term, and the first term of Eq.~(\ref{2m2}) dominates over its second term} so that the potential and its first derivatives with respect to $\phi$ are approximately continuous at the intersection of the first two periods. It is worthy to notice that an integration constant appears when we integrate Eq.~(\ref{rhop}) after mapping it to the energy density and pressure of a given perfect fluid. Therefore, we can always choose the constant properly such that the scalar field $\phi$ is continuous at the connecting point. As a result, we can use the scale factor $a$ as a parametric parameter to plot $V_1(\phi)$ and $V_2(\phi)$, which are shown in Figure \ref{potential} as an example. The scalar field starts with a negative value and rolls down the potential as the universe inflates until it reaches the radiation dominated era.


\section{Scalar Perturbations}
\label{sp}
The accelerated expansion of the universe during the primordial inflationary  era converts the initial quantum fluctuations in the universe into macroscopic cosmological perturbations, which leads to the inhomogeneity we observe nowadays in the CMB \cite{Langlois2010, Mukhanov1992}.
Following the standard approach, we will use gauge invariant quantities that involve the metric perturbations and the scalar field fluctuations \cite{Liddle2000}. For convenience, we will choose the comoving curvature perturbation, $R$, which in addition is conserved on large scales \cite{Wands2000, Lyth2003}.

We expect that a NFTD in the very early universe and just before the inflationary era could give the appropriate corrections to the quadrupole modes of the CMB data as observed nowadays \cite{Komatsu2011}. We will next quantify the quantum cosmological perturbations during that period and obtain the power spectrum of the scalar perturbations.


The scalar perturbations can be described by introducing the variable (see, for example, Ref.~\cite{Bassett2006})
\begin{equation}
 u=z R,
\end{equation}
where $z\equiv \frac{a\dot{\phi}}{H}$. The variable $u$ can be decomposed into Fourier modes, $u_k$, which fulfill the field equation \cite{Bassett2006}
\begin{equation}
\frac{d^2u_k}{d\tau ^2}+\left(k^2-\frac{1}{z}\frac{d^2z}{d\tau ^2}\right)u_k=0.
\label{waveeq}
\end{equation}
The modes $u_k$ can be mapped to the spectrum of the comoving curvature perturbations which reads \cite{Bassett2006}
\begin{equation}
 \frac{2\pi ^2}{k^3}P_R(k)=\frac{\left|u_k\right|^2}{z^2}.
\end{equation}
Given that we are dealing with adiabatic perturbations, the comoving curvature perturbations remain constant on large scales and consequently we can equate the power spectrum at the horizon exit with the power spectrum of the primordial scalar perturbations at the horizon reentry  as observed on the CMB. Therefore, for a given mode $k$, the spectrum is evaluated at the horizon exit; i.e. when $k=a_{\text{cross}} H$, where $a_{\text{cross}}$ stands for the value of the scale factor when the mode exists the horizon.


We next obtain the evolution of the mode function $u_k(\tau)$ for each comoving wave number $k$ in order to obtain the curvature perturbation spectrum. We will tackle this issue numerically rather than using the standard results for slow-roll inflation \cite{Langlois2010, Lidsey1997, Bassett2006}, because those conditions are not fulfilled at very early time when the NFTD is dominant.
It is easier to solve  Eq.~(\ref{waveeq}) numerically by splitting it into two first order differential equations,
\begin{eqnarray}
\begin{cases}
 X^\prime=Y\\
 Y'=-\left(k^2-\frac{z^{\prime\prime}}{z}\right)X,
\end{cases}
\label{wavesf}
\end{eqnarray}
where we have set $X=u_k$.

\begin{figure}[t]
\centering
\includegraphics[width=14cm]{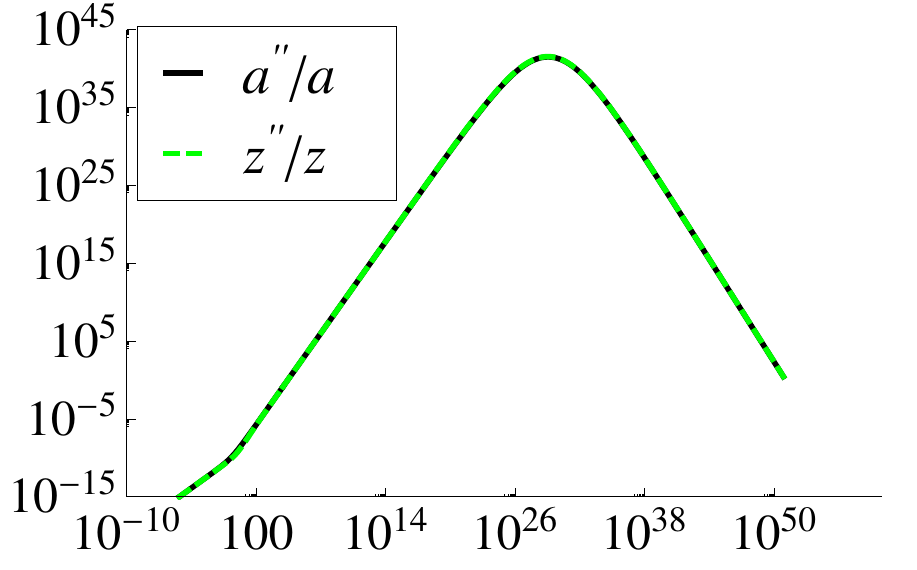}
\centering
\caption{The green dashed curve corresponds to $z^{\prime\prime}/z$ and the black curve corresponds to $a^{\prime\prime}/a.$  As can be seen that the approximation  $z^{\prime\prime}/z \approx a^{\prime\prime}/a$ holds during the NFDW dominated and the power-law inflationary eras.}
\label{dwzpp_z}
\end{figure}
\begin{figure}[t]
\centering
\includegraphics[width=14cm]{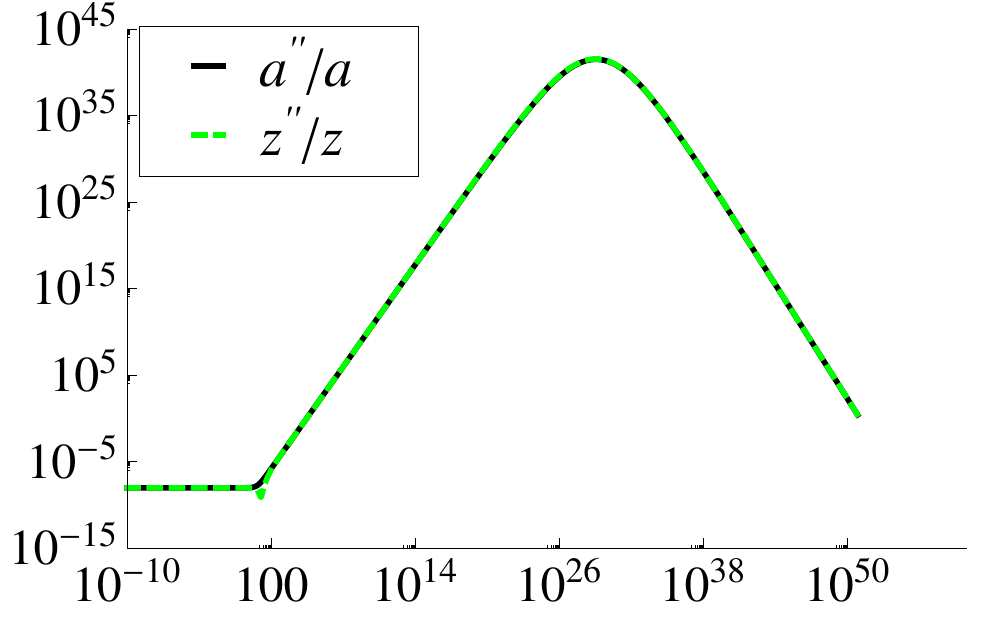}
\centering
\caption{The green dashed curve corresponds to $z^{\prime\prime}/z$ and the black curve corresponds to $a^{\prime\prime}/a.$ As can be noticed that the approximation  $z^{\prime\prime}/z \approx a^{\prime\prime}/a$ holds during the NFCS dominated and the power-law inflationary eras.}
\label{cszpp_z}
\end{figure}

In addition, we need to impose a set of boundary conditions at the time when the wavelength of a given mode $k$ is much smaller than the Hubble radius; that is, $k\gg a H$. Those boundary conditions will depend on the specific NFTD scenario and also on the given scale of the mode, as we will explain shortly.
It is also worthy to stress two things: (i)  the approximation  $z^{\prime\prime}/z \approx a^{\prime\prime}/a$ holds during the NFTD dominance and during the power-law inflationary era
(See Figure \ref{dwzpp_z} and Figure \ref{cszpp_z}), and (ii) the analytical solutions for the power-law inflation \cite{Sa2009} can be used as boundary conditions for the  modes with comoving wave numbers larger than roughly 10$^{-3}$ Mpc$^{-1}$ as explained next. We first recall that the power-law solutions for $X$ and $Y$ are \cite{Lidsey1997}
\begin{align}
	X(\tau)&=\frac{\sqrt{-\pi \tau }}{2}H_{\frac{1}{2}-l}^{(1)}(-k\tau ), \label{pw1}\\
	Y(\tau)&=\frac{-\sqrt{-\pi  \tau}k}{2} H_{\frac{1}{2}-l }^{(1)}(-k\tau )+\frac{l}{2}\sqrt{\frac{-\pi }{\tau}}H_{\frac{1}{2}-l }^{(1)}(-k\tau ), \label{pw2}
\end{align}
where $l$ is the exponent characterizing the power-law expansion in terms of the conformal time $\tau$; i.e. $a \propto \tau^l$. For our model, $l=\left.((1+\beta_2)/(2(1+\beta_3))-1\right.)^{-1}$. For those modes whose $k\geq10^{-3} \, $Mpc$^{-1}$,  we can start the numerical integration of Eq.~(\ref{wavesf}) during the power-law inflationary era where the condition $k\gg a H$ is still satisfied. However, for scales roughly smaller than 10$^{-3}$ Mpc$^{-1}$, we split the boundary conditions imposed on the NFDW and the NFCS separately.%
	\footnote{We will show in the next chapter that, through a more detailed and systematic analysis, the initial vacua have different structures from the approximation made here. As a consequence, the power spectra obtained from the initial vacua adopted in the next chapter also differ from those in this chapter. In particular, we will see that for the case of the preinflationary NFDW dominated era, the large-scale spectrum is actually enhanced, rather than suppressed.} %
	Despite the general solutions (\ref{pw1}) and (\ref{pw2}) are still valid for the NFTD, the exponent describing the power-law expansion, $l$, depends on the specific characters of the NFTD:
\begin{itemize}
\item During NFDW dominant era, $\rho\sim B_1/a^{\beta_1}$ with $\beta_1=1$, which  also implies a power-law expansion because $a(\tau)\propto \tau^{1 / \left( \beta_1/2-1 \right)}$. Thus we can use the solutions (\ref{pw1}) and (\ref{pw2}) as boundary conditions with $l=1/(\beta_1/2-1)$.

\item During the NFCS dominant era, $\rho\sim B_1/a^{\beta_1}$ with $\beta_1=2$. Note that for this value of $\beta_1$ the exponent $l$ in Eqs.~(\ref{pw1}) and (\ref{pw2}) is not well-defined, so we need to fix the initial condition in this case in a different way. It can be shown from Eq.(\ref{fe}) that $a^{\prime\prime}/a$ is a constant if $\rho\propto 1/a^2$. Introducing a new variable $\tilde{k}\equiv\sqrt{k^2-a^{\prime\prime}/a}$, the wave equation (\ref{waveeq}) becomes
\begin{equation}
\frac{d^2u_k}{d\tau ^2}+\tilde{k}^2 u_k=0,
\end{equation}
which has two properly normalized linearly independent solutions,
\begin{equation}
u_k(\tau,k)=\frac{e^{-i\tilde{k}\tau}}{\sqrt{2\tilde{k}}}, \,\,\,\,  \frac{e^{i\tilde{k}\tau}}{\sqrt{2\tilde{k}}}.
\label{csini}
\end{equation}
We then choose the solution with exponent $-i\tilde{k}\tau$ because it reduces to the Minkowski initial condition \cite{Langlois2010}.
\end{itemize}

Finally, it is easier to use the scale factor as the independent variable in the numerical integration of Eq.~(\ref{wavesf}) instead of  the conformal time. The relation between the conformal time $\tau$ and the scale factor $a$ is
\begin{eqnarray}
\tau& =&\frac{\sqrt{3}}{\kappa }\frac{1}{h\left(\beta _1-\zeta \right)}B_1{}^{-1/2}\left(\frac{B_1}{A_2{}^{1\left/\left(1+\beta _3\right)\right.}}\right)^hx^h
\nonumber\\
&\,&_2F_1\left[h, 1-g, h+1; x\right],
\end{eqnarray}
where $x=1-[1+(A_2^{1/(1+\beta _3)}/B_1)a^{\beta _1-\zeta}]^{-1}$, $g=(1-1/(2\zeta ))/(\beta _1-\zeta )$, and $e=(\beta _1/2-1)/(\beta _1-\zeta )$.

\begin{figure}[t]
\centering
\includegraphics[width=14cm]{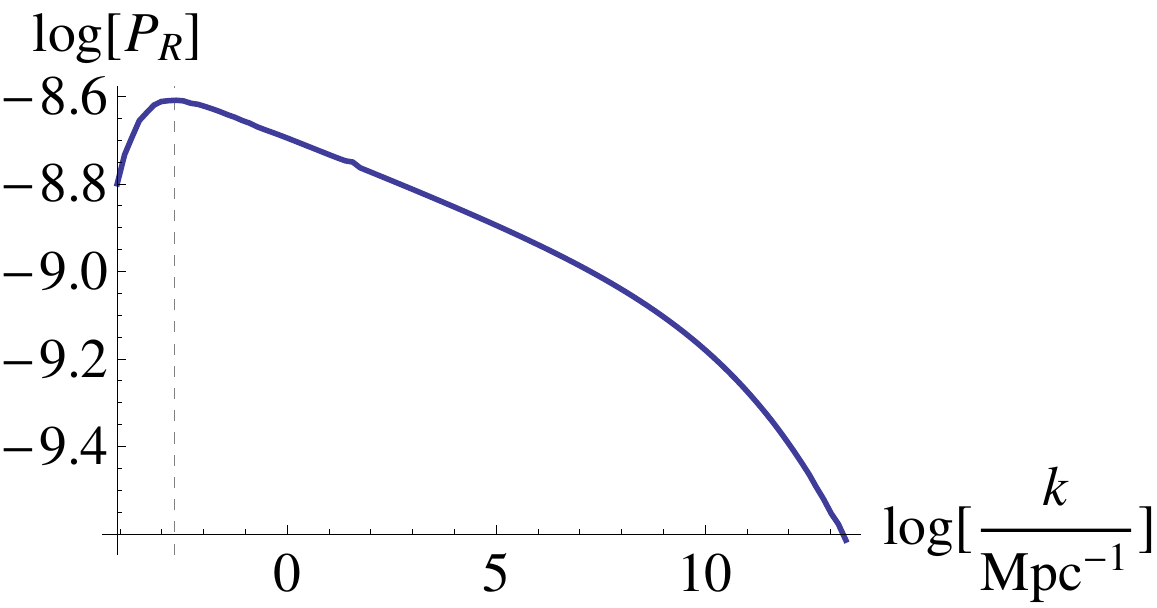}
\centering
\caption{This plot corresponds to the curvature perturbation spectrum for
 the inspired modified GCG model with $\beta_1=1$ (see Eq.~(\ref{2m1})), which describes a NFDW dominated era followed by a power-law inflationary period. We choose $\beta_3=-1.05$. The vertical dashed line corresponds to the pivot scale $k=0.002$ Mpc$^{-1}$.}
\label{DWS}
\end{figure}

\begin{figure}[t]
\centering
\includegraphics[width=14cm]{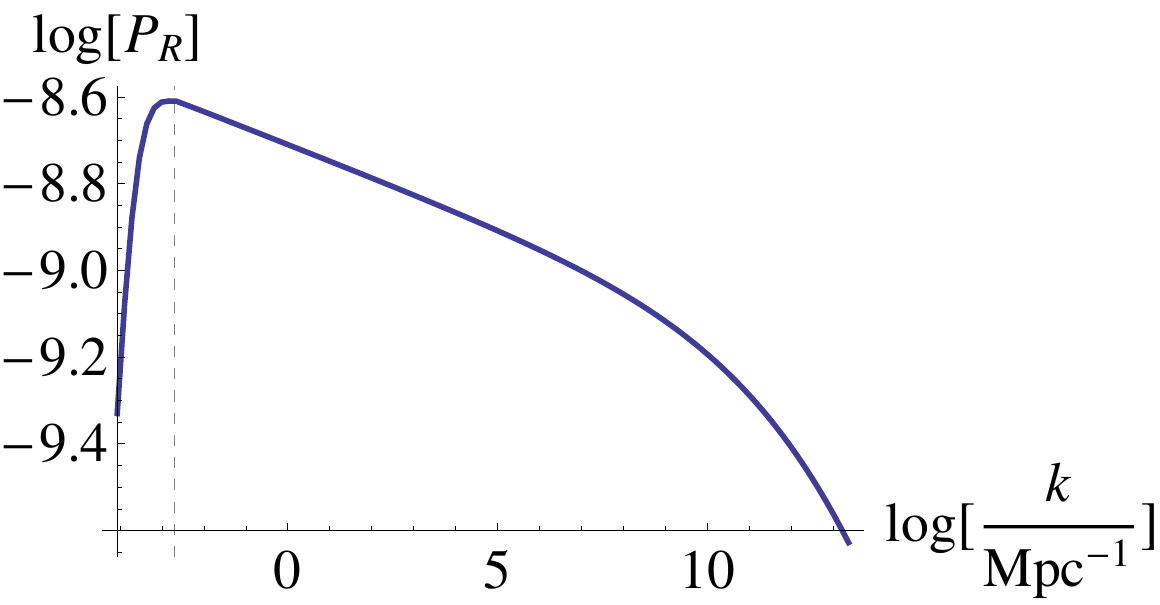}
\centering
\caption{This plot corresponds to the curvature perturbation spectrum for the modified GCG model with $\beta_1=2$ (see Eq.~(\ref{2m1})), which describes a NFCS dominated era followed by a power-law inflationary period. We choose $\beta_3=-1.05$. The vertical dashed line corresponds to the pivot scale $k=0.002$ Mpc$^{-1}$.}
\label{CSS}
\end{figure}

The resulting curvature power spectra are shown in Figure \ref{DWS} and Figure \ref{CSS}, corresponding to the NFDW and the NFCS, respectively. Let us recall that all the parameters used here have been fixed by imposing the observational constraints in the way stated in Sec.~\ref{sec:ModelBuilding}. For modes $k$ such that $10^{-3}\,\textrm{Mpc}^{-1}<k<10^{5}\,\textrm{Mpc}^{-1}$, we obtain a constant slope for the spectrum of the curvature perturbation. This is a simple consequence of the intermediate phase given in Eq.(\ref{2m2}) previously analyzed in Ref.~\cite{Lopez2011, Lopez2012a}, and implies a power-law inflation. Our new and important result is the drop of $P_R$ for the modes whose $k\leq$10$^{-3}$Mpc$^{-1}$, which is helpful in explaining the quadrupole anomaly through an alternative way from those used in Refs.~\cite{Contaldi2003, Boyanovsky2006a, Powell2007, Scardigli2011, Piao2005}. Such a decrease of $P_R$ is a consequence of the NFTD era just before the inflation.


We calculate the CMB temperature anisotropy spectrum by the numerical package CMBFAST \cite{Seljak1996, Zaldarriaga1998, Zaldarriaga2000} with minor modifications to the form of specifying the primordial power spectrum. In the modified version of CMBFAST, the primordial power spectrum is fed into the code as an interpolating function instead of a functional form. This adjustment is made so that rather than accepting only the nearly-scale-invariant spectrum as its initial condition, CMBFAST is now compatible with any general shape of initial spectrum. This feature is essential to our case since the primordial spectra obtained from our scenarios severely deviate from the scale-invariant form in large scales whose wave numbers are smaller than $10^{-3} \, \textrm{Mpc}^{-1}$.

\begin{figure}

\centering

\includegraphics[width=14cm]{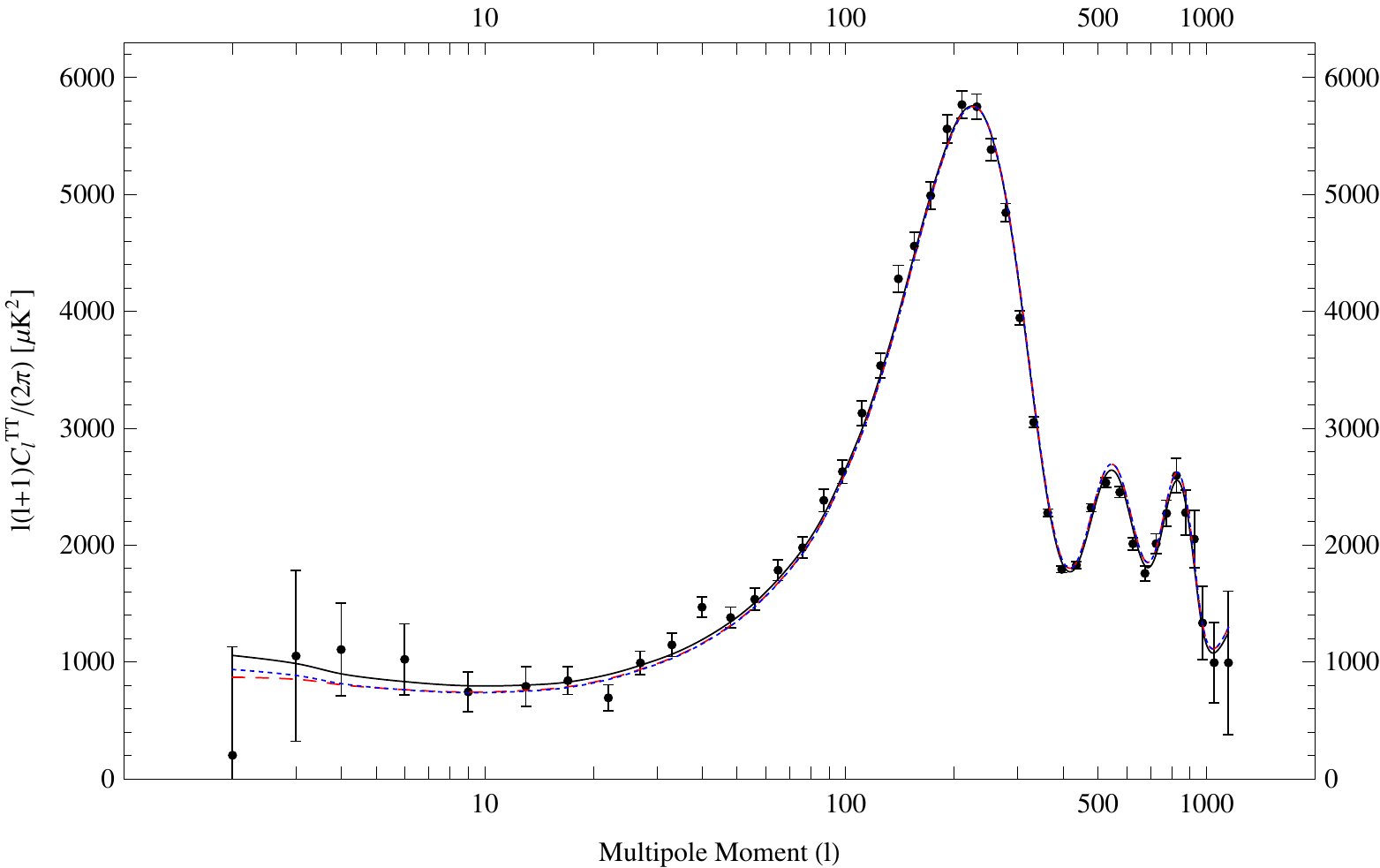}

\caption{
	The CMB temperature anisotropy spectrum. The dots with error bars are the WMAP 7-year data. The solid line is the prediction of standard inflation with a power-law spectrum. The dashed and dotted lines are the spectra of the scenarios of NFCS and NFDW, respectively.
}

\label{fig:CMBSpectrum}

\end{figure}

Figure \ref{fig:CMBSpectrum} shows the CMB spectra generated by the NFTD scenarios along with that generated by the standard inflationary model assuming a power-law initial spectrum. The data of WMAP 7-year observation \cite{Komatsu2011} are also marked in the plot. It can be seen that the preinflation NFTD era alleviates the quadrupole anomaly of the CMB. Note that the NFCS has a stronger effect on reducing the amplitude of the lower modes than the NFDW does. This is a result of the initial slope and the turn-around point of the curvature perturbation spectrum induced by NFDW and NFCS as shown in Figure \ref{DWS} and \ref{CSS}. Also note that regarding the lower modes of the CMB, it is irrelevant that the potentials and the first derivatives of the scalar field with respect to the cosmic time are not rigorously continuous at the connecting point. The reason  is that major contributions to the lower modes of CMB came from the scalar perturbations whose comoving wave numbers are about $10^{-5}$ to $10^{-3} \, \text{Mpc}^{-1}$. These modes had already exited the Hubble radius during the first period so that the power spectrum in this regime is obtained without the need to integrate across the connecting point.


\clearpage


\chapter{Power Spectrum in the Universe with Constant Equation of State}{\label{ch:Scaling}}

	In the previous chapter, we studied two scenarios of the preinflationary era that may potentially account for the power suppression of the long-wavelength CMB spectrum. There are, however, a few things we want to improve. First of all, in the previous treatment, we introduced the generalized Chaplygin gas (GCG) to model the preinflationary and inflationary (as well as the following $\Lambda$CDM) eras. The GCG model is actually a complicated model in the sense that there are many model parameters. Moreover, not any type of such models can fit a given set of observation results. As we have already seen in the previous chapter, a careful model selection is needed to ensure that we can fit the model with the observations. Second of all, the initial conditions in the preinflationary era rely on the solutions \eqref{pw1} and \eqref{pw2} that approximate the background evolution by the power-law expansion. It would be preferable if we can find an exact solution in the general background.%
		\footnote{Note that in the case of NFCS, the initial conditions were fixed in a different way.} %
		Last of all, although we studied two possibilities, there have been many other models proposed to explain the power suppression. What are the common characters that are essential for a model to be viable?	
	
	We conduct a series of investigations to address the above issues in this and the following chapters. In this chapter, we first find the single-field model that gives the background evolution with a given equation-of-state parameter, $w$, and the corresponding general solutions to the curvature perturbations. We then discuss the common assumption on the small-scale behavior of the solution and its causal character. In the end we find the power spectrum in the background with given $w$ and its power-law relation with respect to the wavenumber $k$.

	
	\section{Perturbations with Constant Equation of State}
	
		\label{sec:Solution}
		
		We consider a scalar field $\phi$ in an expanding Friedmann-Lema\^{i}tre-Robertson-Walker (FLRW) universe with the action
		\begin{align}
			\label{eq:Action} S = \int d^4 x \sqrt{-g} P(X, \phi),
		\end{align}
		where the kinetic term
		\begin{align}
			X = -\frac{1}{2} \partial_{\mu} \phi \partial^{\mu} \phi.
		\end{align}
		The energy-momentum tensor can be put into the form of the perfect fluid,
		\begin{align}
			T_{\mu \nu} = P g_{\mu \nu} + (\rho + P) u_{\mu} u_{\nu},
		\end{align}
		by identifying $P$ as the pressure, and the energy density and the velocity as
		\begin{align}
			\rho = &2 X \frac{\partial P}{\partial X} - P, \\
			u_{\mu} = &\frac{\partial_{\mu} \phi}{\sqrt{2 X}}.
		\end{align}
	
	The background evolution with constant equation of motion, $-1 < w \leq 1$, can be modeled by the Lagrangian,
		\begin{align}
			\label{eq:NormalLagrangian} P = X - V(\phi),
		\end{align}
		where, in the homogeneous and isotropic background, $X = \dot{\phi}^2 / 2 \geq 0$, and assuming the potential $V \geq 0$.%
			\footnote{With \eqref{eq:NormalLagrangian}, the evolution of constant $w > -1$ and given initial values $\phi_i$ and $\dot{\phi}_i$ can be realized by the \emph{ad hoc} potential
			\begin{align}
				V(\phi) = \frac{\dot{\phi}_i^2 (1 - w)}{2 (1 + w)} \exp\left[ \sqrt{24 \pi (1 + w)} \; (\phi - \phi_i) \right]. \notag
			\end{align}} %
		The dots denote the time derivatives. If one only requires $\rho = X + V$ to be positive and allows $V$ to be negative, then one can have $w > 1$ when $-X < V < 0$.%
		\footnote{A negative potential with $w > 1$ may be invoked, for example, in the cyclic universe scenario \cite{Steinhardt2002}. It requires special care to treat the perturbations in such models \cite{Battefeld2015}. If we assume the universe is not cyclic, and starts from a big bang followed by a decelerating era (which includes the case $w > 1$), then, as we point out in this work, the initial adiabatic vacuum is acausal. In the cyclic universe scenario, the density perturbations in the post-bounce expanding phase is proposed to be seeded by the perturbations generated in the pre-bounce contracting phase. However, the treatment of perturbations in the contracting phase and across the bounce is still under investigation. See, for example, \cite{Battefeld2015} for a review.}
	
	In the conformal Newtonian gauge, the perturbed FLRW metric is
		\begin{align}
			ds^2 = -( 1 + 2 \Phi ) dt^2 + a^2(t) ( 1 + 2 \Psi ) d\mathbf{x}^2,
		\end{align}
		where $a$ is the scale factor, and $\Phi$ and $\Psi$ are the metric perturbations. The equation of motion of the curvature perturbation, $\mathcal{R}$, is given by the Mukhanov-Sasaki equation in the Fourier space \cite{Mukhanov1985, Sasaki1986},
		\begin{align}
			\label{eq:MSEq} \mathcal{R}'' + 2 \frac{A'}{A} \mathcal{R}' + k^2 \mathcal{R} = 0,
		\end{align}
		where
		\begin{align}
			\mathcal{R} = &\Psi - \frac{H}{\dot{\phi}} \delta \phi, \\
			A = &\frac{a \sqrt{\rho + P}}{H},
		\end{align}
		$k$ is the wavenumber, $H = \dot{a} / a$ is the Hubble parameter, $\delta \phi$ is the field perturbation. The primes denote the derivative with respect to the conformal time $\eta$, defined by $dt = a d\eta$. Introducing the new variable $u = -A \mathcal{R}$, we can get rid of the first-derivative term, turning \eqref{eq:MSEq} into
		\begin{align}
			\label{eq:MukhanovEq} u'' + \left( k^2 - \frac{A''}{A} \right) u = 0.
		\end{align}
	
	If the evolution of the universe is described by a constant $w > -1$, there is a simple relation $A'' / A = a'' / a$ since
		\begin{align}
			\label{eq:ARelation} A = \sqrt{\frac{3 ( 1 + w )}{8 \pi}} \; a.
		\end{align}
		The scale factor evolves as
		\begin{align}
			\label{eq:ScaleFactor} a(\eta) = a_i ( 1 + \alpha \xi )^{1 / \alpha},
		\end{align}
		where $\xi = a_i H_i ( \eta - \eta_i )$, $a_i$ and $H_i$ denote the initial values at $\eta = \eta_i$, and
		\begin{align}
			\label{eq:alpha} \alpha = \frac{1 + 3 w}{2}.
		\end{align}
		Equation \eqref{eq:MukhanovEq} then reads
		\begin{align}
			\label{eq:MukhanovEq2} \frac{d^2 u}{d \xi^2} + \left[ \kappa^2 - \frac{\beta}{( 1 + \alpha \xi )^2} \right] u = 0,
		\end{align}
		with $\kappa = k / a_i H_i$ and $\beta = ( 1 - 3 w ) / 2$. The general solution is
		\begin{align}
			\label{eq:WhittakerSolution} u(\xi) = C_1 M_{0, \mu} \left[ 2 i \kappa \left( \xi + \frac{1}{\alpha} \right) \right] + C_2 W_{0, \mu} \left[ 2 i \kappa \left( \xi + \frac{1}{\alpha} \right) \right],
		\end{align}
		in which $M_{\nu, \mu}(z)$ and $W_{\nu, \mu}(z)$ are the Whittaker functions, and
		\begin{align}
			\label{eq:mu} \mu = \frac{3}{2} \left| \frac{1 - w}{1 + 3 w} \right|.
		\end{align}
		Note that $w = -1/3$ is a singular case.%
		\footnote{The super-horizon spectrum is asymptotically divergent for $w = -1/3$ (or $\mu = 0$). To understand why, first note that in this case the universe does not accelerate nor decelerate ($\ddot{a} = 0$), so the comoving scale of Hubble horizon is constant in time. If $w$ is slightly smaller than $-1/3$, the universe accelerates but slowly. It takes a long time for the horizon to shrink a little. At the meantime the amplitudes of the fluctuations inside the horizon keep decaying, therefore the amplitude of the power spectrum changes much within a small range of $k$.} %
		Corresponding $\alpha$ and $\mu$ for some reference values of $w$ are listed in Table \ref{tab:EquationOfState}.


\begin{deluxetable}{rrrrrrrrrr}
\tablewidth{0pt}
\tablecolumns{10}
\tablecaption{Corresponding values of $\alpha$ and $\mu$ for some reference equation-of-state parameter $w$. The parameter $\alpha = (1 + 3 w) / 2$ is related to the scale factor by \eqref{eq:ScaleFactor}, and $\mu = | 3 ( 1 - w ) / 2 ( 1 + 3 w ) |$ describes the general solution of the perturbation through \eqref{eq:WhittakerSolution}. Note that for the accelerating universe with $w < -1/3$, one has $\alpha < 0$, while for the decelerating universe with $w > -1/3$, one has $\alpha > 0$.}
\tablehead{\vspace{-16.5mm}}
\startdata
$w$	&$-\infty$ &$-1$ &$\displaystyle{-\frac{2}{3}}$	&$\displaystyle{-\frac{1}{3}}$	&$0$ &$\displaystyle{\frac{1}{3}}$ &$\displaystyle{\frac{2}{3}}$ &$1$ &$+\infty$	\\ \\
	
	$\alpha$	&$-\infty$			&$-1$			&$\displaystyle{-\frac{1}{2}}$	&$0$			&$\displaystyle{\frac{1}{2}}$	&$1$				&$\displaystyle{\frac{3}{2}}$	&$2$	&$+\infty$	\\ \\
	
	$\mu$		&$\displaystyle{\frac{1}{2}}$		&$\displaystyle{\frac{3}{2}}$	&$\displaystyle{\frac{5}{2}}$	&$+\infty$	&$\displaystyle{\frac{3}{2}}$	&$\displaystyle{\frac{1}{2}}$	&$\displaystyle{\frac{1}{6}}$	&$0$	&$\displaystyle{\frac{1}{2}}$	\\
\enddata
\label{tab:EquationOfState}
\end{deluxetable}


		If we try to model the slow-roll evolution by assigning $w = -1$, we will end up with $A = 0$ and cannot proceed in the way we did in the previous paragraph. The way around that is to use the attractor solution of the slow-roll era. By writing the density and pressure in terms of field, we have
		\begin{align}
			A = -\frac{\phi'}{H},
		\end{align}
		assuming $\phi' < 0$ without loss of generality. In the attractor regime, $H$ as well as $\dot{\phi} = \phi' / a$ are approximately constant, so we can write
		\begin{align}
			\label{eq:SlowRollA} A = -\frac{\dot{\phi}_i}{H} a,
		\end{align}
		which is proportional to $a$ as it is in \eqref{eq:ARelation}. Also it can be verified by solving the Friedmann equation with constant $H$ that \eqref{eq:ScaleFactor} reproduces the scale factor in the slow-roll case, so the equation of motion \eqref{eq:MukhanovEq2} still holds. We will refer to the slow-roll limit as $w \simeq -1$ in this paper.
	
	For the superinflationary universe with $w < -1$, we model it by the Lagrangian,
		\begin{align}
			\label{eq:PhantomLagrangian} P = -X - V(\phi),
		\end{align}
		with the sign of the kinetic term reversed. With the requirement $\rho = -X + V > 0$, one generally has $V > X > 0$, and the scale factor still evolves as \eqref{eq:ScaleFactor}. By substituting the original definition of $A$ with
		\begin{align}
			A = \frac{a \sqrt{-\rho - P}}{H},
		\end{align}
		it turns out that the equation of motion of the curvature perturbation can still be written in the form of the Mukhanov-Sasaki equation \eqref{eq:MSEq}, and the rest of the analysis follows.

	
	\section{Assumption and Character of the Small-Scale Solution}
	
		\label{sec:InitialCondition}
		
		The common assumption on the initial condition is that the mode solution approaches the Minkowski limit in the short-wavelength limit,
		\begin{align}
			\label{eq:MinkowskiLimit} u(\eta) = \frac{1}{(2\pi)^{3/2} \sqrt{2 k}} e^{-i k \eta} \quad \textrm{(for $k \eta \rightarrow \infty$)}.
		\end{align}
		The Whittaker function that matches this form when $z \gg 1$ is
		\begin{align}
			W_{0, \mu}(2 i z) = e^{- i z} \left[ 1 + \mathcal{O}\left( \frac{1}{z} \right) \right].
		\end{align}
		Therefore, the requirement of matching the adiabatic vacuum picks out the solution,
		\begin{align}
			\label{eq:InitialCondition} u(\xi) = \frac{e^{i \kappa / \alpha}}{(2\pi)^{3/2} \sqrt{2 a_i H_i \kappa}} W_{0, \mu}\left[ 2 i \kappa \left( \xi + \frac{1}{\alpha} \right) \right].
		\end{align}
	
	If at early times of the inflationary history, the universe is initially accelerating and evolves with a constant equation of state, then by assuming the adiabatic vacuum at the sub-horizon limit, we obtain the initial condition \eqref{eq:InitialCondition}. This sub-horizon initial condition, combined with the quasi-de Sitter expansion, predicts the observed nearly scale invariant super-horizon spectrum in the $\Lambda$CDM universe. One of the reasons that make this scenario attractive is that the quantum fluctuations we learn well in the local Minkowski spacetime, after being stretched to the cosmic scale by inflation, also form the seed of the cosmic structure.
	
	When applying the limit \eqref{eq:MinkowskiLimit} to a decelerating universe, this sub-horizon assumption becomes acausal. In the decelerating universe, the Hubble horizon grows faster than the perturbations do, just like in the late-time universe dominated by matter or radiation. The sub-horizon spectrum is therefore formed after the super-horizon modes enter the horizon. The estimation we make to the super-horizon spectrum is then based on that, after the modes enter the horizon, they must fall in the vacuum state at the small-scale limit. Through the analysis in the next section, we will see that the large-scale power suppression caused by the preinflation kinetic era actually originates from the super-horizon spectrum deduced from this picture.

	
	\section{Scaling Relation}
	
		\label{sec:Power}
		
		The power spectrum of $\mathcal{R}$ is defined through the expectation value of $\hat{\mathcal{R}}^2 ( \bold{x}, t )$,
		\begin{align}
			\langle \hat{\mathcal{R}}^2 ( \bold{x}, t ) \rangle = \int \frac{d k}{k} P(k).
		\end{align}
		Expanding $\mathcal{R}$ in terms of the creation and annihilation operators,
			\begin{align}
				\mathcal{R}(\mathbf{x}, t) = \int d^3 \mathbf{k} \left[ a_{\mathbf{k}} \mathcal{R}_k(t) e^{i \mathbf{k}\cdot\mathbf{x}} + a_{\mathbf{k}}^{\dagger} \mathcal{R}_k^*(t) e^{-i \mathbf{k}\cdot\mathbf{x}} \right],
			\end{align}
			and using the commutator
			\begin{align}
				[ a_{\mathbf{k}}, a_{\mathbf{k}'}^{\dagger} ] = \delta^3(\mathbf{k} - \mathbf{k}'),
			\end{align}
			one finds that
		\begin{align}
			P(k) = 4 \pi k^3 |\mathcal{R}_k|^2.
		\end{align}
		Recalling that $\mathcal{R} = - u / A$, we find that, for $w \neq -1$ (so $\alpha \neq -1$), the power spectrum given by the solution \eqref{eq:InitialCondition} is
		\begin{align}
			\label{eq:power2} P = \frac{H_i^2 \kappa^2 | 1 + \alpha \xi |^{-2 / \alpha}}{\pi | 1 + \alpha |} \left| W_{0, \mu}\left[ 2 i \kappa \left( \xi + \frac{1}{\alpha} \right) \right] \right|^2.
		\end{align}
		For the slow-roll case, we approximate $A$ by the slow-roll limit \eqref{eq:SlowRollA}, and the mode function in that limit is then given by \eqref{eq:InitialCondition} with $w \simeq -1$. In terms of the slow-roll parameter,
		\begin{align}
			\epsilon \equiv &\frac{1}{16 \pi} \left( \frac{1}{V} \frac{\partial V}{\partial \phi} \right)^2 = \frac{4 \pi \dot{\phi}^2}{H^2},
		\end{align}
		the power spectrum at the slow-roll limit is given by
		\begin{align}
			\label{eq:power3} P = \frac{H_i^2 \kappa^2 | 1 + \alpha \xi |^{-2 / \alpha}}{\pi \epsilon} \left| W_{0, \frac{3}{2}}\left[ 2 i \kappa \left( \xi + \frac{1}{\alpha} \right) \right] \right|^2.
		\end{align}
	
	Using the small-argument expansion of the Whittaker function \cite{Olver2010}, the super-horizon limits of the power spectrum are found for different ranges of $w$. For $w < -1/3$ and $w > 1$ except $w = -1$, one has
		\begin{align}
			P_{\kappa \ll 1} = \frac{H_i^2 \Gamma^2( 2 \mu )}{\pi \Gamma^2( \mu + \frac{1}{2} ) | 1 + \alpha |} \left| \frac{\alpha}{2} \right|^{2 \mu - 1} \kappa^{-2 \mu + 3}.
		\end{align}
		For $w \simeq -1$, the slow-roll power spectrum \eqref{eq:power3} recovers the familiar scale-invariant spectrum
		\begin{align}
			P_{\kappa \ll 1} = \frac{H_i^2}{\pi \epsilon}.
		\end{align}
		For $-1/3 < w < 1$, the super-horizon power spectrum decays with time, given by
		\begin{align}
			P_{\kappa \ll 1} = \frac{H_i^2 \Gamma^2( 2 \mu )}{\pi \Gamma^2( \mu + \frac{1}{2} ) | 1 + \alpha |} \left| \frac{\alpha}{2} \right|^{2 \mu - 1} \kappa^{-2 \mu + 3} T(\xi),
		\end{align}
		with the time-dependence
		\begin{align}
			T(\xi) = | 1 + \alpha \xi |^{-6 ( 1 - w ) / ( 1 + 3 w )}.
		\end{align}
		For $w = 1$, the spectrum is
		\begin{align}
			P_{\kappa \ll 1} = \frac{H_i^2}{3 \pi^2} \kappa^3 T(\xi),
		\end{align}
		with
		\begin{align}
			T(\xi) = \left| \ln \left[ 2 i \kappa \left( \xi + \frac{1}{\alpha} \right) \right] + \gamma - 2 \ln 2 \right|^2,
		\end{align}
		where $\gamma$ is the Euler-Mascheroni constant.
	
	We can summarize the power-law relations of the super-horizon power spectrum with respect to the normalized wavenumber $\kappa$ by the scaling relation%
		\footnote{This relation is also derived in \cite{Cai2015} as the approximation to the super-horizon spectrum at the end of the multi-stage inflationary evolution. The authors focus on the recursive matrix formalism of the multi-stage preinflationary era, with the assumption that every preinflation era is an accelerating expansion (or decelerating contraction in the bounce inflation scenario).}
		\begin{align}
			\label{eq:ScalingRelation} P \propto \kappa^{-2 \mu + 3},
		\end{align}
		where the correspondence between the case of $w \simeq -1$ and the slow-roll limit is understood. For the convenience of readers, we also state this result in terms of the more familiar parameter, $w$. For $w < -1/3$ and $w > 1$,
		\begin{align}
			P \propto \kappa^{6 ( 1 + w ) / ( 1 + 3 w)}.
		\end{align}
		For $-1/3 < w < 1$,
		\begin{align}
			P \propto \kappa^{12 w / ( 1 + 3 w  )}.
		\end{align}
	
\begin{figure}

\centering

\includegraphics[width = 14 cm]{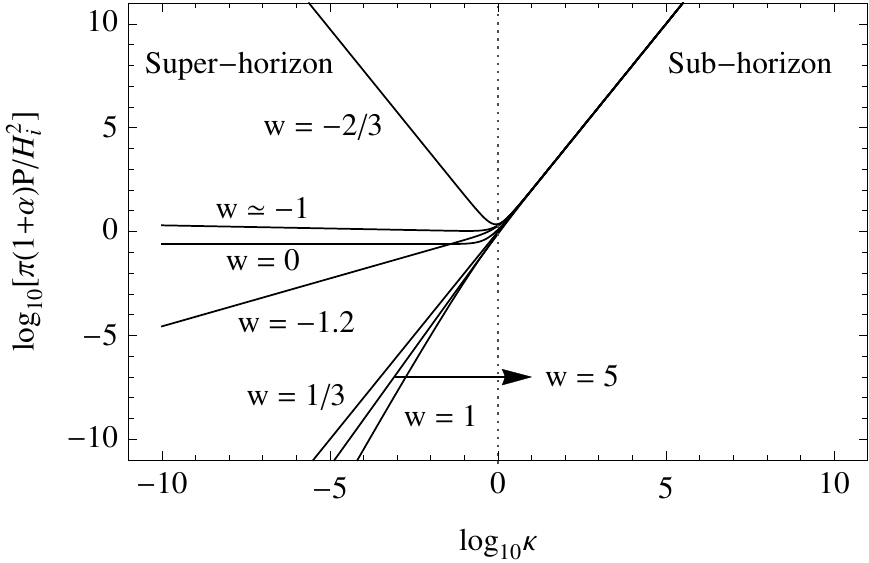}

\caption{The power spectra \eqref{eq:power2} with sample parameters $w = -1.2$, $-2/3$, 0, $1/3$, 1, 5, and at the slow-roll limit $w \simeq -1$. They are plotted with normalizations such that they have the same magnitude at the sub-horizon limit. For $w \simeq -1$ the power spectrum is given by \eqref{eq:power3}, and the normalization is instead $\pi \epsilon P / H_i^2$. The complete relationship between the slope of the super-horizon spectrum and the equation-of-state parameter is described by the scaling relation \eqref{eq:ScalingRelation}, and is plotted in Figure \ref{fig:InitialPower50}.}

\label{fig:InitialPower41}

\end{figure}

\begin{figure}

\centering

\includegraphics[width = 14 cm]{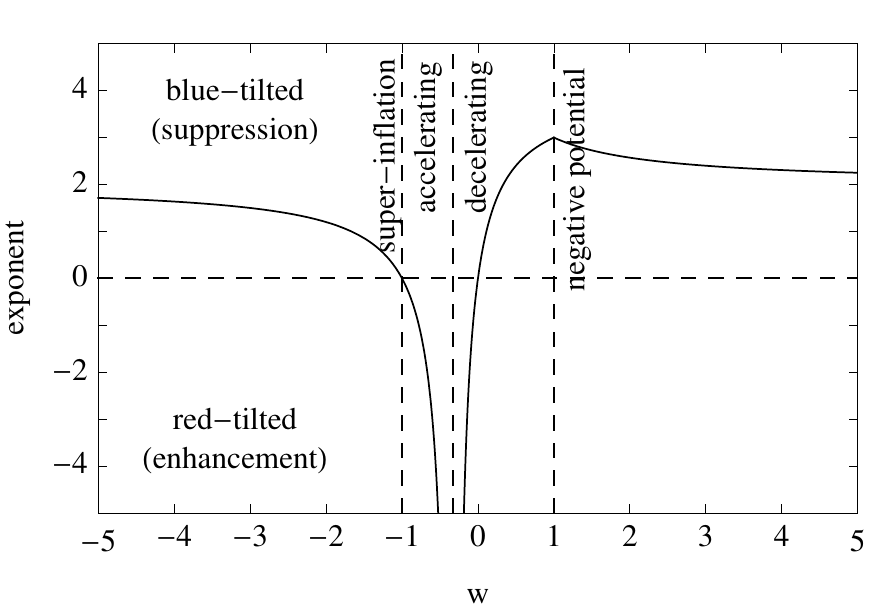}

\caption{The plot of $- 2 \mu + 3$, the exponent of the power-law scaling relation \eqref{eq:ScalingRelation} with respect to the equation-of-state parameter, $w$.}

\label{fig:InitialPower50}

\end{figure}

	The super-horizon behavior of the spectrum can be divided into three types according to the scaling relation. Some typical cases are plotted in Figure \ref{fig:InitialPower41}. For $\mu = 3/2$, the spectrum is scale-invariant. This is the case for $w \simeq -1$ (slow-roll) and $w = 0$. When $\mu < 3/2$, the spectrum is blue-tilted and the power is lower than the scale-invariant spectrum at super-horizon scales. This is attainable from the positive-pressure ($w > 0$) or superinflation era ($w < -1$). In the third case, the super-horizon spectrum is red-tilted, which is achieved when $\mu > 3/2$, or equivalently an era with $-1 < w < 0$ except $w = -1/3$ (Figure \ref{fig:InitialPower50}). Here we reiterate that the spectra obtained are based on the adiabatic vacuum \eqref{eq:MinkowskiLimit}, where the corresponding initial condition for the decelerating universe ($w > -1/3$) is acausal.


\chapter{Evolution of the Power Spectrum}{\label{ch:SingleField}}
	
	As the universe evolves from the preinflationary era to the inflationary era, the equation of state changes accordingly. To understand how the initial spectrum of the preinflationary era is constantly shaped by the expansion of the background geometry, eventually evolving into the super-horizon spectrum, we develop a useful technique to analyze this process: the spectrum evolution. By analyzing the evolution of the spectrum, we find that the character of the long-wavelength spectrum at the end of inflation actually reflects the nature of the initial state in the preinflationary era. If at early times before the onset of inflation, the universe is initially described by some adiabatic vacuum in the background with constant equation of state, the power spectrum is given by \eqref{eq:power2} during that era. As the universe evolves with time, although the power spectrum transforms continuously, the spectral shape at the long-wavelength limit is nevertheless not affected by the evolution; it is preserved in the late-time spectrum.
	
	We demonstrate in this chapter that the spectra can be radically different at large scales for distinct initial vacua. Particularly, if the universe transits from a kinetic era into the inflation era, the large-scale spectrum is suppressed because of the initial adiabatic vacuum assumed in the kinetic era. If the initial vacuum is different---for example, changed by an earlier accelerating era before the kinetic era, as discussed in this section---the large-scale spectrum may even become enhanced.
	
	We compare the evolution of the power spectrum in three cases, modeled phenomenologically by the single field dynamics. The first one is a single slow-roll era (denoted as era C) with Hubble parameter $H_C$. The second one is the slow-roll era (era C) preceded by a kinetic era (era B). In the third case we add one more slow-roll era (era A), with Hubble parameter $H_A$, before the kinetic era (era B), which is again followed by the slow-roll era (era C). When applicable, the quantities at the transition from era A to B are denoted by subscript 1 (so, for example, the scale factor at the transition is equal to $a_1$), and those at the transition from B to C are by subscript 2.
	
	In view of the acausal character of the adiabatic vacuum in the initially decelerating era (the second case with only era B and C), we also analyze the case of having a superinflation era (era S) before the slow-roll era (era C), which also implies power suppression at large scales but is free from the acausal property. Analogously, the quantities at the transition from era S to C are denoted by subscript 2.

	
	\section{Slow-Roll}
	
	\label{subsec:C}
	
	In the slow-roll era (era C), the general solution to the mode function at the slow-roll limit $w \simeq -1$ is
		\begin{align}
			\label{eq:SolutionC} u_C = \; &C_{+} \left( 1 - i \frac{\tilde{a}_C}{\tilde{k}_C} \right) e^{-i \tilde{k}_C / \tilde{a}_C} \notag \\
			&+ C_{-} \left( 1 + i \frac{\tilde{a}_C}{\tilde{k}_C} \right) e^{i \tilde{k}_C / \tilde{a}_C},
		\end{align}
		where $\tilde{k}_C = k / a_2 H_2$ and $\tilde{a}_C = a / a_2 = [ 1 - a_2 H_2 (\eta - \eta_2) ]^{-1}$. Here $a_2$ and $H_2$ can be viewed as quantities at some reference time, $\eta_2$. The notations are chosen for the convenience of later comparison and should not cause confusion. Note that in the slow-roll era, one can approximate $H_C \approx H_2$ as a constant. The normalized power spectrum is given by
		\begin{align}
			\label{eq:PowerC} &\tilde{P}_C = \frac{\tilde{k}_C^3}{\tilde{a}_C^2} \left| \tilde{C}_{+} \left( 1 - i \frac{\tilde{a}_C}{\tilde{k}_C} \right) e^{-i \tilde{k}_C / \tilde{a}_C} \right. \notag \\
			&\quad\quad\quad\quad\quad\quad\quad \left. + \tilde{C}_{-} \left( 1 + i \frac{\tilde{a}_C}{\tilde{k}_C} \right) e^{i \tilde{k}_C / \tilde{a}_C} \right|^2,
		\end{align}
		where $\tilde{C}_{\pm} = \sqrt{a_2 H_2} C_{\pm}$, and
		\begin{align}
			\tilde{P}_C = &\frac{\epsilon P}{16 \pi^2 H_C^2}, \\
			\epsilon = &\left. \frac{4 \pi \dot{\phi}^2}{H^2} \right|_C.
		\end{align}
		are the normalized power spectrum and the slow-roll parameter evaluated in era C, respectively.
	
\begin{figure}

\centering

\includegraphics[width = 14 cm]{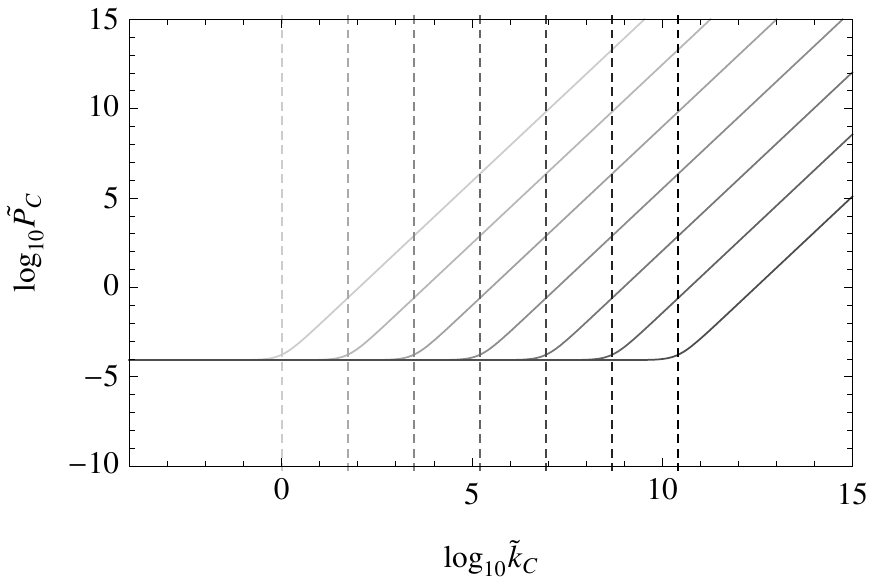}

\caption{Time evolution of the power spectrum in era C in the case of single-stage evolution (only era C). The solid curves are the spectra of $\mathcal{R}$ from early time to late time (from light to dark). The vertical dashed lines denote the comoving horizon size at the corresponding instants (also from light to dark).}

\label{fig:CinC70}

\end{figure}

	In the adiabatic vacuum \eqref{eq:MinkowskiLimit}, only the second term in \eqref{eq:SolutionC} remains, and the mode function reduces to
		\begin{align}
			u_C = \frac{1}{(2 \pi)^{3/2} \sqrt{2 k}} \left( 1 + i \frac{\tilde{a}_C}{\tilde{k}_C} \right) e^{i \tilde{k}_C / \tilde{a}_C}.
		\end{align}
		The power spectrum is
		\begin{align}
			\tilde{P}_C = \frac{1}{16 \pi^3} \left( 1 + \frac{\tilde{k}_C^2}{\tilde{a}_C^2} \right),
		\end{align}
		which recovers the well-known form in the super-horizon limit,
		\begin{align}
			\label{eq:SlowRollSpectrum} P = \frac{1}{\pi} \left( \frac{H_C^2}{\epsilon} \right). \quad \textrm{( $\tilde{k}_C \ll 1$ at $\tilde{a}_C = 1$ )}
		\end{align}
		The comoving horizon size decreases with time, moving toward the right to the small scales in Figure \ref{fig:CinC70}. After the mode exits the horizon, lying on the left-hand side of the horizon scale, the power stays scale-invariant.


	\section{Kinetic---Slow-Roll}
	
	\label{subsec:BC}
	
		In the second case, the energy density is dominated by the kinetic energy at $\eta < \eta_2$, with
		\begin{align}
			H \approx -\sqrt{ \frac{4 \pi}{3} } \frac{\phi'}{a},
		\end{align}
		assuming $\phi' \leq 0$ without lose of generality. In the kinetic era, $w = 1$ and the mode function can be written in terms of the Hankel functions,
		\begin{align}
			\label{eq:SolutionB} u_B = &B_{+} \tilde{a}_B H_0^{(1)}\left( \frac{1}{2} \tilde{k}_B \tilde{a}_B^2 \right) \notag \\
			&+ B_{-} \tilde{a}_B H_0^{(2)}\left( \frac{1}{2} \tilde{k}_B \tilde{a}_B^2 \right),
		\end{align}
		where we denote $\tilde{k}_B = k / a_1 H_1$ and $\tilde{a}_B = a / a_1 = \sqrt{1 + 2 a_1 H_1 ( \eta - \eta_1 )}$. Similarly, for later convenience, we choose $\eta_1 < \eta_2$ to be some reference time for era B. The power spectrum in the kinetic era is given by
		\begin{align}
			\label{eq:PowerB} &\tilde{P}_B = \tilde{k}_B^3 \left| \tilde{B}_{+} H_0^{(1)}\left( \frac{1}{2} \tilde{k}_B \tilde{a}_B^2 \right) \right. \notag \\
			&\quad\quad\quad\quad \left. + \tilde{B}_{-} H_0^{(2)}\left( \frac{1}{2} \tilde{k}_B \tilde{a}_B^2 \right) \right|^2,
		\end{align}
		where $\tilde{B}_{\pm} = \sqrt{a_1 H_1} B_{\pm}$, and
		\begin{align}
			\tilde{P}_B = \frac{3 P}{16 \pi^2 H_1^2}.
		\end{align}
	
	At $\eta > \eta_2$, the universe shifts into the slow-roll stage, and the solution is given by \eqref{eq:SolutionC}. To match the boundary between the eras, we fix the coefficients $C_{+}$ and $C_{-}$ by the continuity of $\mathcal{R}$ and $\mathcal{R}'$. One finds
		\begin{align}
			\label{eq:CoefficientCp} &\tilde{C}_{+} = \frac{e^{i \tilde{k}_C}}{2 \tilde{k}_C} \left\{ \left[ \tilde{k}_C H_{0, C}^{(1)} - ( 1 - i \tilde{k}_C ) H_{1, C}^{(1)} \right] \tilde{B}_{+} \right. \notag \\
			&\quad\quad\quad\quad\quad \left. + \left[ \tilde{k}_C H_{0, C}^{(2)} - ( 1 - i \tilde{k}_C ) H_{1, C}^{(2)} \right] \tilde{B}_{-} \right\}, \\
			\label{eq:CoefficientCm} &\tilde{C}_{-} = \frac{e^{-i \tilde{k}_C}}{2 \tilde{k}_C} \left\{ \left[ \tilde{k}_C H_{0, C}^{(1)} - ( 1 + i \tilde{k}_C ) H_{1, C}^{(1)} \right] \tilde{B}_{+} \right. \notag \\
			&\quad\quad\quad\quad\quad \left. + \left[ \tilde{k}_C H_{0, C}^{(2)} - ( 1 + i \tilde{k}_C ) H_{1, C}^{(2)} \right] \tilde{B}_{-} \right\},
		\end{align}
		where $H_{0, C}^{(1)}$ is the shorthand of $H_0^{(1)}( \tilde{k}_C /2 )$ and so on. The power spectrum in era C is given by \eqref{eq:PowerC} with $\tilde{C}_{+}$ and $\tilde{C}_{-}$ substituted by \eqref{eq:CoefficientCp} and \eqref{eq:CoefficientCm}, respectively.
	
\begin{figure}

\centering

\includegraphics[width = 14 cm]{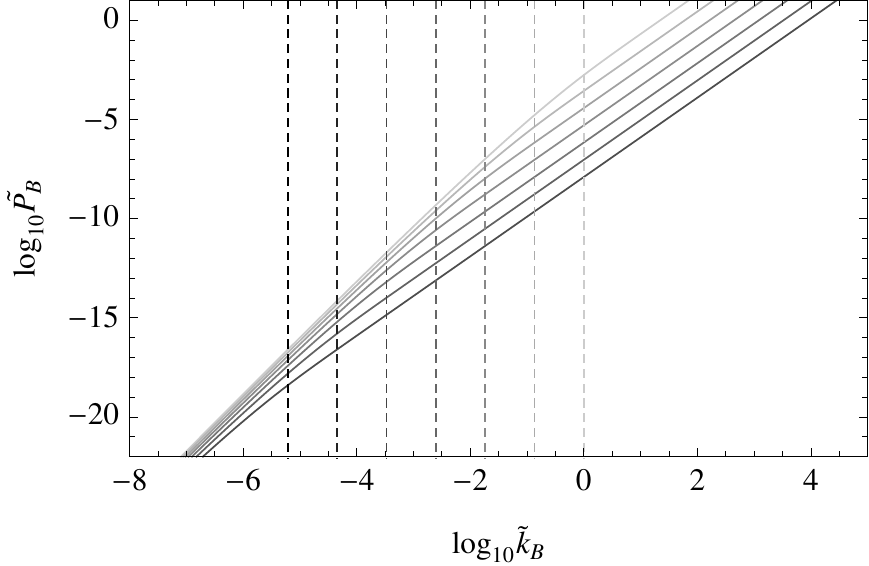}

\caption{Time evolution of the power spectrum in era B in the case of two-stage evolution (era B and C). The solid curves are the spectra of $\mathcal{R}$ from early time to late time (from light to dark). The vertical dashed lines denote the comoving horizon size at the corresponding instants (also from light to dark).}

\label{fig:BCinB70}

\end{figure}

	If the universe is in the adiabatic vacuum in era B, only the second term in \eqref{eq:SolutionB} presents in the mode function,
		\begin{align}
			u_B = \frac{1}{8 \pi} \tilde{a}_B H_0^{(2)}\left( \frac{1}{2} \tilde{k}_B \tilde{a}_B^2 \right),
		\end{align}
		where the normalization is chosen to recover the small-scale limit \eqref{eq:MinkowskiLimit}. In the kinetic era, the comoving horizon size increases with time, moving toward the left to the large scale (Figure \ref{fig:BCinB70}). The power drops after the mode enters the horizon, as in the slow-roll era, but it also decreases at the super-horizon scales before the horizon entry. This is actually a salient feature of the adiabatic vacuum in the decelerating era with $-1/3 < w < 1$.


\begin{figure}

\centering

\includegraphics[width = 14 cm]{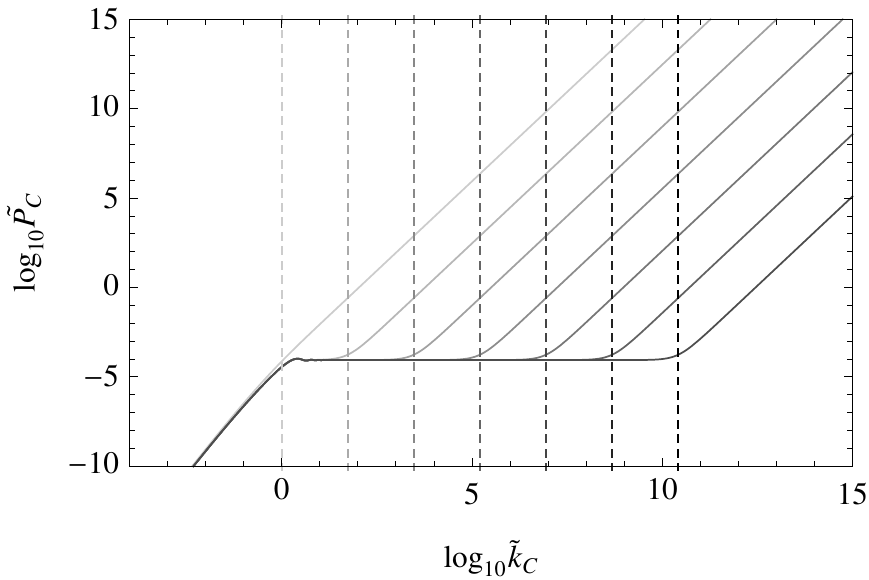}

\caption{Time evolution of the power spectrum in era C in the case of two-stage evolution (era B and C). The solid curves are the spectra of $\mathcal{R}$ from early time to late time (from light to dark). The vertical dashed lines denote the comoving horizon size at the corresponding instants (also from light to dark).}

\label{fig:BCinC70}

\end{figure}

	The evolution in the following slow-roll era (era C) demonstrates how the imprint of the initial vacuum is left at the large scales of the power spectrum at the end of inflation (modeled by era C). In the slow-roll era, the comoving horizon size decreases, going toward the right, and the modes exit the horizon (Figure \ref{fig:BCinC70}). The super-horizon modes have two different types of history. The ones with longest wavelengths have not entered the horizon yet at the end of era B, and stay outside the horizon in era C. These modes preserve the blue-tilted spectrum, and account for the power suppression induced by the kinetic era (cf.~\cite{Contaldi2003}). The other modes with shorter wavelengths enter the horizon in era B, and exit the horizon in era C. They are scale-invariant outside the horizon, as in the case of the single slow-roll scenario. This is because the adiabatic vacua approach the same short-wavelength limit \eqref{eq:MinkowskiLimit} in either the kinetic era or the slow-roll era. Therefore, although the spectrum has a scale-invariant segment due to the slow-roll era, the largest scales of the spectrum reveal the initial vacuum stemming from the kinetic era.


\begin{figure}

\centering

\includegraphics[width = 14 cm]{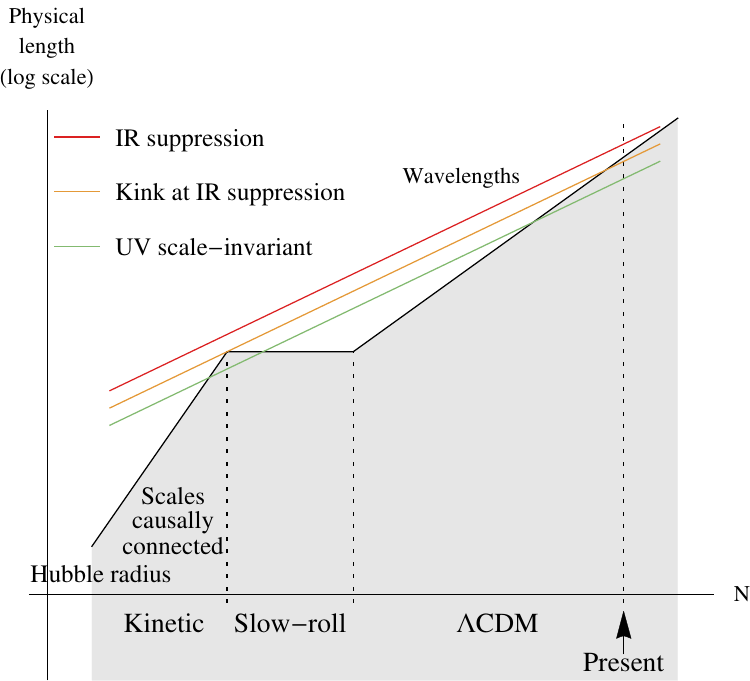}

\caption{Illustration of the evolution of the physical wavelengths of the modes and the Hubble radius with respect to the number of e-fold, $N$. The three parallel straight lines denote the modes with three different wavelengths, long to short from top to bottom (color online). The corresponding features they generate in the power spectrum, Figure \ref{fig:BCinC70}, are labeled in the legends (top to bottom corresponding to long to short wavelengths). The black piecewise-connected lines denote the Hubble radius evolving from the kinetic era to the slow-roll era, and finally into the $\Lambda$CDM era. The shaded region denotes the scales within which are causally connected.}

\label{fig:HorizonExitBC20}

\end{figure}

	The results we obtain crucially depend on the assumption about the initial vacuum of the universe. The prediction of power suppression is challenged by the fact that it originates from the blue-tilted super-horizon initial spectrum in the preinflation decelerating era, in which the super-horizon modes have not been in causal contact throughout the history. We demonstrate this point by an illustration showing the evolution of the Fourier wavelengths and the Hubble horizon (Figure \ref{fig:HorizonExitBC20}). In the decelerating universe, such as the initial kinetic era, the Hubble horizon grows faster than the Fourier wavelengths do. At the end of the initial decelerating era and the beginning of the accelerating inflation, the modes that are about to enter the horizon---those who are the origin of the suppressed large-scale modes today---will soon be expanded and kept outside the horizon by inflation. If the universe starts with the decelerating era before inflation, these mode are then insulated from any sub-horizon dynamics throughout the history. Therefore, the spectrum of the perturbations beyond the horizon size at the end of the decelerating era is not the consequence of causal physics, and the common approach of deducing the initial conditions through requiring the spectrum recover the Minkowski limit at the sub-horizon scale is therefore \emph{a posteriori}. This is of the same footing as the ``horizon problem'' the big bang cosmology faced before the picture of inflation was introduced.


	\section{Slow-Roll---Kinetic---Slow-Roll}
	
	\label{subsec:ABC}
	
		With the additional slow-roll era (era A) prepending the kinetic era (era B), we simply apply solution \eqref{eq:SolutionC} at $\eta < \eta_1$,
		\begin{align}
			\label{eq:SolutionA} u_A = \; &A_{+} \left( 1 - i \frac{\tilde{a}_A}{\tilde{k}_A} \right) e^{-i \tilde{k}_A / \tilde{a}_A} \notag \\
			&+ A_{-} \left( 1 + i \frac{\tilde{a}_A}{\tilde{k}_A} \right) e^{i \tilde{k}_A / \tilde{a}_A},
		\end{align}
		where $\tilde{k}_A = k / a_1 H_1$ and $\tilde{a}_A = a / a_1 = [ 1 - a_1 H_1 ( \eta - \eta_1 ) ]^{-1}$. The mode function of the adiabatic vacuum in era A is
		\begin{align}
			\label{eq:EraAVacuum} u_A = \frac{1}{(2 \pi)^{3/2} \sqrt{2 k}} \left( 1 + i \frac{\tilde{a}_A}{\tilde{k}_A} \right) e^{i \tilde{k}_A / \tilde{a}_A}.
		\end{align}
		Matching the boundary between era A and B, one finds the adiabatic vacuum in era A excites both modes of \eqref{eq:SolutionB} in era B with coefficients
		\begin{align}
			\tilde{B}_{+} = &-\frac{e^{i \tilde{k}_B}}{32 \sqrt{\pi \tilde{k}_B}} \left[ \tilde{k}_B H_{0, B}^{(2)} - ( 1 - i \tilde{k}_B ) H_{1, B}^{(2)} \right], \\
			\tilde{B}_{-} = &\frac{e^{i \tilde{k}_B}}{32 \sqrt{\pi \tilde{k}_B}} \left[ \tilde{k}_B H_{0, B}^{(1)} - ( 1 - i \tilde{k}_B ) H_{1, B}^{(1)} \right].
		\end{align}
		These results can be fed into \eqref{eq:CoefficientCp} and \eqref{eq:CoefficientCm}, identifying $H_1 = H_A$ as the Hubble constant in era A, and obtain the initial coefficients $\tilde{C}_{\pm}$ in era C. The power spectrum in era C is again given by \eqref{eq:PowerC} with the $\tilde{C}_{\pm}$ found.
	
\begin{figure}

\centering

\includegraphics[width = 14 cm]{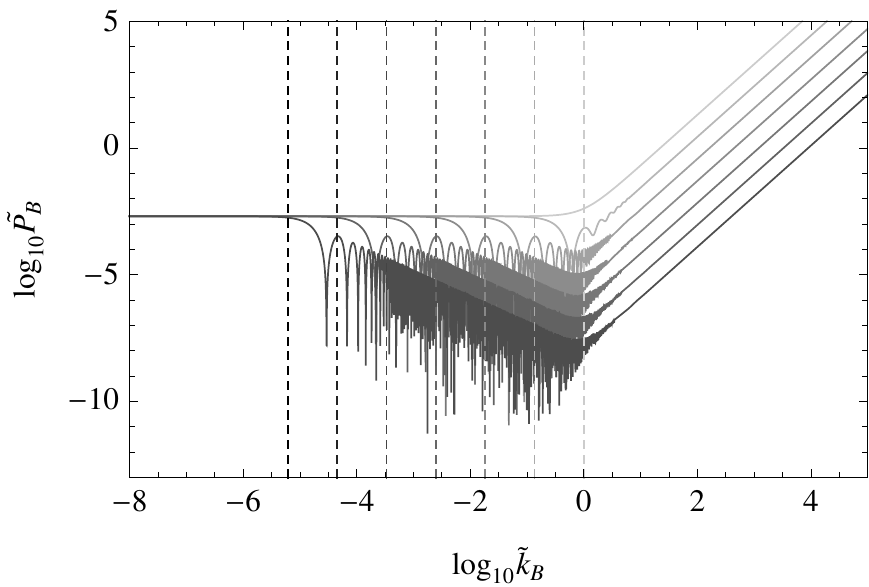}

\caption{Time evolution of the power spectrum in era B in the case of three-stage evolution (era A, B, and C). The solid curves are the spectra of $\mathcal{R}$ from early time to late time (from light to dark). The vertical dashed lines denote the comoving horizon size at the corresponding instants (also from light to dark).}

\label{fig:ABCinB70}

\end{figure}

	With the initial vacuum in the slow-roll era (era A), in which the spectrum evolves in the same way as it does in Figure \ref{fig:CinC70}, the super-horizon spectrum in era B is scale-invariant (Figure \ref{fig:ABCinB70}), different from the blue-tilted spectrum of the adiabatic vacuum in era B (Figure \ref{fig:BCinB70}). Moreover, after the modes enter the horizon, the power decreases and the spectrum becomes red-tilted (Figure \ref{fig:ABCinB70}), opposite to the blue-tilted spectrum in era B without era A (Figure \ref{fig:BCinB70}).


\begin{figure}

\centering

\includegraphics[width = 14 cm]{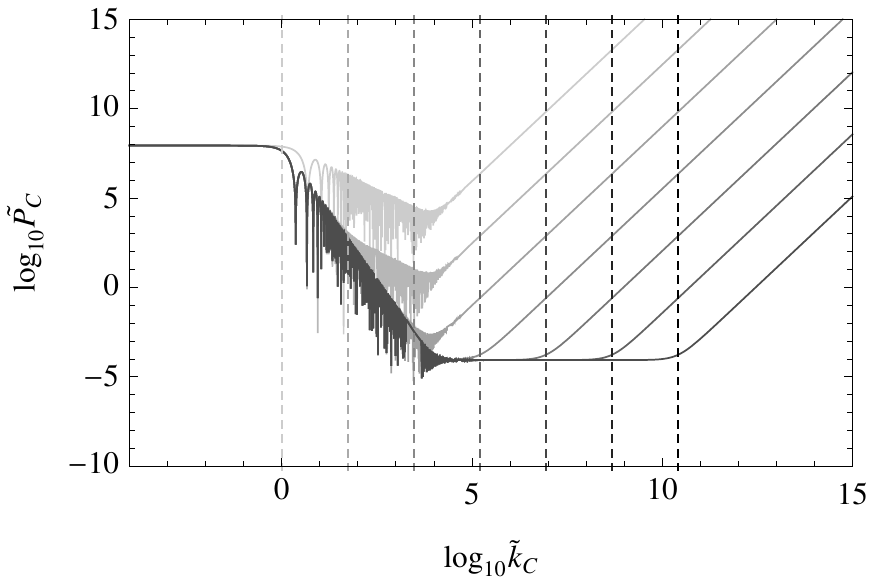}

\caption{Time evolution of the power spectrum in era C in the case of three-stage evolution (era A, B, and C). The solid curves are the spectra of $\mathcal{R}$ from early time to late time (from light to dark). The vertical dashed lines denote the comoving horizon size at the corresponding instants (also from light to dark).}

\label{fig:ABCinC70}

\end{figure}

	In era C, there are three different types of history inherited by the super-horizon modes (Figure \ref{fig:ABCinC70}). The modes with the longest wavelengths do not enter the horizon in era B and are kept outside of the horizon in era C. They preserve the scale-invariant spectrum as the proof of the existence of the initial vacuum in a slow-roll era (era A). For the modes with shorter wavelengths that enter the horizon in era B and exit the horizon in era C, the red-tilted shape of the spectrum persists, denting more as the power decreases inside the horizon. The super-horizon modes with shortest wavelengths exit the horizon for the first time in era C. These modes behave like the ones in era A, leaving the power scale-invariant as they exit the horizon. Note the magnitude of the scale-invariant spectrum generated in era C is lower than that generated in era A, since according to \eqref{eq:SlowRollSpectrum} the mode exiting the horizon from the adiabatic vacuum acquires the power that is proportional to the Hubble expansion rate, which is lower in era C.


\begin{figure}

\centering

\includegraphics[width = 14 cm]{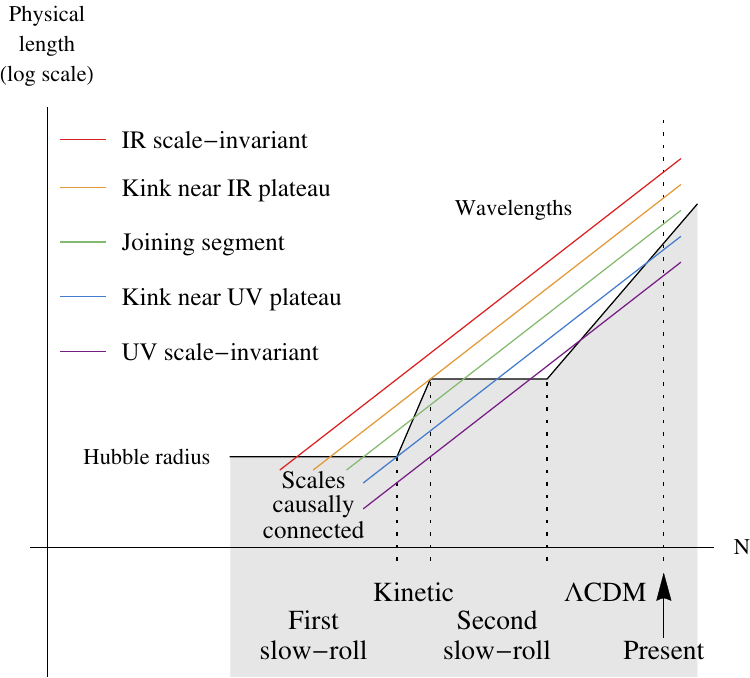}

\caption{Illustration of the evolution of the physical wavelengths of the modes and the Hubble radius with respect to the number of e-fold, $N$. The five parallel straight lines denote the modes with different wavelengths, long to short from top to bottom (color online). The corresponding features they generate in the power spectrum, Figure \ref{fig:ABCinC70}, are labeled in the legends (top to bottom corresponding to long to short wavelengths). The black piecewise-connected lines denote the Hubble radius evolving from the first slow-roll era to the kinetic era, then to the second slow-roll era, finally into the $\Lambda$CDM era. The shaded region denotes the scales within which are causally connected.}

\label{fig:HorizonExitABC40}

\end{figure}

	With the accelerating era A before the kinetic era B, all the modes are initially sub-horizon at the early times and are causal connected (Figure \ref{fig:HorizonExitABC40}). However, although this scenario is free from the acausal initial conditions, the intermediate kinetic era no longer leads to power suppression at the large scales. The accelerating era preceding the kinetic era changes the initial conditions, generating the the scale-invariant spectrum at the large scales. The power spectrum now has two scale-invariant segments: one generated in era A with larger power at the larger scales, and the other generated in era C with lower power at the smaller scales. The intermediate kinetic era in this case generates the spectrum that connects the two scale-invariant segments (Figure \ref{fig:ABCinC70}).


	We obtain three different large-scale behaviors from three different initial vacua and intermediate evolutions. Matching them with the inflation scenarios, we have the following interpretations. In the first case, with a single era C, the super-horizon spectrum is scale-invariant. This corresponds to the picture of inflation with large e-folding numbers (much larger than 60 e-folds). In the second case, the spectrum is suppressed at the large scales if the universe is in the adiabatic vacuum of era B before the onset of era C. This is the scenario of having a fast-roll era before the ``just enough'' inflation (with about 50--60 e-folds). In the third case, if before the fast-roll era (era B) the universe is in another slow-roll era (era A) at the early times, the spectrum is enhanced at the large scales. This corresponds to many supersymmetry or string motivated models that manifest the fast-roll era as a transient between two slow-roll eras \cite{Jain2009, Dudas2012, Yamauchi2011}.

	
	\section{Superinflation---Slow-Roll}
	
	\label{subsec:SC}
	
		Consider the case when $\eta < \eta_2$ the universe is in the superinflation era (era S), which is modeled by the Lagrangian \eqref{eq:PhantomLagrangian}. With $w < -1$, the mode function is
			\begin{align}
				\label{eq:modeS} u_S = &S_+ M_{0, \mu}\left( \frac{2 i \tilde{k}_S}{\alpha} \tilde{a}_S^{\alpha} \right) \notag \\
				&+ S_- W_{0, \mu}\left( \frac{2 i \tilde{k}_S}{\alpha} \tilde{a}_S^{\alpha} \right),
			\end{align}
		where $\tilde{k}_S = k / a_1 H_1$, $\tilde{a}_S = a / a_1 = [ 1 + \alpha a_1 H_1 ( \eta - \eta_1 ) ]^{1/\alpha}$, and $\alpha$ and $\mu$ are given by \eqref{eq:alpha} and \eqref{eq:mu}, respectively. Here $\eta_1 < \eta_2$ also denotes the reference time for era S. The normalized power spectrum is
			\begin{align}
				\label{eq:PowerS} \tilde{P}_S = &\frac{\tilde{k}_S^3}{\tilde{a}_S^2} \left| \tilde{S}_+ M_{0, \mu} \left( \frac{2 i \tilde{k}_S}{\alpha} \tilde{a}_S^{\alpha} \right) \right. \notag \\
				&\quad\quad \left. + \tilde{S}_- W_{0, \mu}\left( \frac{2 i \tilde{k}_S}{\alpha} \tilde{a}_S^{\alpha} \right) \right|^2,
			\end{align}
			where $\tilde{S}_{\pm} = \sqrt{a_1 H_1} S_{\pm}$ and
			\begin{align}
				\label{eq:PowerSDef} \tilde{P}_S = \frac{-(1+\alpha) P_S}{16 \pi^2 H_1^2}.
			\end{align}
		
		The universe goes into the slow-roll era (era C) at $\eta_2$, with mode function given by \eqref{eq:SolutionC}. Using the same matching conditions, the coefficients $\tilde{C}_{\pm}$ are found to be
		\begin{align}
			\label{eq:cp} \tilde{C}_+ = &\frac{\sqrt{\epsilon} e^{-\frac{1}{2} \alpha N_S} e^{i \tilde{k}_C}}{4 \sqrt{-(1+\alpha)} \tilde{k}_C^2} \notag \\
			&\times \left\{ 2 \left[ 2 \tilde{k}_C^2 + 2 i \tilde{k}_C - 1 \right] \tilde{S}_+ M_{0, \mu} \right. \notag \\
			&\quad + 2 \left[ 2 \tilde{k}_C^2 + 2 i \tilde{k}_C - 1 \right] \tilde{S}_- W_{0, \mu} \notag \\
			&\quad + \alpha ( 1 - i \tilde{k}_C ) ( 2 \mu + 1 ) \tilde{S}_+ M_{1, \mu} \notag \\
			&\quad \left. - 2 \alpha ( 1 - i \tilde{k}_C ) \tilde{S}_- W_{1, \mu} \right\}, \\
			\label{eq:cm} \tilde{C}_- = &\frac{\sqrt{\epsilon} e^{-\frac{1}{2} \alpha N_S} e^{-i \tilde{k}_C}}{4 \sqrt{-(1+\alpha)} \tilde{k}_C^2} \notag \\
			&\times \left\{ -2 \tilde{S}_+ M_{0, \mu} \right. \notag \\
			&\quad -2 \tilde{S}_- W_{0, \mu} \notag \\
			&\quad + \alpha ( 2 \mu + 1 ) ( 1 + i \tilde{k}_C ) \tilde{S}_+ M_{1, \mu} \notag \\
			&\quad \left. - 2 \alpha ( 1 + i \tilde{k}_C ) \tilde{S}_- W_{1, \mu} \right\},
		\end{align}
			with all Whittaker functions evaluated at $2 i \tilde{k}_C / \alpha$. The slow-roll parameter in era C is still given by $\epsilon$, and $N_S$ is the number of e-folds from $\eta_1$ to $\eta_2$.
		
\begin{figure}

\centering

\includegraphics[width = 14 cm]{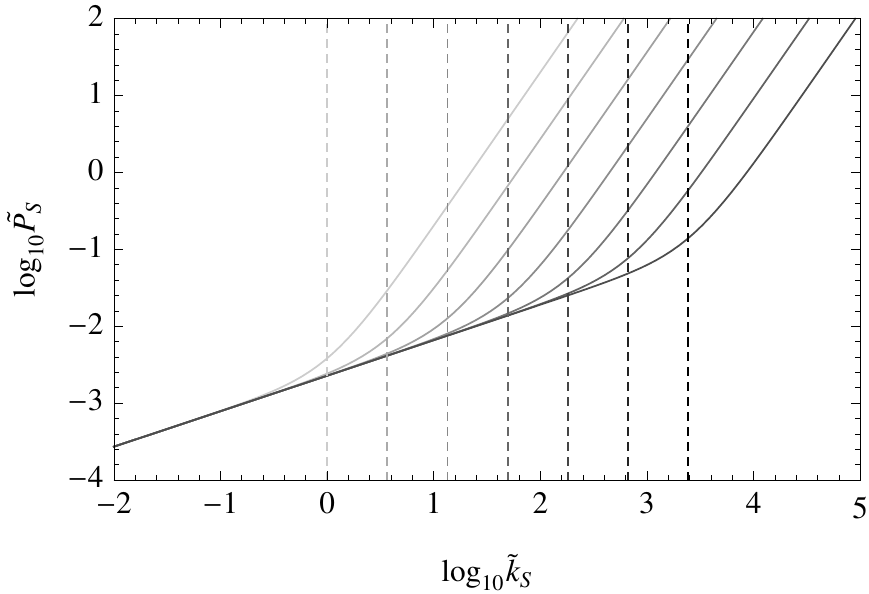}

\caption{Time evolution of the power spectrum in era S in the case of two-stage evolution (era S and C), with $w = -1.2$. The solid curves are the spectra of $\mathcal{R}$ from early time to late time (from light to dark). The vertical dashed lines denote the comoving horizon size at the corresponding instants (also from light to dark).}

\label{fig:SCinS22}

\end{figure}

		The adiabatic vacuum \eqref{eq:InitialCondition} in era S corresponds to the coefficients
			\begin{align}
				\tilde{S}_+ = &0, \\
				\tilde{S}_- = &\frac{\exp\left( \frac{i \tilde{k}_C}{\alpha} e^{-\alpha N_S} \right)}{(2\pi)^{3/2} \sqrt{2 \tilde{k}_C} e^{-\alpha N_S / 2}},
			\end{align}
			where we have used the relation $\tilde{k}_S = \tilde{k}_C e^{-\alpha N_S}$. The power spectrum in era S is plotted in Figure \ref{fig:SCinS22}, taking $w = -1.2$ as an example. Similar to the case of the kinetic era (era B), the super-horizon spectrum is blue-tilted, but in the superinflation era the comoving Hubble radius decreases, and the super-horizon spectrum remains constant in time.
		
\begin{figure}

\centering

\includegraphics[width = 14 cm]{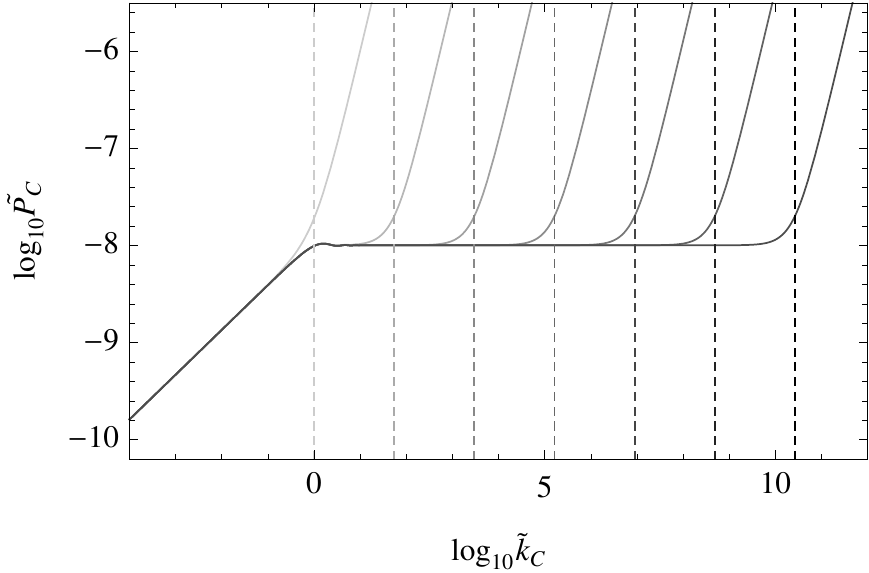}

\caption{Time evolution of the power spectrum in era C in the case of two-stage evolution (era S and C), with $w = -1.2$, $N_S = 6$, and $\epsilon = 0.1$. The solid curves are the spectra of $\mathcal{R}$ from early time to late time (from light to dark). The vertical dashed lines denote the comoving horizon size at the corresponding instants (also from light to dark).}

\label{fig:SCinC260}

\end{figure}

		As the universe enters the slow-roll era (era C), the super-horizon modes have two different types of history (Figure \ref{fig:SCinC260}). The modes with longer wavelengths exit the horizon in era S, retaining the blue-tilted super-horizon spectrum and staying constant in era C. The modes with shorter wavelengths exit the horizon in era C, acquiring the scale-invariant spectrum at horizon crossing.


\clearpage


\chapter{Two-field cascade inflation}{\label{ch:TwoField}}

	After establishing the understanding of the spectrum evolution in multi-stage inflation using the \emph{ad hoc} single-field analysis, in this chapter we calculate the spectrum generated by a two-stage inflation model from the given potential. The purposes are to show that the evolution pattern we obtain in the single-field analysis is also reflected in the two-field dynamics, and to check whether the large-scale power is suppressed due to the coupling between the two fields.


\section{Background Evolution and the Attractor Solution}

	We investigate a simple two-field cascade inflation, which is driven by a heavy scalar field $\psi$ and a light scalar field $\phi$ with the action
		\begin{align}
			\label{eq:action} S = \int d^4 x \sqrt{-g} \left[ -\frac{1}{2} \partial_{\alpha} \phi \partial^{\alpha} \phi -\frac{1}{2} \partial_{\alpha} \psi \partial^{\alpha} \psi - V(\phi, \psi) \right].
		\end{align}
		In large field inflation, the field operates at the super-Plankian scale when the coefficient of the kinetic term is normalized to $1/2$. If the field value is of the order of the Planck mass, $M_P = 1 / \sqrt{G}$, the energy scale of inflation is determined by the mass of the field. The typical evolution can therefore be divided into four stages. The first stage is the inflationary era driven by the heavy field, with fields starting far away from the potential minimum and rolling down alone the hillside of the potential. At the second stage the heavy field falls into the potential minimum and oscillates with damping amplitude. As the energy drops to the scale of the light field mass, the universe enters the third stage in which the light field initiates the other inflation era. Finally at the fourth stage the light field decays, ending the inflation, and the standard $\Lambda$CDM evolution begins.
	
	Consider the cascade inflation realized by the potential
		\begin{align}
			V(\phi, \psi) = \frac{\lambda'}{2} \phi^2 \psi^2 + \frac{1}{2} m^2 \phi^2.
		\end{align}
		The heavy-field inflation is driven by the coupling term $\lambda' \phi^2 \psi^2 / 2$, and the light-field inflation is driven by the mass term $m^2 \phi^2 / 2$. At the stage of heavy-field inflation, the fields follow the attractor solutions, which can be obtained through expressing the equations of motion in terms of the number of e-folds, $N \equiv \ln ( a / a_i )$, where $a_i$ is some initial scale factor. The Friedmann equation is
		\begin{align}
			\label{eq:Friedmann} H^2 = \frac{\frac{8 \pi}{3} V}{1 - \frac{4 \pi}{3} \left[ \left( \frac{d \phi}{d N} \right)^2 + \left( \frac{d \psi}{d N} \right)^2 \right]},
		\end{align}
		where $H = \dot{a} / a$. The dots denote the derivatives with respect to $t$. The equations of motion can be casted into the form
		\begin{align}
			\label{eq:phiEOMinN2} &\frac{d^2 \phi}{d N^2} - 4 \pi \left[ \left( \frac{d \phi}{d N} \right)^2 + \left( \frac{d \psi}{d N} \right)^2 - \frac{3}{4 \pi} \right] \notag \\
			&\quad\quad\quad\quad \times \left( \frac{d \phi}{d N} + \frac{1}{8 \pi V} \frac{\partial V}{\partial \phi} \right) = 0, \\
			\label{eq:psiEOMinN2} &\frac{d^2 \psi}{d N^2} - 4 \pi \left[ \left( \frac{d \phi}{d N} \right)^2 + \left( \frac{d \psi}{d N} \right)^2 - \frac{3}{4 \pi} \right] \notag \\
			&\quad\quad\quad\quad \times \left( \frac{d \psi}{d N} + \frac{1}{8 \pi V} \frac{\partial V}{\partial \psi} \right) = 0.
		\end{align}
		It can be shown that the attractor solutions satisfy
		\begin{align}
			\label{eq:phiAttractor} \frac{d \phi}{d N} = &-\frac{1}{8 \pi V} \frac{\partial V}{\partial \phi}, \\
			\label{eq:psiAttractor} \frac{d \psi}{d N} = &-\frac{1}{8 \pi V} \frac{\partial V}{\partial \psi},
		\end{align}
		provided $\left( d \phi / d N \right)^2 + \left( d \psi / d N \right)^2 - 3 / 4 \pi < 0$ as constrained by \eqref{eq:Friedmann}.%
		\footnote{For $\phi$, the type of the attractor considered here is the horizontal trajectory on the $\phi$--$d \phi / d N$ plane; that is, the one with $(d / d \phi) (d \phi / d N) = 0$. Since $d^2 \phi / d N^2 = (d \phi / d N) \times (d / d \phi) (d \phi / d N)$, we obtain the condition for the attractor solution with non-vanishing $d \phi / d N$ as $d^2 \phi / d N^2 = 0$, which in turn leads to \eqref{eq:phiAttractor}. Similar argument applies to $\psi$.} %
		With potential dominated by $\lambda' \phi^2 \psi^2 / 2$, the attractor solutions are
		\begin{align}
			\label{eq:phiStage1Case1} \phi(N) &= \sqrt{ \phi_i^2 - \frac{1}{2 \pi} ( N - N_i )}, \\
			\label{eq:psiStage1Case1} \psi(N) &= \sqrt{ \psi_i^2 - \frac{1}{2 \pi} ( N - N_i )},
		\end{align}
		where the subscripts $i$ denote the initial values.
		
		At the second stage, $\psi$ exits the slow-roll regime and oscillates at the minimum of the potential with its amplitude damped with time. The potential is still dominated by the coupling term before the next inflation begins. During the oscillatory stage, $\phi$ remains slow roll, while $\psi$ acquires a large kinetic energy that is of the same order of the potential energy, $\dot{\psi}^2 \sim \lambda' \phi^2 \psi^2 \sim H^2$, and the energy density evolves effectively according to $w = 0$ (zero pressure). Ignoring the kinetic energy of $\phi$ in the Friedmann equation, we have
		\begin{align}
			\label{eq:psiFriedmann} \left( \frac{d \psi}{d N} \right)^2 + \frac{\lambda' \phi^2}{H^2} \psi^2 = \frac{3}{4 \pi}.
		\end{align}
		With $A \equiv H / \sqrt{\lambda'} \phi$, we parametrize $\psi$ and $d \psi / d N$ by the amplitude $A$ and the phase $\theta$,
		\begin{align}
			\label{eq:psiPrimeDef} \frac{d \psi}{d N} = &\sqrt{\frac{3}{4 \pi}} \cos \theta, \\
			\label{eq:psiDef} \psi = &\sqrt{\frac{3}{4 \pi}} A \sin \theta.
		\end{align}
		Combining \eqref{eq:psiEOMinN2} and \eqref{eq:psiFriedmann}, we obtain a set of differential equations of $A$ and $\theta$
		\begin{align}
			\label{eq:DEA} \frac{d A}{d N} = &- A \cos^2 \theta \left( 3 + \frac{1}{\phi} \frac{d \phi}{d N} \right), \\
			\label{eq:DETheta} \frac{d \theta}{d N} = &\frac{1}{A} + \cos \theta \sin \theta \left( 3 + \frac{1}{\phi} \frac{d \phi}{d N} \right).
		\end{align}
		Among the two terms contributing to the frequency $d \theta / d N$, the first term $1 / A$ is of the order of $1 / \psi$, since $A = H / \sqrt{\lambda'} \phi \approx \sqrt{\lambda'} \phi \psi / \sqrt{\lambda'} \phi = \psi$. The second term is of order unity as $\phi$ slow rolls. After $\psi$ drops below the Planck mass, $1 / A$ dominates and $\theta$ oscillates rapidly, so we can approximate $\cos^2 \theta$ in \eqref{eq:DEA} by $1 / 2$. With $\phi$ given by the slow-roll solution \eqref{eq:phiStage1Case1}, $A$ and $\theta$ can then be integrated to yield
		\begin{align}
			A(N) &= A_i \left( 1 - \frac{N - N_i}{2 \pi \phi_i^2} \right)^{-1/4} e^{-\frac{3}{2} ( N - N_i )}, \\
			\theta(N) &= \theta_i + \frac{N - N_i}{A_i} - \frac{( N - N_i )^2}{16 \pi \phi_i^2 A_i}.
		\end{align}
	
	As the field $\psi$ attenuates, gradually the energy density from the coupling term is taken over by the $m^2 \phi^2 / 2$ term. The universe enters the third stage, in which the other inflation driven by the light-field begins. This stage is effectively described by the single-field inflation, with the same attractor solution \eqref{eq:phiStage1Case1} for $\phi$.


\section{Perturbations and Initial Conditions}
	
	Here we consider the perturbation to the homogeneous background in the conformal Newtonian gauge. The field perturbations are
		\begin{align}
			\phi(\mathbf{x}, t) = \bar{\phi}(t) + \delta \phi(\mathbf{x}, t), \\
			\psi(\mathbf{x}, t) = \bar{\psi}(t) + \delta \psi(\mathbf{x}, t),
		\end{align}
		where $\bar{\phi}(t)$ and $\bar{\psi}(t)$ denote the background solutions. The equations of motion of the perturbations in the Fourier space are
		\begin{align}
			\label{eq:deltaPhiEOM} &\delta \ddot{\phi} + 3 H \delta \dot{\phi} + \left( \frac{k^2}{a^2} + \frac{\partial^2 V}{\partial \phi^2} \right) \delta \phi = \notag \\
			&\quad\quad\quad\quad\quad\quad -\frac{\partial^2 V}{\partial \phi \partial \psi} \delta \psi -2 \frac{\partial V}{\partial \phi} \Psi + 4 \dot{\bar{\phi}} \dot{\Psi}, \\
			\label{eq:deltaPsiEOM} &\delta \ddot{\psi} + 3 H \delta \dot{\psi} + \left( \frac{k^2}{a^2} + \frac{\partial^2 V}{\partial \psi^2} \right) \delta \psi = \notag \\
			&\quad\quad\quad\quad\quad\quad -\frac{\partial^2 V}{\partial \phi \partial \psi} \delta \phi -2 \frac{\partial V}{\partial \psi} \Psi + 4 \dot{\bar{\psi}} \dot{\Psi}, \\
			&\label{eq:PsiEOM} \dot{\Psi} + H \Psi = 4 \pi ( \dot{\bar{\phi}} \delta \phi + \dot{\bar{\psi}} \delta \psi ).
		\end{align}
		Note that the energy-momentum tensor of action \eqref{eq:action} has no velocity perturbations either anisotropic inertia, so the Einstein equation gives $\Phi = \Psi$. The curvature perturbation in two-field system is
		\begin{align}
			\mathcal{R} = -\Psi - H \frac{ \dot{\bar{\phi}} \delta \phi + \dot{\bar{\psi}} \delta \psi }{ \dot{\bar{\phi}}^2 + \dot{\bar{\psi}}^2 }.
		\end{align}
	
	The initial conditions for the sub-horizon modes are set in the era of the heavy-field inflation by the method of iteration. We first neglect the metric perturbations $\Psi$ and find the solutions to $\delta \phi$ and $\delta \psi$ at the short-wavelength limit. The initial conditions for the sub-horizon field perturbations are set as the normalized positive-frequency solutions. We then feed the initial $\delta \phi$ and $\delta \psi$ back into the Einstein equations and obtain the initial conditions of $\Psi$. At the end we perform the consistency check to see whether the initial $\Psi$ obtained are indeed much smaller than $\delta \phi$ and $\delta \psi$ at the short-wavelength limit.
	
	To solve the field perturbations, first note that the equations of motion \eqref{eq:deltaPhiEOM} and \eqref{eq:deltaPsiEOM} can be put into a simpler form by substitute $t$ by the conformal time $\eta$, and introducing the new variables $w = a \delta \phi$ and $q = a \delta \psi$. Neglecting the metric perturbation, the equations for $w$ and $q$ are
		\begin{align}
			w'' + \left( k^2 + \frac{\partial^2 V}{\partial \phi^2} - \frac{a''}{a} \right) w = &- \frac{\partial^2 V}{\partial \phi \partial \psi} a^2 q, \\
			q'' + \left( k^2 + \frac{\partial^2 V}{\partial \psi^2} - \frac{a''}{a} \right) q = &- \frac{\partial^2 V}{\partial \phi \partial \psi} a^2 w,
		\end{align}
		where the primes denote the derivatives with respect to the conformal time $\eta$. For $k \gg a H$ the two equations decouple and reduce to the equations of harmonic oscillators
		\begin{align}
			w'' + k^2 w &= 0, \\
			q'' + k^2 q &= 0.
		\end{align}
		After the quantization, the normalized positive-frequency mode functions are
		\begin{align}
			\label{eq:InitialDeltaPhi} \delta \phi (\eta) = &\frac{e^{-i k \eta}}{(2 \pi)^{3/2} \sqrt{2 k} \; a}, \\
			\label{eq:InitialDeltaPsi} \delta \psi (\eta) = &\frac{e^{-i k \eta}}{(2 \pi)^{3/2} \sqrt{2 k} \; a},
		\end{align}
		which are the initial conditions for the sub-horizon field perturbations. Feeding \eqref{eq:InitialDeltaPhi} and \eqref{eq:InitialDeltaPsi} into the Einstein equation
		\begin{align}
			\label{eq:PsiRelation} &\Psi = \frac{1}{\dot{\bar{\phi}}^2 + \dot{\bar{\psi}}^2 - \frac{k^2}{4 \pi a^2}} \left[ \dot{\bar{\phi}} \delta \dot{\phi} + \dot{\bar{\psi}} \delta \dot{\psi} \right. \notag \\
			&\quad\quad \left. + \left( 3 H \dot{\bar{\phi}} + \frac{\partial V}{\partial \phi} \right) \delta \phi + \left( 3 H \dot{\bar{\psi}} + \frac{\partial V}{\partial \psi} \right) \delta \psi \right],
		\end{align}
		one obtains the initial conditions for the sub-horizon metric perturbations.
	
	To justify that the metric perturbations are negligible when finding solutions to the field perturbations, in this paragraph we are going to show that $\Psi$ obtained by \eqref{eq:PsiRelation} is much smaller than $\delta \phi$ and $\delta \psi$ at the short-wavelength limit in the era of heavy-field inflation. The following discussion in this paragraph assumes $k \gg a H$. First note that from \eqref{eq:phiStage1Case1} and \eqref{eq:psiStage1Case1} one has $\dot{\bar{\phi}} \sim H / \bar{\phi}$ and $\dot{\bar{\psi}} \sim H / \bar{\psi}$. The metric perturbation \eqref{eq:PsiRelation} therefore goes like
		\begin{align}
			\label{eq:PsiApprox1} \Psi \sim &\frac{1}{\left( \frac{k}{a H} \right)^2} \left[ \frac{\delta \dot{\phi}}{H \bar{\phi}} + \frac{\delta \dot{\psi}}{H \bar{\psi}} \right. \notag \\
			&\left. + \left( \frac{1}{\bar{\phi}} + \frac{\lambda' \bar{\phi} \bar{\psi}^2}{H^2} \right) \delta \phi + \left( \frac{1}{\bar{\psi}} + \frac{\lambda' \bar{\phi}^2 \bar{\psi}}{H^2} \right) \delta \psi \right],
		\end{align}
		where we have omitted all the constant coefficients and used the fact that the potential is dominated by $\lambda' \phi^2 \psi^2 / 2$ during the heavy-field inflation. From solutions \eqref{eq:InitialDeltaPhi} and \eqref{eq:InitialDeltaPsi} one has
		\begin{align}
			\label{eq:deltaPhiDot} \delta \dot{\phi} = - H \delta \phi \left( 1 + \frac{i k}{a H} \right) \sim \frac{k}{a} \delta \phi
		\end{align}
		and similarly for $\delta \dot{\psi}$. The first two terms in the bracket of \eqref{eq:PsiApprox1} then go like $( k / a H ) \cdot ( \delta \phi / \bar{\phi} )$ and $( k / a H ) \cdot ( \delta \psi / \bar{\psi} )$. Using the Friedmann equation we know that $\lambda' \bar{\phi} \bar{\psi}^2 / H^2 \sim 1 / \bar{\phi}$ and $\lambda' \bar{\phi}^2 \bar{\psi} / H^2 \sim 1 / \bar{\psi}$, therefore the last two terms in the bracket of \eqref{eq:PsiApprox1} go like $\delta \phi / \bar{\phi}$ and $\delta \psi / \bar{\psi}$. Since during the heavy-field inflation $\bar{\phi}$ and $\bar{\psi}$ is of order $\mathcal{O}(1)$ in Planck units, one has
		\begin{align}
			\Psi \sim \frac{1}{\left( \frac{k}{a H} \right)} \left( \frac{\delta \phi}{\bar{\phi}} + \frac{\delta \psi}{\bar{\psi}} \right).
		\end{align}
		Therefore the initial $\Psi$ is indeed much smaller than the initial $\delta \phi$ and $\delta \psi$ at the limit of $k \gg a H$ in the era of heavy-field inflation.


\section{Numerical Solution and the CMB Spectrum}

	The CMB spectrum is found by three steps of numerical calculations. We first obtain the background dynamics by solving \eqref{eq:phiEOMinN2} and \eqref{eq:psiEOMinN2}. The reheating energy scale is assumed to be $7.19 \times 10^{-4} M_P \sim 10^{16} \textrm{GeV}$. The evolution of perturbations is then solved by integrating \eqref{eq:deltaPhiEOM}, \eqref{eq:deltaPsiEOM}, and \eqref{eq:PsiEOM}, with the initial conditions set by \eqref{eq:InitialDeltaPhi}, \eqref{eq:InitialDeltaPsi}, and \eqref{eq:PsiRelation}. The spectrum of the curvature perturbations is found by evolving each Fourier mode until it reaches the steady value after the horizon exit. The resulting spectrum is then fed into \texttt{CAMB} \cite{Lewis2000, Lewis2012}, which is modified to accept arbitrary initial spectrum represented by an interpolating function, to calculate the spectrum of the CMB temperature fluctuations.
	
	There are two parameters and four initial conditions in our model. The two parameters are the coupling constant, $\lambda'$, and the mass of the light field, $m$. The four initial conditions are the initial values and derivatives of the fields $\phi$ and $\psi$. We assume that the light field drives about the last 60 e-folds of inflation, so the mass $m$ is determined by the Hubble scale during inflation deduced by the observation. The initial derivatives of $\phi$ and $\psi$ are set to zero, leaving the system released from rest and evolving into the attractor solutions.
	
\begin{figure}

\center

\includegraphics[width = 14 cm]{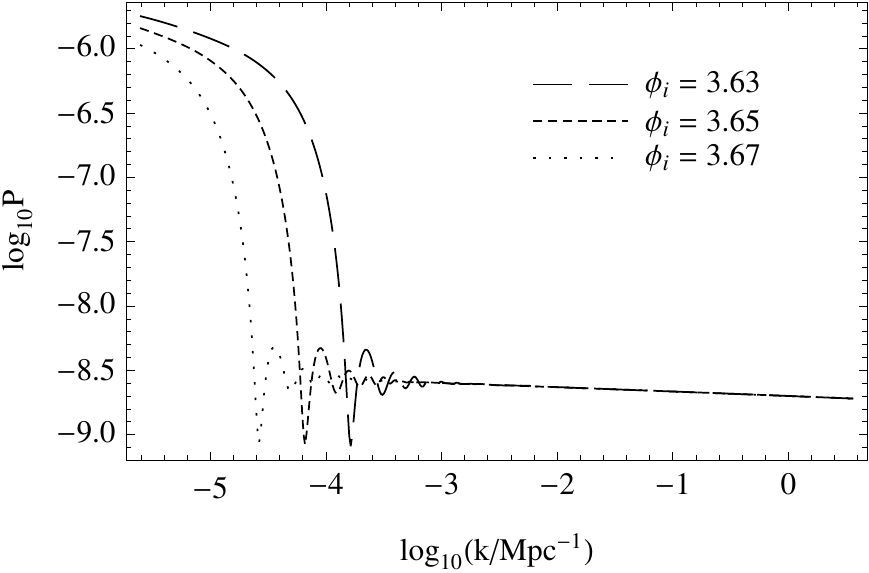}

\caption{The spectra of the curvature perturbations with $\phi_i = 3.63 M_P$, $3.65 M_P$, and $3.67 M_P$. The other parameters are held fixed as $\lambda' = 10^{-9}$, $\psi_i = 1.60 M_P$, and $m = 1.22 \times 10^{-6} M_P$.}

\label{fig:ChiSpectrum560}

\end{figure}

\begin{figure}

\center

\includegraphics[width = 14 cm]{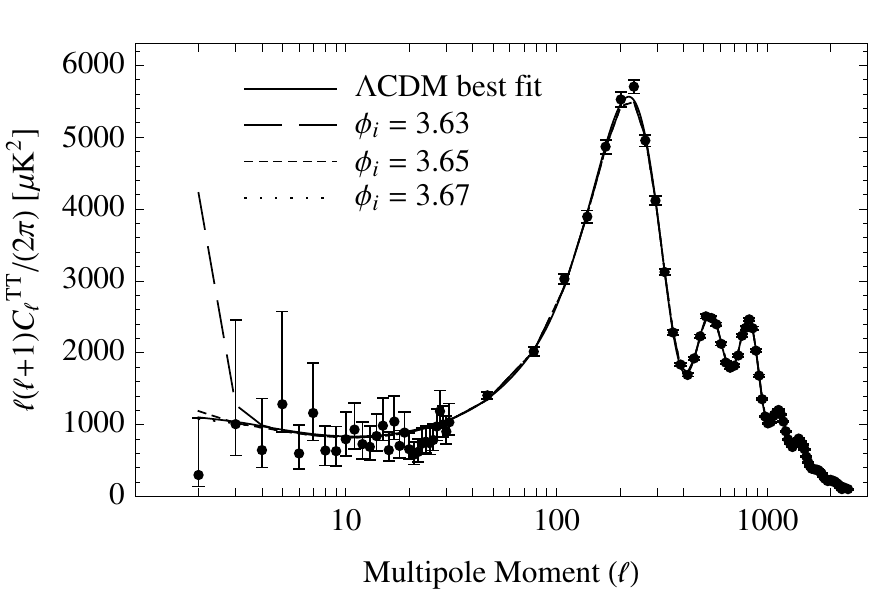}

\caption{The CMB temperature-temperature correlation (TT) spectra with $\phi_i = 3.63 M_P$, $3.65 M_P$, and $3.67 M_P$. The dots with error bars are the Planck 2013 data. The other parameters are held fixed as $\lambda' = 10^{-9}$, $\psi_i = 1.60 M_P$, and $m = 1.22 \times 10^{-6} M_P$.}

\label{fig:CMB553ScanPhi}

\end{figure}

	The initial value of the light field $\phi$ determines the behavior of the system in two ways. First, it determines the energy scale of the heavy-field inflation as well as that of the oscillatory period, since at the beginning of the oscillatory stage $\psi \sim M_P$ and $H \sim \sqrt{\lambda'} \phi$. Second, it determines the number of e-folds of the light-field inflation. With larger initial $\phi$, the light-field inflation begins at a higher energy scale and lasts longer. In this case, only those modes with larger wavelengths that are affected by the heavy-field inflation and the oscillatory stage (Figure \ref{fig:ChiSpectrum560}). Therefore, less deviations are manifested in the present-day CMB spectrum since those large modes have not entered the Hubble horizon today (Figure \ref{fig:CMB553ScanPhi}).


\begin{figure}

\center

\includegraphics[width = 14 cm]{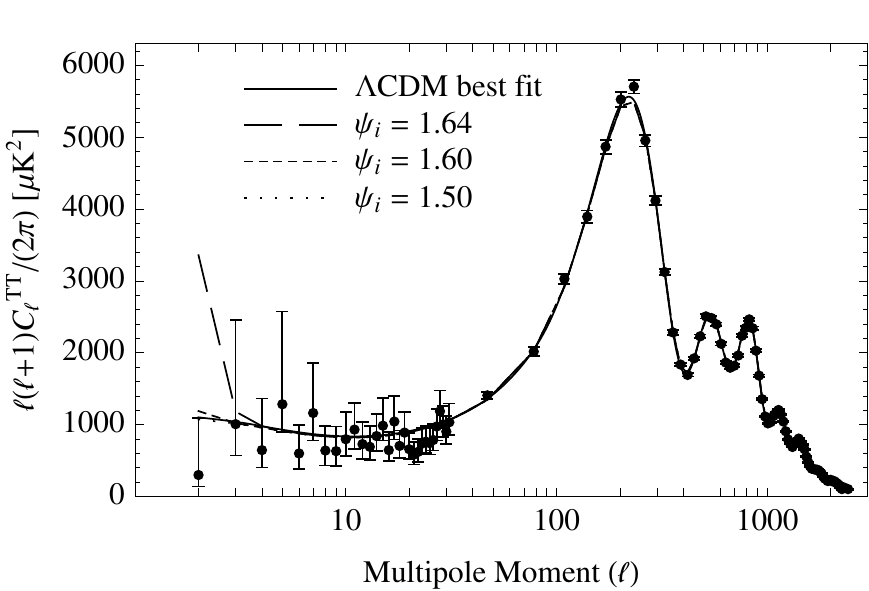}

\caption{The CMB TT spectra with $\psi_i = 1.50 M_P$, $1.60 M_P$, and $1.64 M_P$. The dots with error bars are the Planck 2013 data. The other parameters are held fixed as $\lambda' = 10^{-9}$, $\phi_i = 3.65 M_P$, and $m = 1.22 \times 10^{-6} M_P$.}

\label{fig:CMB562ScanPsi}

\end{figure}

	Raising the value of initial $\psi$ affects the spectrum by making the heavy-field inflation longer and, while holding the initial $\phi$ fixed, the light-field inflation shorter, without changing the duration of the transition era. The larger the initial $\psi$ is, more visible $k$ modes and $\ell$ modes are affected (Figure \ref{fig:CMB562ScanPsi}).


\begin{figure}

\center

\includegraphics[width = 14 cm]{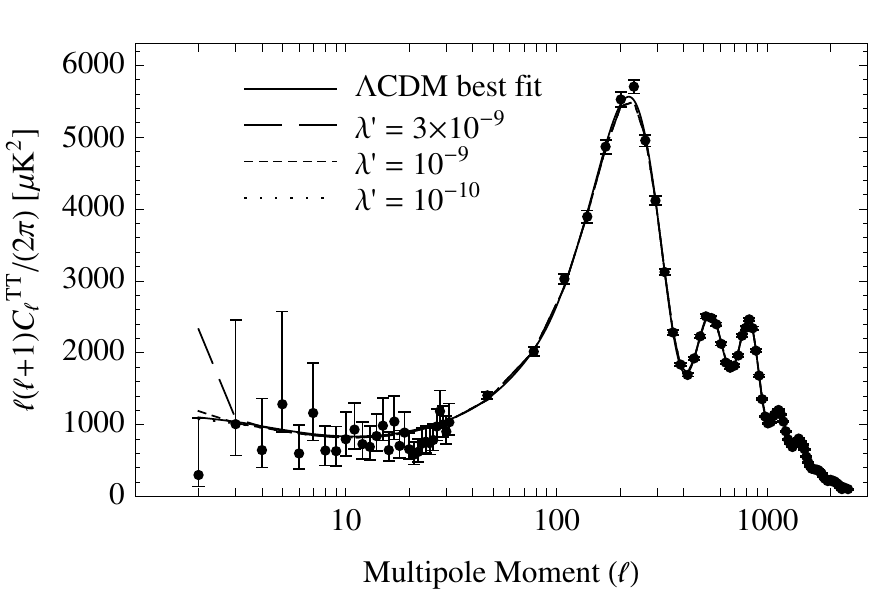}

\caption{The CMB TT spectra with $\lambda' = 10^{-10}$, $10^{-9}$, and $3 \times 10^{-9}$. The dots with error bars are the Planck 2013 data. The other parameters are held fixed as $\phi_i = 3.65 M_P$, $\psi_i = 1.60 M_P$, and $m = 1.22 \times 10^{-6} M_P$.}

\label{fig:CMB522ScanLambdaPrime}

\end{figure}

	The coupling $\lambda'$ determines the strength of the interaction between the two fields. With stronger interactions, it takes longer for the transient oscillation to settle, and therefore more modes with short wavelengths are affected while other conditions held fixed (Figure \ref{fig:CMB522ScanLambdaPrime}).


	We see from both the spectra of the curvature perturbations (Figure \ref{fig:ChiSpectrum560}) and CMB (Figure \ref{fig:CMB553ScanPhi}) that the large-scale spectrum is determined by the initial era with which the universe begins: The large-scale spectrum reflects the scale-invariant shape of the super-horizon spectrum in the initial slow-roll (heavy-field inflation) era. Although the oscillatory stage behaves like $w = 0$, the spectrum evolution is qualitatively similar to that shown in Subsec.~\ref{subsec:ABC}. A quantitative analysis of the spectrum evolution involving the zero-pressure ($w = 0$) era is given in the next section, showing that the \emph{ad hoc} single-field analysis does capture the pattern of the spectrum evolution and is in agreement with the numerical results.


\section{Spectrum Evolution Involving a Zero-Pressure Era}
	
	Consider the universe consists of three successive eras: a first slow-roll era (era A), an intermediate zero-pressure ($w = 0$) era (era B), and a second slow-roll era (era C). Much parallel to the discussion in Subsec.~\ref{subsec:ABC}, we denote the quantities at the transitions from era A to B and from era B to C by subscripts 1 and 2, respectively.
	
	In era A, the mode function of the initial adiabatic vacuum is given by \eqref{eq:EraAVacuum}. In era B, we first keep the discussion general for $w > -1$. The mode function $u$ is given by \eqref{eq:WhittakerSolution},
			\begin{align}
				\label{eq:modeM} u_B = &B_+ M_{0, \mu}\left( \frac{2 i \tilde{k}_B}{\alpha} \tilde{a}_B^{\alpha} \right) \notag \\
				&+ B_- W_{0, \mu}\left( \frac{2 i \tilde{k}_B}{\alpha} \tilde{a}_B^{\alpha} \right),
			\end{align}
			where $\tilde{k}_B = k / a_1 H_1$ and $\tilde{a}_B = a / a_1 = [ 1 + \alpha a_1 H_1 ( \eta - \eta_1 ) ]^{1/\alpha}$. Parameters $\alpha$ and $\mu$ are given by \eqref{eq:alpha} and \eqref{eq:mu}, respectively. The normalized power spectrum is
			\begin{align}
				\label{eq:PowerM} \tilde{P}_B = &\frac{\tilde{k}_B^3}{\tilde{a}_B^2} \left| \tilde{B}_+ M_{0, \mu} \left( \frac{2 i \tilde{k}_B}{\alpha} \tilde{a}_B^{\alpha} \right) \right. \notag \\
				&\quad\quad \left. + \tilde{B}_- W_{0, \mu}\left( \frac{2 i \tilde{k}_B}{\alpha} \tilde{a}_B^{\alpha} \right) \right|^2,
			\end{align}
			where $\tilde{B}_{\pm} = \sqrt{a_1 H_1} B_{\pm}$ and
			\begin{align}
				\label{eq:PowerMDef} \tilde{P}_B = \frac{(1+\alpha) P_B}{16 \pi^2 H_1^2}.
			\end{align}
			In era C, the mode function is \eqref{eq:SolutionC}, and the normalized power spectrum is \eqref{eq:PowerC}. For simplicity, we assume that the slow-roll parameter $\epsilon$ in era A and C has the same value.
	
	By matching at the boundary $\eta = \eta_1$, we obtain
		\begin{align}
			\tilde{B}_+ = &\Delta_B \left[ (-2 i \tilde{k}_B^2 + 2 \tilde{k}_B + i) W_{0,\mu} \right. \notag \\
			&\quad\quad \left. + \alpha (\tilde{k}_B + i) W_{1, \mu} \right], \\
			\tilde{B}_- = &\Delta_B \left[ -(-2 i \tilde{k}_B^2 + 2 \tilde{k}_B + i) M_{0,\mu} \right. \notag \\
			&\quad\quad \left. + \frac{2 \mu + 1}{2} \alpha (\tilde{k}_B + i) M_{1, \mu} \right],
		\end{align}
		where
		\begin{align}
			\Delta_B = &\frac{\sqrt{\alpha + 1} e^{i \tilde{k}_B}}{2 (\pi \tilde{k}_B)^{3/2} \alpha \sqrt{\epsilon}} \notag \\
			&\times \frac{1}{(2 \mu + 1) M_{1,\mu} W_{0, \mu} + 2 M_{0,\mu} W_{1, \mu}},
		\end{align}
		and the Whittaker functions are evaluated at $2 i \tilde{k}_B / \alpha$. Matching at $\eta = \eta_2$ gives
		\begin{align}
			\label{eq:cp2} \tilde{C}_+ = &\frac{\sqrt{\epsilon} e^{-\frac{1}{2} \alpha N_B} e^{i \tilde{k}_C}}{4 \sqrt{1+\alpha} \; \tilde{k}_C^2} \notag \\
			&\times \left\{ 2 \left[ 2 \tilde{k}_C^2 + 2 i \tilde{k}_C - 1 \right] \tilde{B}_+ M_{0, \mu} \right. \notag \\
			&\quad + 2 \left[ 2 \tilde{k}_C^2 + 2 i \tilde{k}_C - 1 \right] \tilde{B}_- W_{0, \mu} \notag \\
			&\quad + \alpha ( 1 - i \tilde{k}_C ) ( 2 \mu + 1 ) \tilde{B}_+ M_{1, \mu} \notag \\
			&\quad \left. - 2 \alpha ( 1 - i \tilde{k}_C ) \tilde{B}_- W_{1, \mu} \right\}, \\
			\label{eq:cm2} \tilde{C}_- = &\frac{\sqrt{\epsilon} e^{-\frac{1}{2} \alpha N_B} e^{-i \tilde{k}_C}}{4 \sqrt{1+\alpha} \; \tilde{k}_C^2} \notag \\
			&\times \left\{ -2 \tilde{B}_+ M_{0, \mu} \right. \notag \\
			&\quad -2 \tilde{B}_- W_{0, \mu} \notag \\
			&\quad + \alpha ( 2 \mu + 1 ) ( 1 + i \tilde{k}_C ) \tilde{B}_+ M_{1, \mu} \notag \\
			&\quad \left. - 2 \alpha ( 1 + i \tilde{k}_C ) \tilde{B}_- W_{1, \mu} \right\},
		\end{align}
			with the Whittaker functions evaluated at $2 i \tilde{k}_C / \alpha$. One also has the relation $\tilde{k}_B = \tilde{k}_C e^{-\alpha N_B}$.
	
\begin{figure}

\centering

\includegraphics[width = 14 cm]{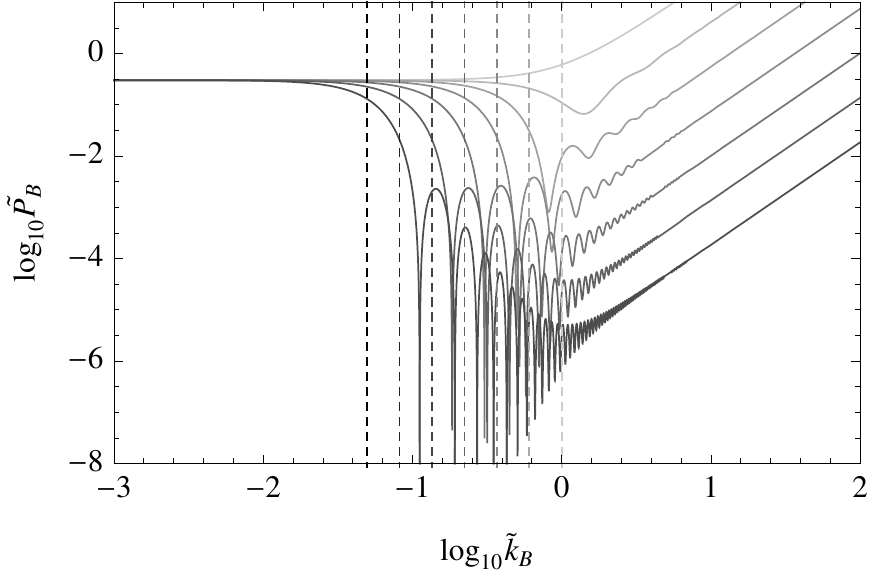}

\caption{Time evolution of the power spectrum in era B in the case of three-stage evolution (era A, B, and C), in which $w = 0$ in era B. The solid curves are the spectra of $\mathcal{R}$ from early time to late time (from light to dark). The vertical dashed lines denote the comoving horizon size at the corresponding instants (also from light to dark).}

\label{fig:AMCinM40}

\end{figure}

\begin{figure}

\centering

\includegraphics[width = 14 cm]{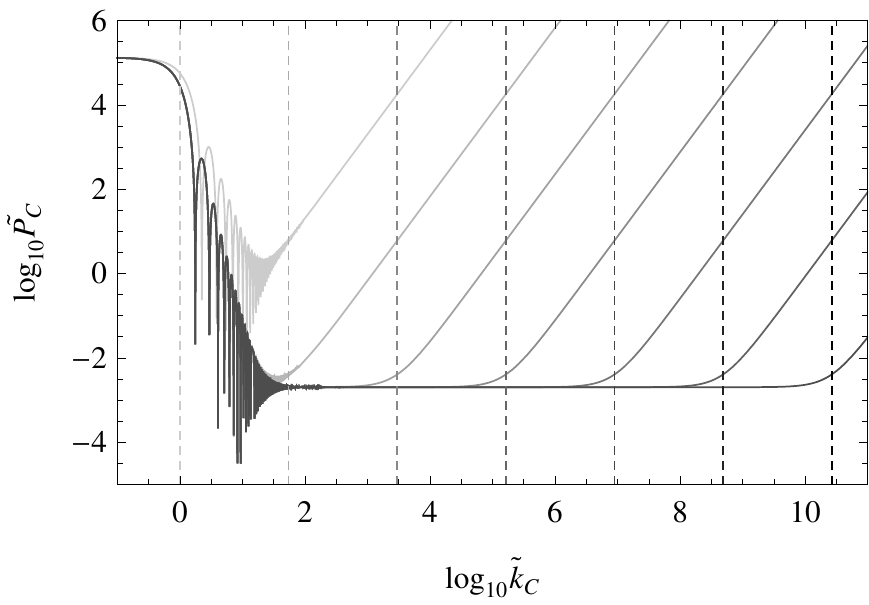}

\caption{Time evolution of the power spectrum in era C in the case of three-stage evolution (era A, B, and C), in which $w = 0$ in era B. The solid curves are the spectra of $\mathcal{R}$ from early time to late time (from light to dark). The vertical dashed lines denote the comoving horizon size at the corresponding instants (also from light to dark).}

\label{fig:AMCinC271}

\end{figure}

	Setting $w = 0$ in era B, we find the spectrum evolution as Figure \ref{fig:AMCinM40}. Similar to the case of the intermediate kinetic era in Subsec.~\ref{subsec:ABC}, the super-horizon modes inherits the scale-invariant spectrum in the initial slow-roll era (era A), and the modes that enter the horizon during the zero-pressure era (era B) have the red-tilted spectrum. Entering into the second slow-roll era (era C), as shown in Figure \ref{fig:AMCinC271}, the modes that enter the horizon in era B are expelled out of the horizon again, leading to the steep red-tilted spectrum connecting the two plateaus. The right plateau is formed by the modes that are sub-horizon during eras A and B and then exit the horizon in era C. Comparing to the spectrum of the two-field model (Figure \ref{fig:ChiSpectrum560}), which also has a zero-pressure era between two inflationary eras, we see that the \emph{ad hoc} single-field analysis agrees with the numerical result and largely captures the physics of the spectrum formation.


\clearpage


\chapter{No-Boundary Wave Function as the Initial Condition of Inflation}{\label{ch:HH}}

	The two candidates we found in the previous analysis---a preinflationary superinflation era and a positive-pressure era---both have their difficulties as the realistic initial condition to the universe. On the one hand, the phantomness of the superinflationary era requires a gravitational theory that is self-consistent while breaking the energy conditions, which is hard to construct. On the other hand, as a decelerating phase, the positive-pressure era does not have a canonical choice of the vacuum state such as the Bunch-Davies vacuum. These difficulties may imply that, instead of struggling in the realm of the semi-classical models, the power suppression can be more naturally addressed by a quantum gravity theory.

	In this chapter, we first describe the Hartle-Hawking no-boundary wave function, which is a solution to the Wheeler-DeWitt equation that forms the foundation of quantum gravity, in the minisuperspace model \cite{Hartle2008a, Hartle2008}. Especially, we focus on the background-level solution. We then review the dynamics of the matter field and metric perturbations introduced by Halliwell and Hawking \cite{Halliwell1985}. By using the steepest decent approximation, we can calculate the expectation values of perturbations. In the end of this chapter we compute the power spectra of the inflaton field perturbations and show that, with a moderate mass scale, the massive inflaton can induce the suppression of the long-wavelength spectrum.


\section{Minisuperspace model}
	
	The ADM metric \cite{Arnowitt2008} for the homogeneous closed universe is
		\begin{align}
			ds^2 &= \sigma^2 \left[ -(\bar{N}^2 - \bar{N}_i \bar{N}^i) d\lambda^2 + 2 \bar{N}_i dx^i d\lambda \right. \notag \\
			&\quad \left. + \bar{h}_{i j} dx^i dx^j \right],
		\end{align}
		where $\sigma$ is a constant normalization, and
		\begin{align}
			\bar{N} &= N_0(\lambda), \\
			\bar{N}_i &= 0, \\
			\bar{h}_{i j} &= a^2(\lambda) \bar{\gamma}_{i j}, \\
			\bar{\gamma}_{i j} dx^i dx^j &= d\chi^2 + \sin^2\chi (d\theta^2 + \sin^2\theta d\varphi^2) = d\Omega_3^2.
		\end{align}
		The action for a scalar field in the close universe is
		\begin{align}
			I &= \frac{1}{16 \pi} \int d^4x \sqrt{-g} R \notag \\
			&\quad + \int d^4x \sqrt{-g} \left[ -\frac{1}{2} \partial^{\mu} \Phi \partial_{\mu} \Phi - V(\Phi) \right],
		\end{align}
		where
		\begin{align}
			%
			V(\Phi) = V_0 + \frac{1}{2} m^2 \Phi^2.
		\end{align}
		Defining the variables
		\begin{align}
			\phi &= \sqrt{\frac{4 \pi}{3}} \; \Phi, \\
			\tilde{V}(\Phi) &= \frac{8 \pi \sigma^2}{3} V(\Phi),
		\end{align}
		and integrating over the compact geometry, the action can be expanded as
		\begin{align}
			I[N_0, a, \phi] &= \frac{3 \pi \sigma^2}{4} \int d\lambda \; N_0 \left\{ -a \left( \frac{a'}{N_0} \right)^2 + a \right. \notag \\
			&\quad \left. + a^3 \left[ \left( \frac{\phi'}{N_0} \right)^2 - \tilde{V}(\Phi) \right] \right\},
		\end{align}
		where the primes denote the derivatives against $\lambda$. It is convenient to further define
		\begin{align}
			\tilde{V}_0 &= \frac{8 \pi \sigma^2}{3} V_0, \\
			\tilde{m} &= \sigma m,
			%
			%
		\end{align}
		so that
		\begin{align}
			%
			\tilde{V} = \tilde{V}_0 + \tilde{m}^2 \phi^2.
		\end{align}
	
	The no-boundary wave function can be written as the path integral,
		\begin{align}
			\label{eq:WaveFunction} \Psi = \int \mathcal{D} \hat{a} \mathcal{D} \hat{\phi} \mathcal{D} \hat{N} \; e^{-\frac{1}{\hbar} \hat{I}[ \hat{a}, \hat{\phi}, \hat{N} ]},
		\end{align}
		where $\hat{a}$, $\hat{\phi}$, and $\hat{N}$ are the corresponding fields in the Euclidean metric,
		\begin{align}
			ds^2 = \sigma^2 \left[ \hat{N}_0^2 d\lambda^2 + \hat{a}^2 d\Omega_3^2 \right],
		\end{align}
		obtained from the Lorentzian one by substituting $N_0$ by $-i \hat{N}_0$ and adding hats to other fields for clarity. The Euclidean action is taken as $\hat{I} = -i I|_{N_0 = -i \hat{N}_0}$:
		\begin{align}
			\hat{I} &= \frac{1}{16 \pi} \int d^4 x \sqrt{+\hat{g}} \hat{R} \notag \\
			&\quad + \int d^4 x \sqrt{+\hat{g}} \left[ -\frac{1}{2} \partial^{\mu} \hat{\Phi} \partial_{\mu} \hat{\Phi} - V(\hat{\Phi}) \right].
		\end{align}
		After integration, we have
		\begin{align}
			\hat{I}[\hat{N}_0, \hat{a}, \hat{\phi}] &= \frac{3 \pi \sigma^2}{4} \int d\lambda \; \hat{N}_0 \left\{ -\hat{a} \left( \frac{\hat{a}'}{\hat{N}_0} \right)^2 - \hat{a} \right. \notag \\
			&\quad \left. + \hat{a}^3 \left[ \left( \frac{\hat{\phi}'}{\hat{N}_0} \right)^2 + \tilde{V}(\hat{\Phi}) \right] \right\}.
		\end{align}
	
	Doing variation with respect to $\hat{N}_0$, we obtain the Hamiltonian constraint,
		\begin{align}
			%
			\label{eq:EuclideanHamiltonianConstraint} \dot{\hat{a}}^2 - 1 + \hat{a}^2 \left[ - \dot{\hat{\phi}}^2 + \tilde{V}(\hat{\Phi}) \right] = 0,
		\end{align}
		where dots denote derivatives against $\tau$, which is defined by
		\begin{align}
			d\tau = \hat{N}_0 d\lambda.
		\end{align}
		Using the steepest descent approximation, the wave function is dominated by the extreme path $(\hat{a}_{\textrm{ext}}(\tau), \hat{\phi}_{\textrm{ext}}(\tau))$ that satisfies
		\begin{align}
			\frac{\delta \hat{I}}{\delta \hat{a}} &= 0, \\
			\frac{\delta \hat{I}}{\delta \hat{\phi}} &= 0,
		\end{align}
		which are
		\begin{align}
			\label{eq:EuExt1} \ddot{\hat{a}} + 2 \hat{a} \dot{\hat{\phi}}^2 + \hat{a} \tilde{V}(\hat{\Phi}) &= 0, \\
			\label{eq:EuExt2} \ddot{\hat{\phi}} + 3 \frac{\dot{\hat{a}}}{\hat{a}} \dot{\hat{\phi}} - \frac{1}{2} \frac{\partial \tilde{V}}{\partial \hat{\phi}} &= 0,
		\end{align}
		respectively. Note that the Hamiltonian constraint \eqref{eq:EuclideanHamiltonianConstraint} is used when deriving the equations above.
	
	To solve $\hat{a}(\tau)$, we consider the case in which $\dot{\hat{\phi}}^2$ is negligible, and combine \eqref{eq:EuclideanHamiltonianConstraint} and \eqref{eq:EuExt1} to obtain
		\begin{align}
			\label{eq:aEqSlowRoll} \hat{a} \ddot{\hat{a}} - \dot{\hat{a}}^2 + 1 = 0.
		\end{align}
		The ``no-boundary'' boundary condition at $\tau = 0$ sets $\hat{a}(0) = 0$. To keep \eqref{eq:EuExt2} finite, we also require $\dot{\hat{\phi}}(0) = 0$. Then by the Hamiltonian constraint \eqref{eq:EuclideanHamiltonianConstraint} we know $\dot{\hat{a}}(0) = 1$. The only free initial conditions left are the real and imaginary parts of $\hat{\phi}(0)$.
	
	Equation \eqref{eq:aEqSlowRoll} has four solutions,
		\begin{align}
			\hat{a}(\tau) &= \pm \frac{1}{H_0} \sin\left[ H_0 ( \tau - \tau_0 ) \right], \\
			\hat{a}(\tau) &= \pm \frac{1}{H_0} \sinh\left[ H_0 ( \tau - \tau_0 ) \right].
		\end{align}
		For physical solutions we should pick the plus sign. By requiring $\hat{a}(0) = 0$ we have $\tau_0 = 0$, and automatically we have consistently $\dot{\hat{a}}(0) = 1$. In order to connect to the Lorentzian space, which requires $\hat{a}'(\tau_{\textrm{connect}}) = 0$, the qualified solution is
		\begin{align}
			\label{eq:scaleAEu} \hat{a}(\tau) = \frac{1}{H_0} \sin( H_0 \tau ).
		\end{align}
	
	The solution connects to the Lorentzian space at $\tau_{\textrm{connect}} = \pi / 2 H_0$. In Lorentzian space, we define
		\begin{align}
			dt = N_0 d\lambda,
		\end{align}
		therefore $d\tau = i dt$. We can then describe the Euclidean trajectory by $\tau = 0$ to $\pi / 2 H_0$, and the Lorentzian one by the complex contour
		\begin{align}
			\tau = \frac{\pi}{2 H_0} + i t
		\end{align}
		with $t > 0$. We then have the Lorentzian solution
		\begin{align}
			\label{eq:scaleALo} a(t) = \frac{1}{H_0} \cosh( H_0 t ).
		\end{align}


\section{Basis functions on the closed universe}

	The scalar harmonics are the solutions to
		\begin{align}
			\partial_i \partial^i Q_n(\vec{x}) + ( n^2 - 1 ) Q_n(\vec{x}) = 0, \quad\quad ( n = 1, 2, 3\dots ).
		\end{align}
		and each $Q_n(\vec{x})$ is a finite sum over $l$ and $m$ (with $l = 0, 1, \dots, n - 1$ and $m = -l, -l + 1, \dots, l$) like the familiar spherical harmonics $Y_{l m}(\theta, \varphi)$ on $S^2$. We write $Q_n(\vec{x})$ as
		\begin{align}
			Q_{n l m}(\chi, \theta, \varphi) = \Pi_{n l}(\chi) Y_{l m}(\theta, \varphi).
		\end{align}
		In closed universe, the Laplacian is
		\begin{align}
			\vec{\nabla}^2 = \frac{1}{\sin^2 \chi} \left[ \sin^2 \chi \frac{\partial^2}{\partial \chi^2} +2 \sin \chi \cos \chi \frac{\partial}{\partial \chi} + \frac{1}{\sin \theta} \frac{\partial}{\partial \theta} \left( \sin \theta \frac{\partial}{\partial \theta} \right) + \frac{1}{\sin^2 \theta} \frac{\partial^2}{\partial \varphi^2} \right].
		\end{align}
		One can recognize that the last two terms in the bracket are actually the operator of spherical harmonics. The Fock harmonics $\Pi_{n l}$ satisfy
		\begin{align}
			\frac{d^2 \Pi_{n l}}{d \chi^2} + 2 \cot \chi \frac{d \Pi_{n l}}{d \chi} + \left[ ( n^2 - 1 ) -\frac{l(l+1)}{\sin^2 \chi} \right] \Pi_{n l} = 0.
		\end{align}
		The solutions are
		\begin{align}
			\Pi_{n l}(\chi) &= \sqrt{\frac{2}{n \pi} \frac{(n-l-1)!}{(n+l)!}} (\sin \chi)^l \left( \frac{d}{d \cos \chi} \right)^{l+1} \cos (n \chi), \\
			Y_{l m}(\theta, \varphi) &= \sqrt{\frac{2 l + 1}{4 \pi} \frac{(l-m)!}{(l+m)!}} P_l^m(\cos \theta) e^{i m \varphi},
		\end{align}
		with orthonormal inner products
		\begin{align}
			\int_0^{\pi} d \chi \int_0^{\pi} d \theta \int_0^{2 \pi} d \varphi \; \sin^2 \chi \; \sin \theta \; \left[ \Pi_{n' l'}(\chi) Y_{l' m'}(\theta, \varphi) \right]^* \left[ \Pi_{n l}(\chi) Y_{l m}(\theta, \varphi) \right] = \delta_{n n'} \delta_{l l'} \delta_{m m'}.
		\end{align}
		Similar definitions exist for other basis functions.


\section{Perturbation spectrum from the wave function}
	
	The perturbations to the spatial part of the metric in the $S^3 \times R$ closed universe can be organized as
		\begin{align}
			h_{i j} &= a^2 \gamma_{i j}, \notag \\
			\gamma_{i j} &= \bar{\gamma}_{i j} + \epsilon_{i j},
		\end{align}
		where $\epsilon_{i j}$ denotes
		\begin{align}
			\label{eq:epsilon} \epsilon_{i j} &= \sum_{n, l, m} \left[ \sqrt{6} q_{n l m} \frac{1}{3} \bar{\gamma}_{i j} Q_{n l m} + \sqrt{6} b_{n l m} ( P_{i j} )_{n l m} \right. \notag \\
			&\quad\quad\quad\quad + \sqrt{2} c^o_{n l m} ( S^o_{i j} )_{n l m} + \sqrt{2} c^e_{n l m} ( S^e_{i j} )_{n l m} \notag \\
			&\quad\quad\quad\quad \left. + 2 d^o_{n l m} ( G^o_{i j} )_{n l m} + 2 d^e_{n l m} ( G^e_{i j} )_{n l m} \right],
		\end{align}
		and
		\begin{align}
			P_{i j} = \frac{1}{n^2 - 1} \nabla_i \nabla_j Q + \frac{1}{3} \bar{\gamma}_{i j} Q.
		\end{align}
		Here the covariant derivatives are with respect to $\bar{\gamma}_{i j}$. The first, second, and third lines of \eqref{eq:epsilon} denote the scalar, vector, and tensor perturbations, respectively. Suppressing the spherical coordinate indices, $n$, $l$, $m$, the coefficients, $q$, $b$, $c^o$, $c^e$, $d^o$, $d^e$, are time dependent, while the basis, $Q$, $P_{i j}$, $S^o_{i j}$, $S^e_{i j}$, $G^o_{i j}$, $G^e_{i j}$, are space dependent.
	
	The perturbations to the lapse and the shift functions are
		\begin{align}
			N &= N_0 \left[ 1 + \sum_{n, l, m} \frac{1}{\sqrt{6}} g_{n l m} Q_{n l m} \right], \\
			N_i &= a \sum_{n, l, m} \left[ \frac{1}{\sqrt{6}} k_{n l m} (P_i)_{n l m} + \sqrt{2} j_{n l m} (S_i)_{n l m} \right],
		\end{align}
		where
		\begin{align}
			P_i = \frac{1}{n^2 - 1} \nabla_i Q.
		\end{align}
		Finally, the perturbation to the scalar field is 
		\begin{align}
			\Phi = \sqrt{\frac{3}{4 \pi}} \phi + \sum_{n, l, m} \sqrt{\frac{3 \pi}{2}} f_{n l m} Q_{n l m}.
		\end{align}
		Among the perturbations, $q$, $b$, $g$, $k$, and $f$ are the scalar ones, $j$, $c^o$ and $c^e$ are the vector ones, and $d^o$ and $d^e$ are the tensor ones.
	
	The action can be expanded around the background fields to the second order as the sum of the eigenmodes \cite{Halliwell1985},
		\begin{align}
			I &= I_0(a, \bar{\phi}, N_0) \notag \\
			&\quad + \sum_{n, l, m} I_{n l m}(a, \bar{\phi}, N_0; q_{n l m}, \dots, k_{n l m}).
		\end{align}
		Choosing the gauge in which $q_{n l m} = b_{n l m} = 0$, the constraint equations can be obtained by variating the quadratic part of the perturbation action with respect to $g_{n l m}$ and $k_{n l m}$,
		\begin{align}
			\label{eq:gnlm} g_{n l m} &= 3 \frac{(n^2-1) H \dot{\phi} f_{n l m} + \dot{\phi} \dot{f}_{n l m} + \tilde{m}^2 \phi f_{n l m}}{(n^2 - 4) H^2 + 3 \dot{\phi}^2}, \\
			k_{n l m} &= 3 ( n^2 - 1 ) N_0 a \notag \\
			&\hspace{-1cm} \times \frac{H \dot{\phi} \dot{f}_{n l m} + H \tilde{m}^2 \phi f_{n l m} - 3 \dot{\phi} ( -H^2 + \dot{\phi}^2 ) f_{n l m}}{(n^2 - 4) H^2 + 3 \dot{\phi}^2}.
		\end{align}
		Here the dots denote derivatives against the Lorentzian time $t$. The equation of motion for $f_{n l m}$ can be obtained by the variation with respect to $f_{n l m}$,
		\begin{align}
			\label{eq:fEOM} &\quad\; \ddot{f}_{n l m} + 3 H \dot{f}_{n l m} + \left( \tilde{m}^2 + \frac{n^2 - 1}{a^2} \right) f_{n l m} \notag \\
			&= -2 \tilde{m}^2 \phi g_{n l m} + \dot{\phi} \dot{g}_{n l m} - \frac{\dot{\phi} k_{n l m}}{N_0 a}.
		\end{align}
	
	The amplitude of the perturbations of the scalar field, $\delta \Phi$, can be obtained through calculating the expectation value with the no-boundary wave function, focusing on the part relevant for $f_{n l m}$. Using the steepest descent approximation, the Euclidean action $\hat{I}$ in the wave function receives contributions mostly from the solution to the equations of motion, evaluated to be
		\begin{align}
			\hat{I} &\approx \frac{a^3}{2 i N} \left( f_{n l m} \frac{d f_{n l m}}{d \tau} - \frac{d \phi}{d \tau} g_{n l m} f_{n l m} \right) \notag \\
			&= \frac{1}{2} a^3 \left( f_{n l m} \frac{d f_{n l m}}{d t} - \frac{d \phi}{d t} g_{n l m} f_{n l m} \right).
		\end{align}
		We therefore have
		\begin{align}
			\Psi[f_{n l m}] \approx B_{n l m} \exp\left[ -\frac{1}{2} a^3 \left( f_{n l m} \dot{f}_{n l m} - \dot{\phi} g_{n l m} f_{n l m} \right) \right],
		\end{align}
		where the dots denote derivatives against $t$. The normalization can be fixed by requiring
		\begin{align}
			&|B_{n l m}|^2 \int_{-\infty}^{\infty} df_{n l m} \notag \\
			&\quad \left| \exp\left[ -\frac{1}{2} a^3 \left( f_{n l m} \dot{f}_{n l m} - \dot{\phi} g_{n l m} f_{n l m} \right) \right] \right|^2  = 1.
		\end{align}
		The expectation of the field perturbations averaged over the space is given by
		\begin{align}
			\langle \delta \Phi^2(t, \vec{x}) \rangle &= \frac{1}{2 \pi^2} \int d\chi d\theta d\varphi \; \sin^2\chi \sin\theta \notag \\
			&\quad\;\; \times \frac{3 \pi}{2} \sum_{n l m} \sum_{n' l' m'} \langle f_{n l m} f_{n' l' m'} \rangle Q_{n l m}  Q_{n' l' m'} \notag \\
			&= \frac{3}{4 \pi} \sum_{n l m} \langle f_{n l m}^2 \rangle,
		\end{align}
		where
		\begin{align}
			\label{eq:f2Integral} &\langle f_{n l m}^2 \rangle = |B_{n l m}|^2 \int_{-\infty}^{\infty} df_{n l m} \notag \\
			&\quad f_{n l m}^2 \left| \exp\left[ -\frac{1}{2} a^3 \left( f_{n l m} \dot{f}_{n l m} - \dot{\phi} g_{n l m} f_{n l m} \right) \right] \right|^2.
		\end{align}
		Defining the power spectrum, $P(n)$, as
		\begin{align}
			\langle \delta \Phi^2(t, \vec{x}) \rangle &= \sum_n \frac{1}{n} P(n),
		\end{align}
		with $l$ and $m$ summed over, we then find
		\begin{align}
			\label{eq:PowerSpectrum} P(n) = \frac{3 n}{4 \pi} \sum_{l, m} \langle f_{n l m}^2 \rangle.
		\end{align}
		Note that if $\langle f_{n l m}^2 \rangle$ depends only on $n$, the summation over $l$ and $m$ can be immediately carried out, leaving
		\begin{align}
			P(n) = \frac{3 n^3}{4 \pi} \langle f_{n}^2 \rangle.
		\end{align}
	
	To evaluate the expectation value $\langle f_{n l m}^2 \rangle$, we adopt the proposal of \cite{Laflamme1987}. We replace $\dot{f}_{n l m}$ by a combination of the canonical variable $f_{n l m}$ and its c-number value $\tilde{f}_{n l m}$,
		\begin{align}
			\dot{f}_{n l m} \rightarrow \frac{\dot{\tilde{f}}_{n l m}}{\tilde{f}_{n l m}} f_{n l m}.
		\end{align}
		When $\dot{\phi}$ or the metric perturbations $g_{n l m}$ are negligible, the wave function is then
		\begin{align}
			\Psi[ f_{n l m} ] = B_{n l m} \exp\left( -\frac{a^3 \dot{\tilde{f}}_{n l m}}{2 \tilde{f}_{n l m}} f_{n l m}^2 \right).
		\end{align}
		The normalization is evaluated as
		\begin{align}
			| B_{n l m} |^2 = \sqrt{\frac{a^3 \dot{\tilde{f}}_{n l m}}{\pi \tilde{f}_{n l m}}}.
		\end{align}
		The expectation value can then be found to be
		\begin{align}
			\label{eq:f2Exp} &\langle f_{n l m}^2 \rangle = \frac{\tilde{f}_{n l m}}{2 a^3 \dot{\tilde{f}}_{n l m}}.
		\end{align}


\section{Effect of mass on the power spectrum}
\label{sec:Solution}

	In the Euclidean space, we consider the scale factor solution \eqref{eq:scaleAEu} and a constant scalar field in the background. Neglecting the metric perturbations $g_n$ (we suppress the indices $l$ and $m$ in this section, since the equation of motion does not depend on them), we calculate the field perturbations $f_n$ (we ignore the tilde that denotes the c-number solution wherever no confusion arises) by numerically solving the equation of motion
		\begin{align}
			\label{eq:fEOMNoMetricEu} \frac{d^2 \hat{f}_n}{d \tau^2} + 3 H_0 \cot(H_0 \tau) \frac{d \hat{f}_n}{d \tau} - \left[ \tilde{m}^2 + \frac{(n^2 - 1) H_0^2}{\sin^2 (H_0 \tau)} \right] \hat{f}_n = 0,
		\end{align}
		where we use the hat to emphasize that it is the solution in the Euclidean space. In order to keep equation \eqref{eq:fEOMNoMetricEu} finite, we require that both $\hat{f}_n(\tau)$ and $\hat{f}_n'(\tau)$ vanish at $\tau = 0$. More precisely, we adopt the following ansatz as the initial condition for numerical calculations:
		\begin{align}
			\hat{f}_n(\tau_i) &= \frac{1}{2} \epsilon \tau_i^2, \\
			\dot{\hat{f}}_n(\tau_i) &= \epsilon \tau_i,
		\end{align}
		where $\tau_i \ll 1$ is the initial Euclidean time from which we start to integrate the differential equations, and $\epsilon$ is an arbitrary parameter. Note that since the expectation value $\langle f_n^2 \rangle$ depends only on the ratio $\tilde{f}_n / \dot{\tilde{f}}_n$, the power spectrum is independent of the choice of $\epsilon$. In our numerical calculation, we set $\tau_i = 10^{-4}$ and $\epsilon = 1$, and evolve the Euclidean system from $\tau_i$ to $\tau_f = \pi / 2 H_0$.
	
	In the Lorentzian spacetime, we use the analytical solution \eqref{eq:scaleALo} for the scale factor and a constant scalar field in the background to model the slow-roll inflation. The equation of motion for the field perturbation reads
		\begin{align}
			\label{eq:fEOMNoMetricLo} \frac{d^2 f_n}{d t^2} + 3 H_0 \tanh(H_0 t) \frac{d f_n}{d t} + \left[ \tilde{m}^2 + \frac{(n^2 - 1) H_0^2}{\cosh^2 (H_0 t)} \right] f_n = 0.
		\end{align}
		The boundary conditions connecting the Euclidean and Lorentzian solutions are the Cauchy-Riemann conditions \cite{Hwang2012, Hwang2012a, Hwang2014, Hwang2015, Chen2016a}
		\begin{align}
			\textrm{Re}\{ f(t_i) \} &= \textrm{Re}\{ \hat{f}(\tau_f) \}, \\
			\textrm{Im}\{ f(t_i) \} &= \textrm{Im}\{ \hat{f}(\tau_f) \}, \\
			\textrm{Re}\{ \dot{f}(t_i) \} &= -\textrm{Im}\{ \dot{\hat{f}}(\tau_f) \}, \\
			\textrm{Im}\{ \dot{f}(t_i) \} &= \textrm{Re}\{ \dot{\hat{f}}(\tau_f) \},
		\end{align}
		where we set $t_i = 0$ to be the initial time of integration in the Lorentzian space. We then solve the system from $t_i$ to the horizon-exit time,
		\begin{align}
			t_{\textrm{exit}} = \frac{1}{H_0} \sinh^{-1} n,
		\end{align}
		for mode $n$. Note that for each mode, the expectation value \eqref{eq:f2Exp}, hence the power spectrum \eqref{eq:PowerSpectrum}, is evaluated at its horizon-exit time.%
			\footnote{The power spectrum in general does not conserve at super-horizon scales, given that we are calculating the spectrum of the inflaton field perturbations in a closed FLRW universe, adopting the wave function interpretation to deduce the expectation values of the perturbations. The relation between this formulation and the approach of traditional quantum field theory still requires further clarification, which we leave to the future works.}


	The Hubble parameter in the Lorentzian space is
		\begin{align}
			H(t) = H_0 \tanh(H_0 t).
		\end{align}
		Therefore $H_0$ corresponds to the Hubble constant during the exponentially growing period. To fix the value of $H_0$, we consider the Hamiltonian constraint in Lorentzian space,
		\begin{align}
			\frac{\dot{a}^2}{a^2} = H^2 = \frac{8 \pi \sigma^2}{3} V(\Phi) - \frac{1}{a^2} + \frac{4 \pi}{3} \dot{\Phi}^2.
		\end{align}
		For the case that the scalar field is massless, the constant potential, $V(\Phi) = V_0$, drives the exponential growth of the scale factor $a$. The Hubble parameter is approximately
		\begin{align}
			H \approx \sqrt{ \frac{8 \pi \sigma^2}{3} V_0 }.
		\end{align}
		We choose the normalization of the metric to be
		\begin{align}
			\label{eq:sigmaConvention} \sigma^2 = \frac{1}{V_0}.
		\end{align}
		Therefore, during the exponential growth, $H \approx H_0 \approx \sqrt{8 \pi / 3}$.


	For the case of massive scalar field, during the exponential expanding period, the Hubble parameter is approximately
		\begin{align}
			H \approx \sqrt{ \frac{8 \pi}{3} \left( 1 + \frac{m^2 \Phi^2}{2 V_0} \right) }.
		\end{align}
		Note that $\tilde{m} = m / \sqrt{V_0}$ with the choice of $\sigma$ as \eqref{eq:sigmaConvention}.

\begin{figure}

\centering

\includegraphics[width = 14 cm]{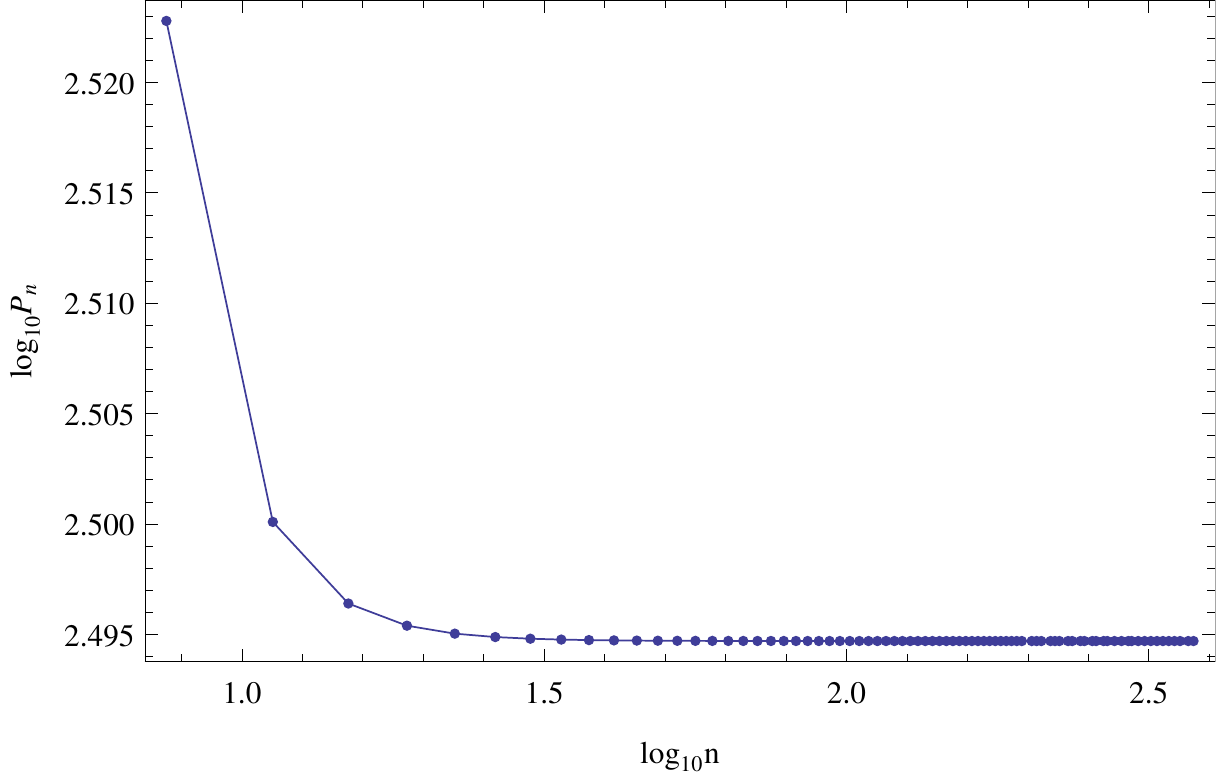}

\caption{The power spectrum obtained by numerically solving the perturbations with $\tilde{m} = 0$, $H_0 = \sqrt{8 \pi / 3}$.}

\label{fig:spectrum_0_1}

\end{figure}
\begin{figure}

\centering

\includegraphics[width = 14 cm]{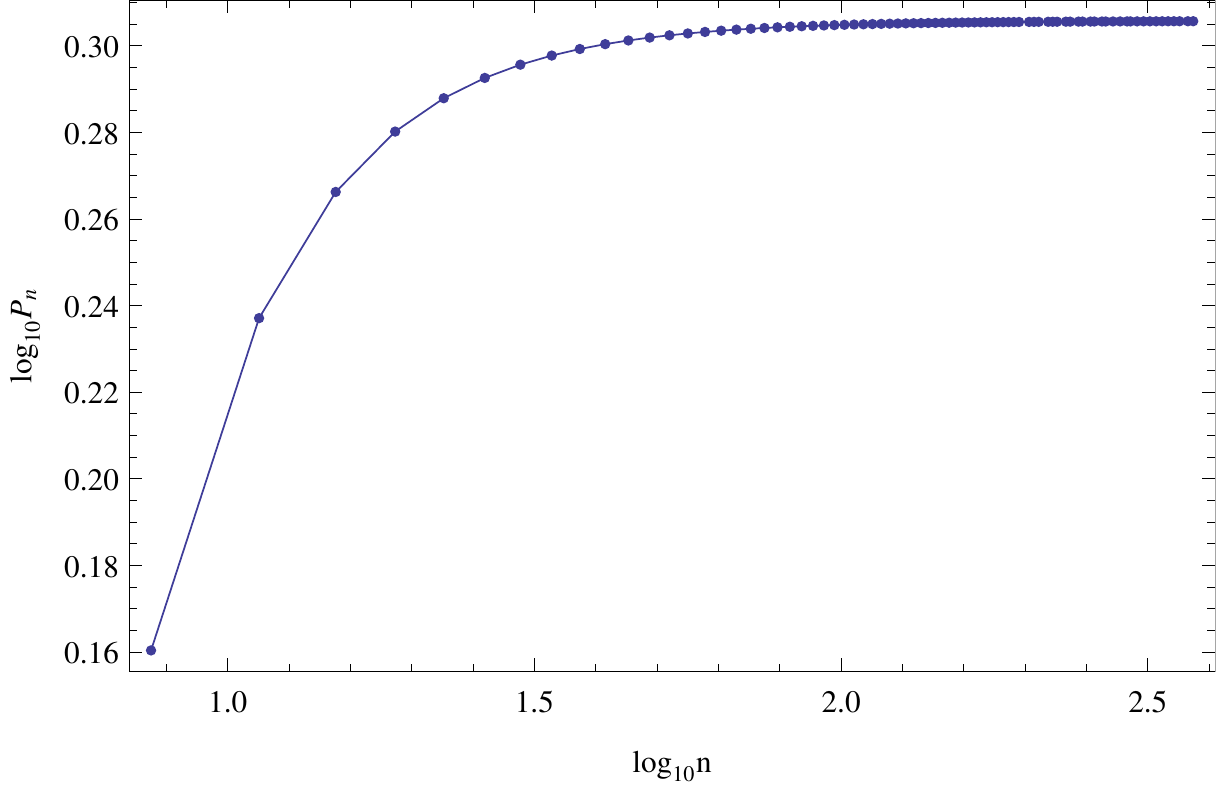}

\caption{The power spectrum obtained by numerically solving the perturbations with $\tilde{m} = 1000 \sqrt{0.1}$, $H_0 = \sqrt{8 \pi / 3}$.}

\label{fig:spectrum_1000_1}

\end{figure}
\begin{figure}

\centering

\includegraphics[width = 14 cm]{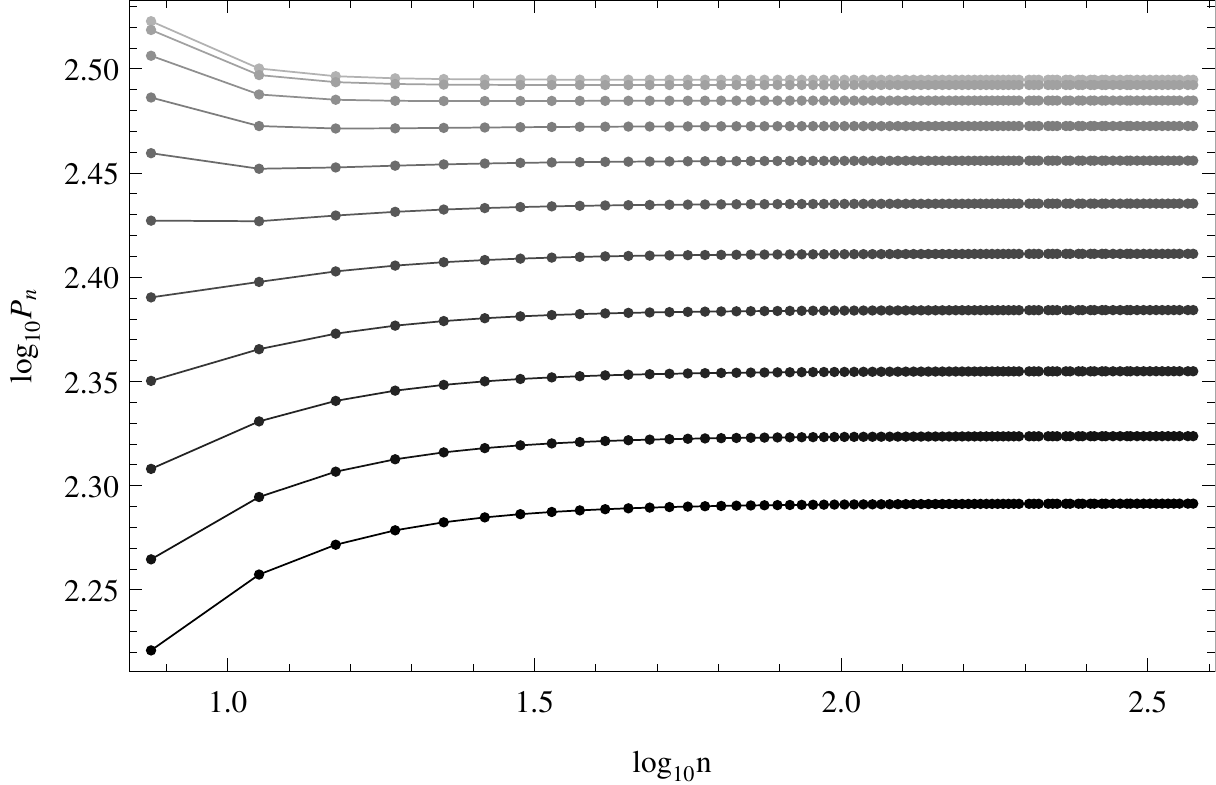}

\caption{The power spectrum obtained by numerically solving the perturbations with $\tilde{m} = \sqrt{0.1} \times \{ 0, 1, \dots, 10 \}$, from top to bottom. All spectra are plotted with $H_0 = \sqrt{8 \pi / 3}$.}

\label{fig:spectrum_m_1}

\end{figure}

	Figure \ref{fig:spectrum_0_1} shows the power spectrum in the massless case with $H_0 = \sqrt{8 \pi / 3}$. We see that while the power spectrum is scale-invariant in the small scales, it is enhanced in the large scales. Figure \ref{fig:spectrum_1000_1} is the power spectrum for a large mass $\tilde{m} = 1000 \sqrt{0.1}$ with $H_0 = \sqrt{8 \pi / 3}$. Opposed to the massless case, we see that in this massive case the large-scale spectrum is suppressed. In Figure \ref{fig:spectrum_m_1} we show the spectra corresponding to a range of masses, holding $H_0 = \sqrt{8 \pi / 3}$. We can observe the trend that, as the mass increases, the large-scale spectrum turns from being enhanced to being suppressed. We find that roughly the power is enhanced when $\tilde{m}$ is greater than $0.5 H_0$, and suppressed when $\tilde{m}$ is less than $0.5 H_0$.


\begin{figure}

\centering

\includegraphics[width = 14 cm]{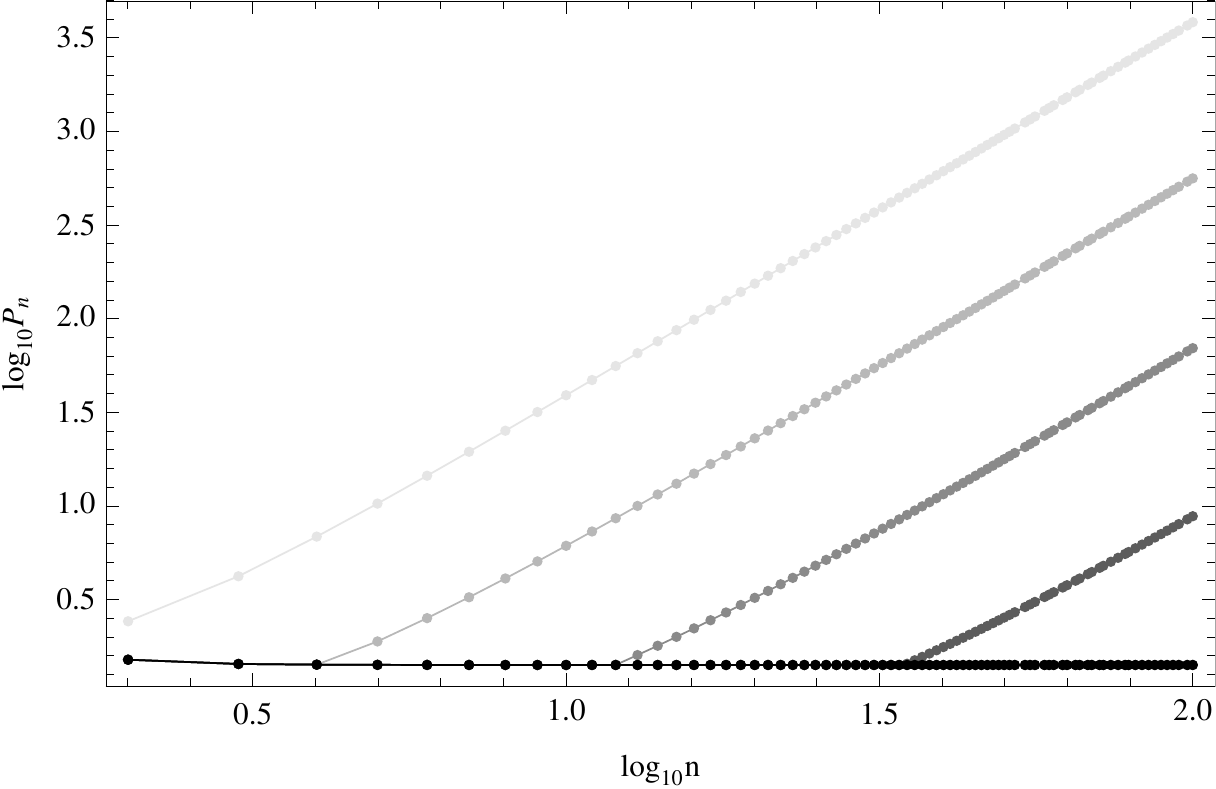}

\caption{The time evolution of power spectrum in the case of $\tilde{m} = 0$, $H_0 = \sqrt{8 \pi / 3}$. The darker curves correspond to the spectra at later times. The lightest curve is the initial Lorentzian spectrum at time $t_i$. For each $n$ mode, the power is evaluated up to its horizon crossing time.}

\label{fig:power_0_exit}

\end{figure}

\begin{figure}

\centering

\includegraphics[width = 14 cm]{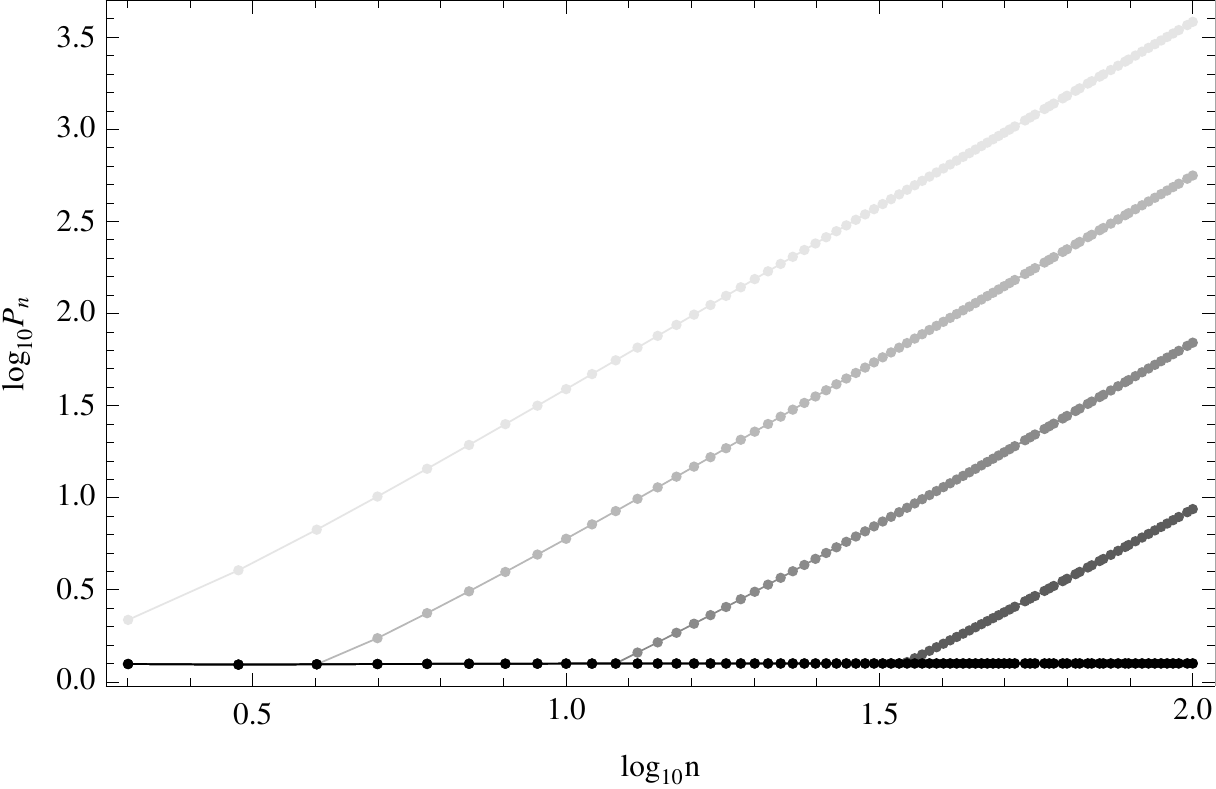}

\caption{The time evolution of power spectrum in the case of $\tilde{m} = 0.5 H_0$, $H_0 = \sqrt{8 \pi / 3}$. The darker curves correspond to the spectra at later times. The lightest curve is the initial Lorentzian spectrum at time $t_i$. For each $n$ mode, the power is evaluated up to its horizon crossing time.}

\label{fig:power_0p5H_exit}

\end{figure}

\begin{figure}

\centering

\includegraphics[width = 14 cm]{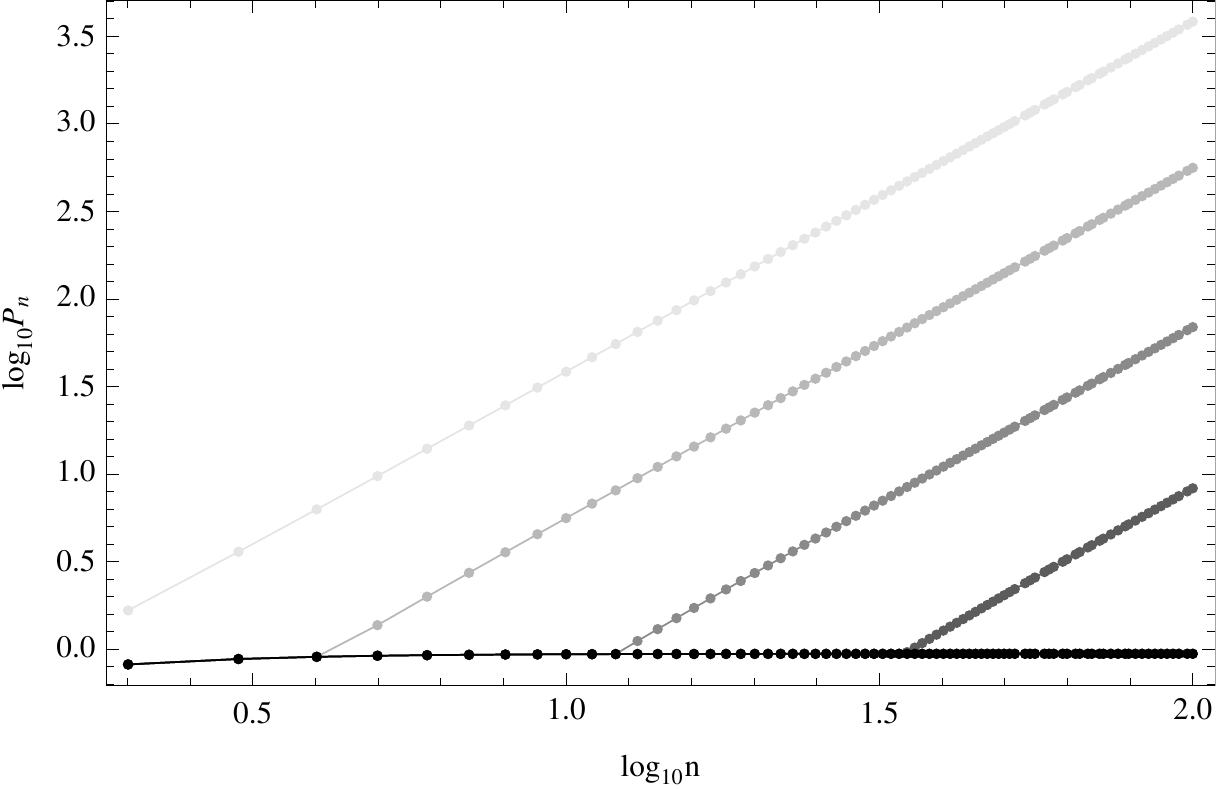}

\caption{The time evolution of power spectrum in the case of $\tilde{m} = H_0$, $H_0 = \sqrt{8 \pi / 3}$. The darker curves correspond to the spectra at later times. The lightest curve is the initial Lorentzian spectrum at time $t_i$. For each $n$ mode, the power is evaluated up to its horizon crossing time.}

\label{fig:power_H_exit}

\end{figure}

\begin{figure}

\centering

\includegraphics[width = 14 cm]{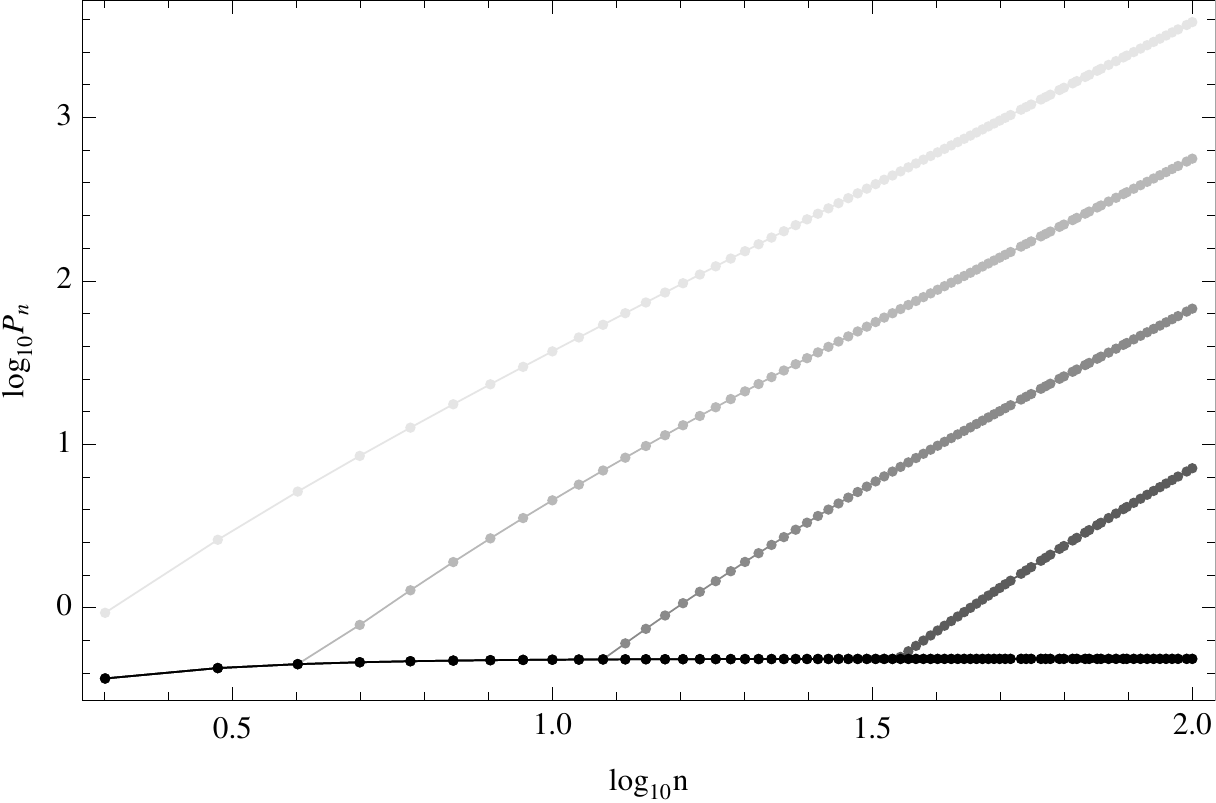}

\caption{The time evolution of power spectrum in the case of $\tilde{m} = 2 H_0$, $H_0 = \sqrt{8 \pi / 3}$. The darker curves correspond to the spectra at later times. The lightest curve is the initial Lorentzian spectrum at time $t_i$. For each $n$ mode, the power is evaluated up to its horizon crossing time.}

\label{fig:power_2H_exit}

\end{figure}

	To find out the mechanism that leads to this transition from enhancement to suppression as the mass increases, we first study the time evolution of the power spectrum in the Lorentzian space. The time evolution of spectrum in the massless case is given in Figure \ref{fig:power_0_exit}. For massive case, the time evolution of the spectra for the cases of $\tilde{m} = 0.5 H_0$, $H_0$, and $2 H_0$ is given in Figures \ref{fig:power_0p5H_exit} to \ref{fig:power_2H_exit}. 


	Through the spectrum evolution, we find that the power enhancement or suppression are reflected in the initial spectra in the Lorentzian space. At the small scales, before the horizon exit the slopes of the spectra are close to that of the spectrum of the Bunch-Davis vacuum. At the horizon crossing, the small-scale spectra are nearly scale-invariant. At the large scales, we see that at the horizon crossing the spectra is enhanced or suppressed determined by the mass of the scalar field as we showed before. Moreover, we note that even before the horizon crossing, already in the initial spectra in the Lorentzian space there are corresponding power enhancement or suppression relative to the small-scale Bunch-Davis vacuum. The origin of the power enhancement or suppression therefore lies on the Lorentzian initial condition, or, equivalently, on the Euclidean final spectrum.
	
	To find out the effect of mass on the Euclidean final spectrum, we note that the Euclidean equation of motion \eqref{eq:fEOMNoMetricEu} can also be analytically solved, yielding the solution
		\begin{align}
			\hat{f}_n(\tau) = A \frac{P_{\nu}^n[ \cos (H_0 \tau) ]}{\sin (H_0 \tau)},
		\end{align}
		where $A$ is an overall coefficient that has no effect on the final Euclidean spectrum,
		\begin{align}
			\nu = \frac{-1 + \sqrt{9 - 4 \tilde{m}^2 / H_0^2}}{2},
		\end{align}
		and we have picked the solution that is consistent with the no-boundary initial condition. When $\tilde{m}^2 / H_0^2 > 9 / 4$, $\nu$ and $\hat{f}_n(\tau)$ become complex. The power spectrum at the beginning of the Lorentzian time can be evaluated using the Euclidean solution at $\tau = \pi / 2 H_0$ through the boundary conditions. When evaluating the ratio $\tilde{f}_n / \dot{\tilde{f}}_n$ with complex $\tilde{f}_n$, we interpret it as the amplitude $| \tilde{f}_n / \dot{\tilde{f}}_n |$. We then have the initial power spectrum in the Lorentzian space as
		\begin{align}
			P(n) = \frac{3 n^3 H_0^2}{8 \pi} \left| \frac{P_{\nu}^n(0)}{P_{\nu}^n{}'(0)} \right|.
		\end{align}
		In the large-mass limit, we can intuitively understand the power suppression of the initial power spectrum induced by the mass term in the following way. In such a limit, the solution to the equation of motion \eqref{eq:fEOMNoMetricEu} roughly consists of an exponentially growing mode, $\exp(\tilde{m} \tau)$, and an exponentially decaying mode, $\exp(-\tilde{m} \tau)$. Hence, the amplitude $| \tilde{f}_n / \dot{\tilde{f}}_n |$ is roughly of the order of $1 / \tilde{m}$, which is suppressed by $\tilde{m} = m / \sqrt{V_0}$. Note that the large-mass limit actually lies beyond the linear regime of perturbations, and the purpose of considering it is only to provide an intuitive understanding. As shown in Figure \ref{fig:spectrum_m_1}, the long-wavelength spectrum is already suppressed as $\tilde{m}^2 / H_0^2$ is as small as roughly $0.1 \sqrt{6} / \sqrt{8 \pi / 3} \approx 0.43$. Therefore, it only requires a moderate mass to induce the effect of suppression.

\clearpage


\chapter{Conclusions and Discussions}{\label{ch:Conclusions}}

	In this thesis we explore the initial conditions to the inflationary universe using the long-wavelength suppression of the CMB spectrum as an important observational clue. We start by considering the toy model to the preinflationary era that consists of the NFTD of two kinds: (i) a NFDW that induces an accelerating rate smaller than that of the standard slow-roll inflation, and (ii) a NFCS that describes a universe expanding in a constant rate. We model such a  matter content for the very early universe as a kind of the generalized Chaplygin gas described by Eq.~(\ref{2m}) (see also Eq.~(\ref{1.1})), which gives a smooth transition from a NFTD dominated era to a de Sitter-like phase or a power-law inflationary era. We constrain our model using the  WMAP7 data for the power spectrum of the scalar perturbations, $P_s=2.45\times10^{-9}$, and the spectral index, $n_s=0.963$, at the pivot scale, $k_0=0.002$ Mpc$^{-1}$ \cite{Komatsu2011}. After fixing the parameters of the model and imposing the approximate initial vacuum states for the perturbations, we obtain the curvature power spectra for the cases of a NFDW and a NFCS, as shown in Figures \ref{DWS} and \ref{CSS}. The most important feature of the spectra is the drop of the power for the long-wavelength modes with $k \leq 0.002 $ Mpc$^{-1}$. Nevertheless, through a more detailed and systematic analysis presented later, the initial vacua are found to have different structures from the approximated ones used here. In particular, we will see that for the case of the preinflationary NFDW dominated era, the large-scale spectrum is actually enhanced, rather than suppressed.

	We then systematically investigate the power spectrum in the general background geometry. We study the curvature perturbations of a scalar field in the FLRW universe parameterized by the equation of state parameter $w$, and find that the large-scale spectrum at the end of inflation reflects the super-horizon spectrum of the initial state. We show that if the universe begins in the superinflation era ($w < -1$) or that with positive pressure ($w > 0$), the large-scale curvature perturbation spectrum is suppressed due to the blue-tilted super-horizon spectrum in the initial era. We first find the scaling relation of the super-horizon spectrum for a scalar field in the FLRW background with constant equation of state. At the large scales, the spectrum is blue-tilted for positive-pressure ($w > 0$) or superinflation ($w < -1$) era, and red-tilted for the era with $-1 < w < 0$, except the singular case with $w = -1/3$. In the slow-roll ($w \simeq -1$) and zero-pressure ($w = 0$) background, the super-horizon spectrum is scale-invariant. We also point out that the conclusions are drawn from assuming the mode function approaches the Minkowski limit at small scales. Although being natural in the accelerating universe, this assumption becomes \emph{a posteriori} in the decelerating universe since the sub-horizon modes are evolved from the super-horizon modes, which are initially across causally disconnected regions before entering the horizon.
	
	We develop the method of spectrum evolution to analyze how the expansion of the background geometry transforms the initial spectrum in the preinflationary era into the super-horizon spectrum in the inflationary era. By analyzing three scenarios:~a single slow-roll era, a slow-roll era preceded by a kinetic era, and two successive slow-roll eras connected by a kinetic era, we show the following two facts. First, the large-scale power suppression in the model with a single preinflation kinetic era stems from the blue-tilted super-horizon spectrum of the initial kinetic era. Second, the additional slow-roll era preceding the kinetic era changes the super-horizon initial spectrum, so the large-scale power is enhanced, rather than suppressed. These results show that the large-scale spectrum depends sensitively on the initial vacuum. In the universe beginning with the positive-pressure era, as we pointed out earlier, the super-horizon modes are initially across causally disconnected regions, and the well-motivated assumption on the initial state is still lacking. Some investigations about the effect of the different initial vacuum on the spectrum have been carried out in the literature \cite{Sriramkumar2005, Boyanovsky2006, Holman2008, Agullo2011}. We also explore the case that is free from the acausal issue: a superinflation era preceding the slow-roll era, and show that the large-scale spectrum is suppressed due to the initially blue-tilted spectrum in the superinflation era.
	
	To show that the pattern of evolution we obtain through the single-field analysis also applies to the two-field systems, we calculate the curvature perturbation and CMB spectra of a two-stage inflation model from the given two-field potential. We show that the large-scale power is enhanced due to the initial spectrum set in the first accelerating era, and the effect of the intermediate decelerating era on the spectrum is connecting the two plateaus generated in the two accelerating eras, which agrees with the picture obtained through the \emph{ad hoc} single-field analysis.
	
	In view of the shortcomings of the positive-pressure era and the superinflation era as consistent initial conditions to the universe, we explore other possibilities provided by the quantum gravitational theory. Particularly, we investigate the power spectrum of perturbations with the no-boundary wave function \cite{Hartle1983} as the initial condition of the inflationary universe. We have relied on very conservative approaches, including the canonical quantization \cite{DeWitt1967}, the Euclidean path integral approach and the steepest descent approximation \cite{Hartle1983}, and the use of instantons at the background as well as perturbation levels \cite{Halliwell1985}, which are consistent with traditional techniques of quantum field theory in several regimes \cite{Hwang2013}. We find that the inflationary universe is approximately scale-invariant at the short-wavelength scales, while the power spectrum of the pure de Sitter space is enhanced at the long-wavelength scales. We also find that the power spectrum can be either enhanced or suppressed due to the detailed choice of the potential; for example, the mass term of the inflaton field. In particular, as long as the inflaton is moderately massive, the long-wavelength spectrum is suppressed. This opens a possibility that the power suppression is indeed a hint to that our universe starts from an instanton with a massive inflaton field that approximates the Hartle-Hawking wave function.
	
	This line of exploration definitely needs more work. It will be interesting to see more detailed calculations for other realistic inflationary scenarios; e.g., the Starobinsky-type inflation models \cite{Starobinsky1980}. Also, we investigated for compact and homogeneous instantons, but there are other instantons that also explain the origin of our universe; e.g., the Coleman-De Luccia instantons \cite{Coleman1980} or the Euclidean wormholes \cite{Chen2016b, Kang2017, Chen2017}. Another brave question is: what is the relation between the big bounce model of the loop quantum cosmology \cite{Ashtekar2015} and the Hartle-Hawking wave function \cite{Hartle1983}? Both approaches explain the power suppression, but it is yet unclear which one is more suitable as the model of the beginning of our universe. We leave these interesting issues as future research topics.

\clearpage




\addcontentsline{toc}{chapter}{\bf Bibliography}
\bibliographystyle{JHEP}
\bibliography{Cosmology}

\clearpage

\end{document}